\documentclass[superscriptaddress,prl,twocolumn,tightenlines,10pt,a4paper]{revtex4}%

\usepackage{amsthm}

\usepackage{bbm}
\usepackage{pbox}
\usepackage{amsmath}
\usepackage{amsfonts}
\usepackage{amssymb}
\usepackage{graphicx}
\usepackage{subfigure}
\usepackage{savesym}
\usepackage{amsmath}
\usepackage{txfonts}
\usepackage{multirow}
\usepackage{epstopdf}
\usepackage{color}
\usepackage{float}
\usepackage{dsfont}
\usepackage{hyperref}

\usepackage{pgffor}
\usepackage{pdfpages}

\usepackage[T1]{fontenc}

\usepackage{braket}
\begin{document}

\title{Efficient tomography of a quantum many-body system}

\author{B. P. Lanyon\normalfont\textsuperscript{$\dagger$}}
\email{ben.lanyon@uibk.ac.at}

	\affiliation{Institut f\"ur Quantenoptik und Quanteninformation,\\
	\"Osterreichische Akademie der Wissenschaften, Technikerstr. 21A, 6020 Innsbruck,
	Austria}
	\affiliation{
	Institut f\"ur Experimentalphysik, Universit\"at Innsbruck,
	Technikerstr. 25, 6020 Innsbruck, Austria}

\author{C. Maier}
\thanks{These authors contributed equally to this work.}

	\affiliation{Institut f\"ur Quantenoptik und Quanteninformation,\\
	\"Osterreichische Akademie der Wissenschaften, Technikerstr. 21A, 6020 Innsbruck,
	Austria}
	\affiliation{
	Institut f\"ur Experimentalphysik, Universit\"at Innsbruck,
	Technikerstr. 25, 6020 Innsbruck, Austria}
	
\author{M. Holz\"apfel}

	\affiliation{Institut f\"ur Theoretische Physik and IQST, Albert-Einstein-Allee 11, Universit\"at Ulm, 89069 Ulm, Germany}
	
\author{T. Baumgratz}

	\affiliation{Institut f\"ur Theoretische Physik and IQST, Albert-Einstein-Allee 11, Universit\"at Ulm, 89069 Ulm, Germany}
	\affiliation{Clarendon Laboratory, Department of Physics, University of Oxford, Oxford OX1 3PU, United Kingdom}
\affiliation{Department of Physics, University of Warwick, Coventry CV4 7AL, United Kingdom}

\author{C. Hempel}

	\affiliation{
	Institut f\"ur Experimentalphysik, Universit\"at Innsbruck,
	Technikerstr. 25, 6020 Innsbruck, Austria}

	\affiliation{ARC Centre for Engineered Quantum Systems, School of Physics, The University of Sydney, Sydney, New South Wales 2006, Australia.}
	
\author{P. Jurcevic}
 
	\affiliation{Institut f\"ur Quantenoptik und Quanteninformation,\\
	\"Osterreichische Akademie der Wissenschaften, Technikerstr. 21A, 6020 Innsbruck,
	Austria}
	\affiliation{
	Institut f\"ur Experimentalphysik, Universit\"at Innsbruck,
	Technikerstr. 25, 6020 Innsbruck, Austria}

\author{I.~Dhand}

		\affiliation{Institut f\"ur Theoretische Physik and IQST, Albert-Einstein-Allee 11, Universit\"at Ulm, 89069 Ulm, Germany}

\author{A. S. Buyskikh}

	\affiliation{Department of Physics and SUPA, University of Strathclyde, Glasgow G4 0NG, UK}

\author{A. J. Daley}

	\affiliation{Department of Physics and SUPA, University of Strathclyde, Glasgow G4 0NG, UK}

\author{M. Cramer}

	\affiliation{Institut f\"ur Theoretische Physik and IQST, Albert-Einstein-Allee 11, Universit\"at Ulm, 89069 Ulm, Germany}
	\affiliation{Institut f\"ur Theoretische Physik, Leibniz Universit\"at Hannover, Hannover, Germany}

\author{M. B. Plenio}

	\affiliation{Institut f\"ur Theoretische Physik and IQST, Albert-Einstein-Allee 11, Universit\"at Ulm, 89069 Ulm, Germany}	

\author{R. Blatt}

\affiliation{Institut f\"ur Quantenoptik und Quanteninformation,\\
	\"Osterreichische Akademie der Wissenschaften, Technikerstr. 21A, 6020 Innsbruck,
	Austria}
	\affiliation{
	Institut f\"ur Experimentalphysik, Universit\"at Innsbruck,
	Technikerstr. 25, 6020 Innsbruck, Austria}

\author{C. F. Roos}

	\affiliation{Institut f\"ur Quantenoptik und Quanteninformation,\\
	\"Osterreichische Akademie der Wissenschaften, Technikerstr. 21A, 6020 Innsbruck,
	Austria}
	\affiliation{
	Institut f\"ur Experimentalphysik, Universit\"at Innsbruck,
	Technikerstr. 25, 6020 Innsbruck, Austria}

\date{\today}

\begin{abstract}

\noindent Quantum state tomography (QST) is the gold standard technique for obtaining an estimate for  the state of small quantum systems in the laboratory \cite{PhysRevA.40.2847}. 
Its application to systems with more than a few constituents (e.g. particles) soon becomes impractical as the effort required grows exponentially in the number of constituents. 
Developing more efficient techniques is particularly pressing as precisely-controllable quantum systems that are well beyond the reach of QST are emerging in laboratories. 
Motivated by this, there is a considerable ongoing effort to develop new characterisation tools for quantum many-body systems  
 \cite{Cramer:2010fk, %cramer MPS tomo
PhysRevLett.105.150401, %Eisert compressed sensing
Flammia:2011, %flammia DFE
daSilva:2011, %Silva DFE
Baumgratz:2013, %scalable reconstruction of rhos..
baumgratzMaxLikeli2013, %scalable max likelihood
PhysRevLett.105.250403, %permutationally invariant
PhysRevLett.111.147205, %Demler ramsey sepctroscopy
Senko430, %Monroe spectroscopy
PhysRevLett.115.100501, %our spectroscopy
PhysRevLett.106.100401,  %exp. demo of compressed sensing
Steffens:2015}. %Towards experimental quantum-field tomography with ultracold atoms
Here we demonstrate Matrix Product State (MPS) tomography \cite{Cramer:2010fk}, which is theoretically proven to allow the states of a broad class of quantum systems to be accurately estimated with an effort that increases efficiently with constituent number. 
We first prove that this broad class includes the out-of-equilbrium states produced by 1D systems with finite-range interactions, up to any fixed point in time.
We then use the technique to reconstruct the dynamical state of a trapped-ion quantum simulator comprising up to 14 entangled spins (qubits): a size far beyond the reach of QST. 
Our results reveal the dynamical growth of entanglement and description complexity as correlations spread out during a quench: a necessary condition for future beyond-classical performance. 
MPS tomography should find widespread use to study large quantum many-body systems and to benchmark and verify quantum simulators and computers. 

\end{abstract}

\maketitle

\newpage

\noindent An MPS \cite{Schollwock:2011qd} is an efficient representation of a quantum state that makes use of the presence of short-ranged quantum correlations in typical states to avoid expressing the wave function in a basis that spans the full Hilbert space.
While the MPS description can be exact given a large enough matrix dimension (exponentially large in the number of system components), for a broad class of entangled many-body states it offers an accurate description with a number of parameters that increases only polynomially in system components. 
The complexity of an MPS is determined by the amount of entanglement in the system it describes, as quantified in \cite{RevModPhys.82.277, PhysRevLett.100.030504}. 
If the entanglement grows, the MPS can be expanded to maintain an accurate description. The MPS formalism underpins some of the most successful classical algorithms for describing the states and dynamics of interacting many-body quantum systems \cite{Schollwock:2011qd}. In this work we demonstrate how it simplifies the goal of characterising the state of a quantum system in the laboratory. \

MPS tomography recognises both that the kinds of states typically found in physical systems can be efficiently described as an MPS, and that the information required to identify them in the laboratory is accessible locally; that is, by making measurements only on subsets of particles that lie in the same neighbourhood. In such cases, the number of measurements required to identify the state scales only linearly in system components and the processing time scales only polynomially  \cite{Cramer:2010fk, Baumgratz:2013}. Crucially, MPS tomography makes no prior assumptions about the form of the state, underlying dynamics, Hamiltonian or temperature, because the state estimate can be certified: an assumption-free lower bound on the fidelity with the lab state is provided \cite{Cramer:2010fk}. 

States particularly well suited to MPS tomography include those where there is a maximum distance over which significant correlations exist between the constituents (locally-correlated states). Examples of such states include the 2D cluster states---universal resource states for quantum computing---as well as the ground states of 1D systems with short-range interactions (where particles interact far more strongly with their neighbours, than those farther away) \cite{fannes1992, PhysRevB.73.085115, Brandao2013}. 
We find that MPS tomography is also well-suited to characterise out-of-equilibrium states produced after finite evolution times in systems with finite-ranged interactions (most naturally-occurring interaction mechanisms have this short-range character). In such a setting, Lieb-Robinson bounds imply exponentially decreasing correlations with distance, ensuring the existence of an efficient MPS representation of the state (corollary 3 of \cite{Brandao2013}, see also \cite{Brandao:2015qy}). 
%That such an MPS representation can also be found and certified by local measurements for 1D systems is proven in the Supplementary Material. The underlying intuition is now described. 
Once such an MPS representation has been found using MPS tomography~\cite{Cramer:2010fk,baumgratzMaxLikeli2013}, it can be certified by local measurements for 1D systems, as is proven in the Supplementary Material. The underlying intuition is now described. 

Consider an $N$-component quantum system in a simple product state (or other locally-correlated state). Interactions are then turned on (a quench), causing the system to evolve into many-body entangled states. In the presence of finite-range interactions (e.g. nearest-neighbour only), information and correlations spread out in the system with a strict maximum group velocity \cite{Lieb1972, Nachtergaele2010, Cheneau2012}. Therefore, after a finite evolution time there is  a maximum distance over which correlations extend in the system (the \emph{correlation length, $L$}), beyond which correlations decay exponentially in distance. 
The information required to describe the state is largely contained in the \emph{local reductions}: the reduced states (density matrices) of all groups of neighbouring particles contained within $L$. In 1D systems, such locally-correlated states can be described by a compact MPS and, to identify the total $N$-component MPS, all the experimentalist need do is perform the measurements required to reconstruct the local reductions (see Supp. Mat.). Each local reduction can be determined by full QST, requiring measurements in at most $3^L$ bases. Since the number of local reductions increases only linearly in $N$ for a 1D system, the measurement number is efficient in this parameter. For 2D systems it is not yet known if a general efficient MPS description of locally-correlated states always exists \cite{Brandao2013}. 

After estimating the local reductions, the experimental work is done. The estimates are passed to a classical algorithm which finds an MPS estimate %for the total $N$-component state 
in a time polynomial in $N$ \cite{Cramer:2010fk, baumgratzMaxLikeli2013}. Finally, a certificate for the overlap between the MPS estimate and the lab state is efficiently calculated (Supp. Mat.). The correlation length $L$ need not be known \emph{a priori}. 
%, although there may well be practical ways to determine it independently. 
If the certified fidelity is deemed not high enough after measurements on any chosen number of sites ($k$), then one can try again, this time making measurements over larger $k$. 
Therefore, we have a technique to obtain a reliable estimate for the ground and dynamical states of `local' quantum systems, that is efficient in system-component number $N$.  \footnote{In the case of system interactions that decay slower than exponential with distance, information and correlations are not restricted to travel at a strict maximal velocity and there is no a priori guarantee that a locally correlated state is generated or, more generally, that the reductions over some length uniquely define the global state. However, one can still carry out MPS tomography and see if it does provide a useful (certified) description. Indeed, our experiments involve interactions that fall of slower than exponential, and MPS tomography still provides a useful description.}. 
A conceptual example of the generation and characterisation of locally-correlated states in 1D is presented in Figure 1.

There is a connection then, between the interaction-range in a quantum system and the ability to guarantee an efficient characterisation of its dynamic states, as the system size is scaled up. Our strategy is not restricted to 1D systems or systems with strictly finite-interactions. While the detailed conditions under which an efficient MPS (or PEPS) description is known to exist or not to exist are not well known, it is a strength of our algorithm that it comes with a certificate that bounds the quality of the estimate and importantly, alerts us to a failure of the reconstruction if we have chosen a block that was too small. 
How slowly interactions can fall off with distance, before the picture of a local propagation of correlations breaks down has recently been extensively studied \cite{hastings2006, PhysRevLett.111.260401, PhysRevLett.111.207202, PhysRevLett.113.030602, PhysRevX.3.031015}.

 \begin{figure}[t]
\vspace{0mm}
  \begin{center}
    \includegraphics[width=1\columnwidth]{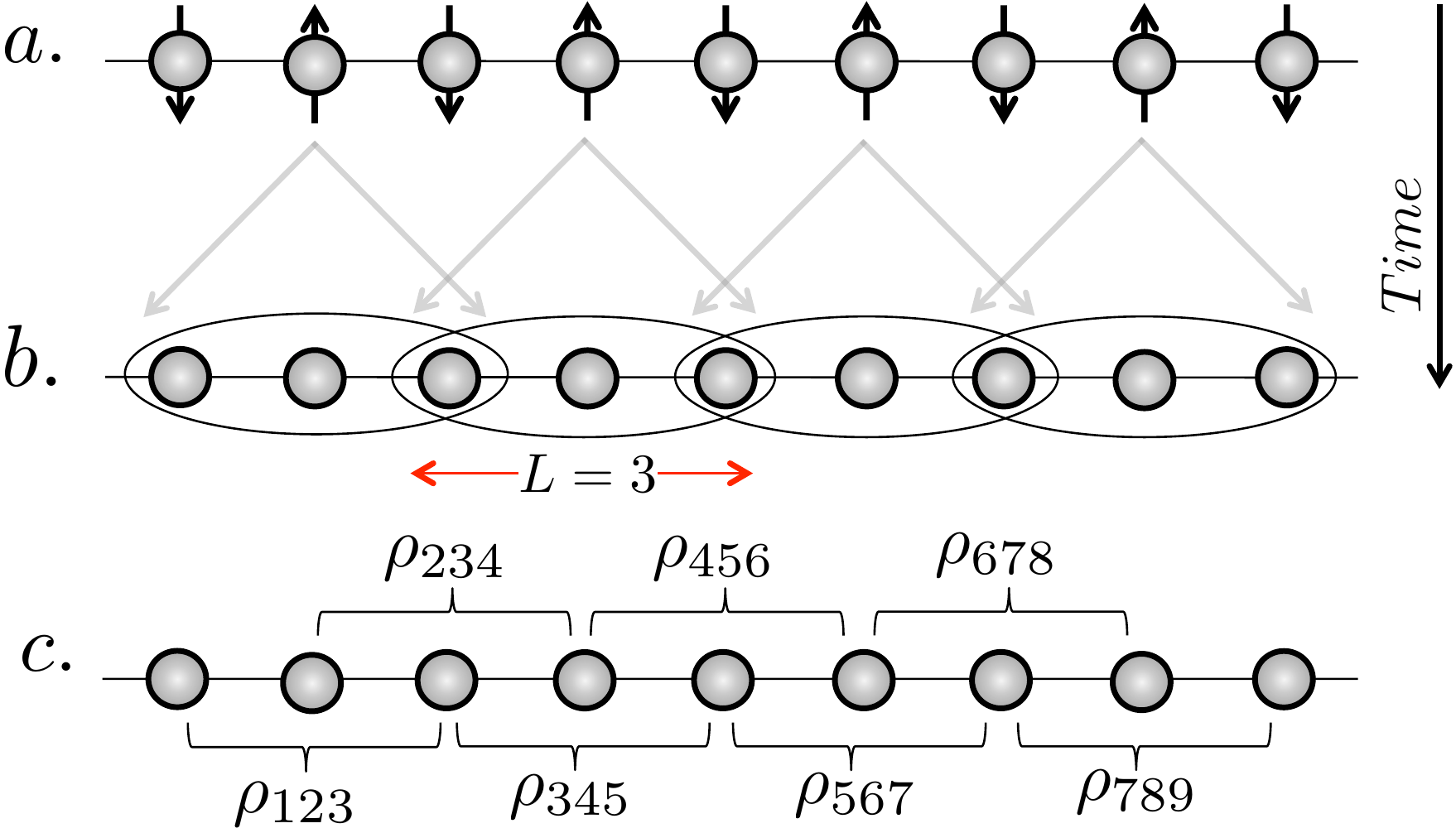}
 %   \vspace{-2mm}
   \caption{\textbf{Generation and characterisation of locally-correlated quantum states}. 
 \textbf{a}. Quantum spins fixed on a 1D lattice initialised into some separable pure state. Finite-range spin-spin interactions are then abruptly turned on. In the subsequent dynamics, quantum correlations spread out with a maximum group velocity, producing light-like cones (grey arrows, only a few are shown) and a locally-correlated entangled state. 
\textbf{b}. After the particular evolution time shown, quantum correlations have spread to neighbouring spin triplets (not all shown). The established correlation length is $L=3$. The total $N$-spin state can be accurately described by a compact MPS, efficient in N. The correlation length increases linearly in time. 
\textbf{c}. To identify the state in the laboratory, the experimentalist need only perform sufficient measurements to reconstruct all $N-L+1$ neighbouring spin triplet reduced density matrices. The experimental effort therefore increases linearly in spin number $N$.  
Generalisation to higher spatial dimensions and to mixed-state estimates using matrix product operators \cite{Baumgratz:2013,baumgratzMaxLikeli2013} are possible, although no general certification method is currently known for mixed states \cite{Kim2014}.
}
   \label{figure1}
  \vspace{-8mm}
  \end{center}
 \end{figure}

MPS tomography is not generally efficient in the system evolution (quench) time. For finite range interactions, the correlation length $L$ can increase at most linearly in time as entanglement grows and spreads out in the system, demanding exponentially growing measurements to estimate each local reduction \cite{PhysRevLett.97.150404, PhysRevLett.97.050401}. 
This puts practical limits on the evolution time until which the system state can be efficiently characterised via MPS tomography: once correlations have spread out over the whole system the effort becomes the same as full QST.  
Ultimately, this ``failure'' of MPS tomography during quench dynamics due to entanglement growth is exactly what is desired in a quantum simulator (or quantum computer):  
if it is possible to reconstruct the state of a pure quantum system all the way through its dynamical evolution, then it cannot be doing anything beyond the capabilities of a classical computer \cite{PhysRevLett.91.147902}. 
MPS tomography is therefore a powerful tool to benchmark quantum dynamics and to verify evolution towards classically-intractable regimes. 
A signature of the latter would be that, as the system evolves, the size of the local reductions required to obtain an accurate pure MPS description would continue to increase. 

Our experimental system (quantum simulator) consists of a string of trapped $^{40}$Ca$^{+}$ ions. In each ion $j{=}1\dots N$, two electronic states represent a spin-1/2 particle. Under the influence of laser-induced forces, the spin interactions are well described by an `XY' model in a large transverse field, with Hamiltonian $H_{XY}{=}\hbar \sum_{i<j}J_{ij}(\sigma_i^+\sigma_j^-{+}\sigma_i^-\sigma_j^+)+B\sum_{j}\sigma_j^z$. Here $J_{ij}$ is an $N \times N$ spin-spin coupling matrix, $\sigma_i^{+}$ ($\sigma_i^{-}$) is the spin raising (lowering) operator for spin $i$ and $\sigma_j^z$ is the Pauli $Z$ matrix for spin $j$. All spins down $\ket{\downarrow_z}^{\otimes N}$ is the ground state, spins pointing up $\ket{\uparrow_z}$ are the quasiparticle excitations in the system \cite{PhysRevLett.115.100501}. Interactions reduce approximately with a power-law $J_{ij}\propto1/|i-j|^{\alpha}$ with distance $|i-j|$. Here  $1.1{<}\alpha{<}1.6$, for which the predominant feature of spreading wave packets of correlations is evident \cite{jurcevicquasi, PhysRevLett.115.100501, PhysRevLett.111.207202}. 

MPS tomography is applied to quench dynamics, starting from the initial antiferromagnetic N\'eel-ordered product state $\ket{\Phi(0)}=\ket{\uparrow,\downarrow,\uparrow,\downarrow,\dots}$. This highly excited initial state ($N/2$ excitations) leads to the emergence of locally-correlated entangled states involving all $N$ particles and evolves in a subspace whose size, contrary to those of low-excitation subspaces \cite{jurcevicquasi}, grows exponentially with $N$.
After preparing $\ket{\Phi(0)}$ with a spatially-steerable laser, focused on a single ion, spin interactions are abruptly turned on (a quench) and then off after a desired evolution time t, freezing the generated state and allowing for spin measurement. The ideal model state is $\ket{\Phi(t)}=\exp(-i H_{XY}t)\ket{\Phi(0)}$.
Through repeated state preparation and measurement, estimates of the expectation values for local $k$-spin observables are obtained. 
For example, to estimate each of the $N{-}2$ local reductions of neighbouring $k=3$ sites (spin triplets),  measurements in $3^k=27$ different bases are carried out.
The results are input into a combination of two efficient MPS tomography algorithms~\cite{Cramer:2010fk, baumgratzMaxLikeli2013}, which output an initial MPS estimate for the simulator state $\rho_{lab}$. 
Finally, a \emph{certified} MPS state estimate $\ket{\Psi^{k}_{c}}$ is found. The lower bound on the fidelity of this state with the actual state in the laboratory is given by $F_c^k$, i.e. $\bra{\Psi^k_{c}}\rho_{lab}\ket{\Psi^k_{c}}\geq F^k_{c}$ (see Supp. Mat.).

The largest application of full QST was for an 8 qubit W-state \cite{Haffner:2005jk}, for which measurements  were made in 6561 different bases taken over a period of ten hours \footnote{Entangled W-states \cite{Haffner:2005jk} and entangled GHZ-states \cite{PhysRevLett.106.130506} are very simple to describe: the number of non-zero probability amplitudes is small in both cases, and remains small for any system size. The entangled states generated in the quench experiments presented here are far more complicated: the number of non-zero probability amplitudes is significantly larger and grows exponentially in system size.}. 
We begin experiments with 8 spin (qubit) quench dynamics, and reconstruct 8-spin entangled states via MPS tomography, using measurements in only 27 bases taken over a period of around ten minutes. Local measurements are performed to reconstruct all $k$-local reductions of individual spins ($k=1$), neighbouring spin pairs ($k=2$) and spin triplets ($k=3$), at various simulator evolution times. The results of these measurements directly reveal important properties. Single-site `magnetisation' shows how spin excitations disperse and then partially refocus (Figure 2a). In the first few ms, strong entanglement is seen to develop in all neighbouring spin pairs and triplets, then later reducing, first in pairs then in triplets, consistent with correlations spreading out across larger numbers of spins in the system (Figure 2c-d).

 \begin{figure}[t]
%\vspace{-5mm}
  \begin{center}
        \includegraphics[width=1\columnwidth]{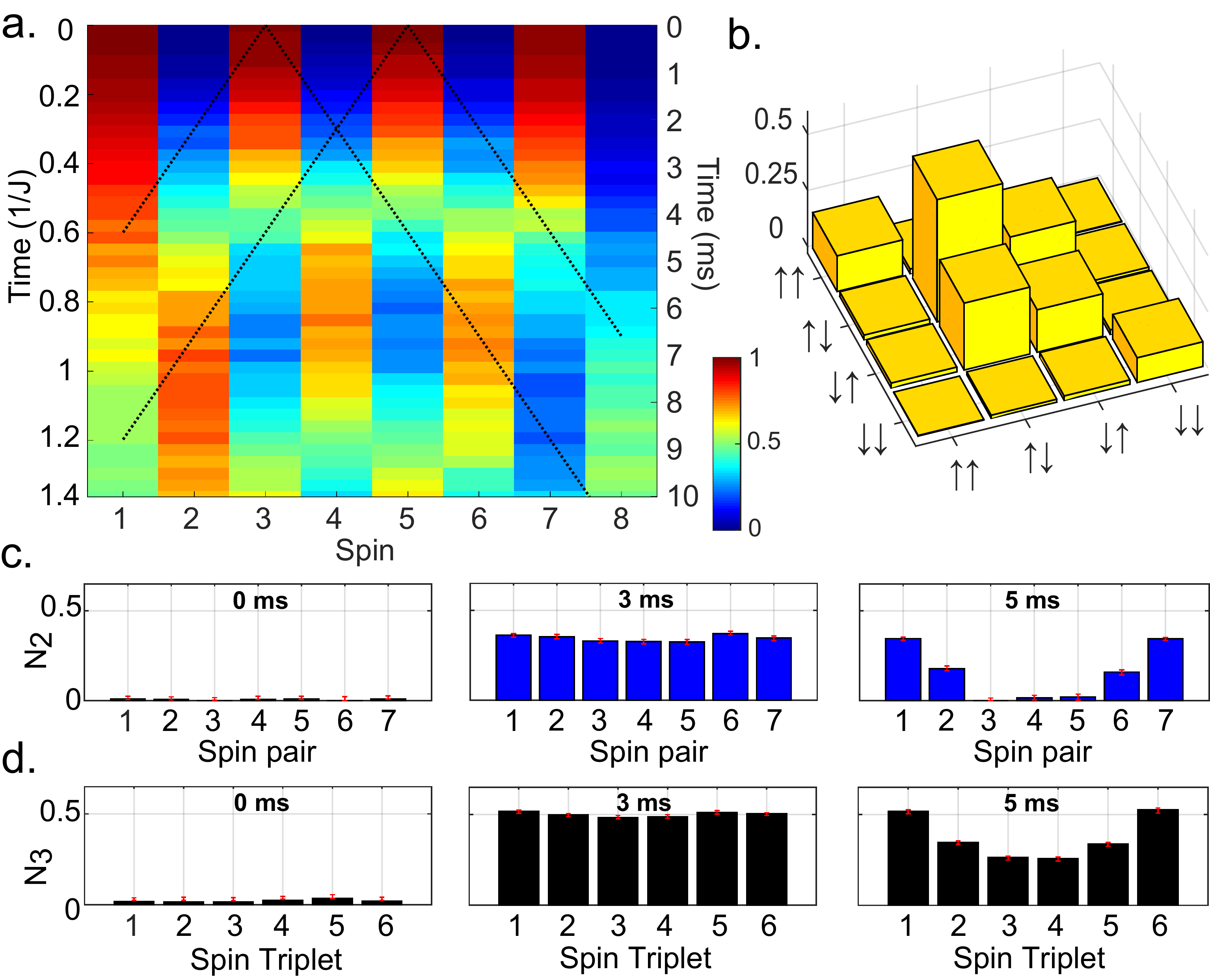}
   \vspace{-7mm}
   \caption{\textbf{Local measurement results for an 8 spin system}. 
   \textbf{a}. Single site magnetisation: Probability of finding a spin up at each site, during quench dynamics. The interaction range $\alpha\approx 1.6$. Lefthand time axis is renormalised by the average nearest-neighbour $J$  couplings. Two light-like cones are shown, exemplifying an estimate for the maximum speed at which correlations spread (see Supp. Mat.). 
   \textbf{b}. Density matrix (absolute value) of spins 3 \& 4 at time of $3$ ms, reconstructed via QST. The state is entangled, with a bipartite negativity of $N_2=0.31 \pm{0.01}$ and a fidelity with an ideal theoretical model of over $0.99$. 
   \textbf{c.-d.} Entanglement in all neighbouring spin pairs (\textbf{c}.) and spin triplets (\textbf{d.}) at three evolution times, as labelled: values calculated from measured density matrices (e.g. panel \textbf{b.}). The entanglement measure is bipartite negativity $N_2$ (tripartite negativity $N_3$) for spin pairs (triplets). $N_3$ is the geometric mean of all three bipartite negativity splittings. 
} 
   \label{figure2}
\vspace{-7mm}
  \end{center}
 \end{figure}

Fidelity lower bounds $F^k_c$ from MPS tomography during the 8-spin quench are shown in Fig 3a. 
The results closely match an idealised model: MPS tomography applied to the exact local reductions of the ideal states $\ket{\Phi(t)}$. The differences between model and data are largely due imperfect knowledge of local reduction due to the finite number of measurements used in experiments (Projection noise, see Supp. Mat.). 
Measurements on $k{=}1$ sites at $t=0$ provides a certified MPS state reconstruction $\ket{\Psi^1_c}$, with $F^1_c =  0.98\pm{0.01}$ and $| \langle \Psi^1_c|\Phi(0)\rangle |^2=0.98$, proving that the system is initially well described by a pure product N\'eel state (Figure 3a). 
The fidelity lower bounds based on single-site measurements rapidly degrade as the simulator evolves, falling to 0 by $t=2$~ms. 
Nevertheless, an accurate pure-state description is still achieved by measuring on larger ($k=2$) and larger ($k=3$) reduced sites (Figure 3a). 
The model fidelity bounds $F^3_c$ begin to drop after $t=2$~ms, consistent with the time at which the information wavefronts are expected to reach next-nearest-neighbours (light-like cones, Figure 2a), allowing for correlations beyond 3 sites to develop.
Measurements on $k=3$ sites reveal an MPS description with more than 0.8 fidelity up to $t = 3$~ms, before rapidly dropping to 0 at 6~ms. This is consistent with the model and the entanglement properties measured  directly in the local reductions (Figure 2b-c): At $t=3$~ms entanglement in spin triplets maximises, before reducing to almost zero at 6ms as correlations have spread out to include more distant spins. In this case, 3-site local reductions are not sufficient to uniquely distinguish the global state. Note, even if $F^k_c=0$, the MPS estimates $\ket{\Psi^k_c}$ can still be an accurate description of the lab state ($F^k_c$ are only lower bounds).  

The data in Figure~3a clearly reveal the generation and spreading-out of entanglement during simulator evolution up to 3 - 4~ms, and are consistent with this behaviour continuing beyond this time. To confirm this, it would be necessary to measure on increasingly large numbers of sites, demanding measurements that grow exponentially in $k$. That the amount of entanglement in the simulator is growing in time can be seen from the inset in figure 3a: the half-chain entropies of the certified MPSs $\ket{\Psi^3_c}$ are seen to grow as expected for a sudden quench, closely following that in ideal model states $\ket{\Phi(t)}$. For all times at which $F^3_c>0$ (except $t=0$), the pure MPS-reconstructed states $\ket{\Psi^3_c}$ are non-separable across all partitions. 

Figure 3b-c compares spin-spin correlations (`correlation matrices') present in $\ket{\Psi^3_c}$ at $t=3$~ms ($F^3_c>0.84\pm{0.05}$), with those measured directly in the lab. The certified MPS captures the strong pairwise correlations in the simulator state and even correctly predicts the sign and spatial profile of correlations beyond next-nearest neighbour: that is, of state properties beyond those measured to construct it (beyond $k=3$).  See Supp. Mat. for extended results. 

We implement a 14 spin quench: a system size well beyond the reach of full QST.  
Local measurements of the quench (Figure 4) reveal that strong entanglement, in pairs and triplets, develops right across the system. 
MPS tomography of the initial spin state identifies an accurate product state description $\ket{\Psi^1_{c}}$ with a fidelity of at least $F^1_c=0.88\pm{0.07}$ with the simulator state ($|\langle \Psi^1_{c}| \Phi(0)\rangle |^2=0.96$), using only single-site measurements. 
Spin-spin interactions are slightly longer-range in the 14-spin quench, than for 8 ($\alpha_{14}\approx1.3$ compared with $\alpha_{8}\approx1.6$), meaning that long-range correlations should develop faster. An idealised model predicts that 3-site measurements still provide an accurate certified description up to $t_{14}=0.36$ 1/J (4~ms), before rapidly failing at later times due to correlation spreading. 
 In the experiment, a 14-spin MPS description $\ket{\Psi^3_{c}}$ is achieved at $t_{14}$, using 3-site measurements, with a certified minimum fidelity $F^3_c=0.39\pm{0.08}$ (an idealised model of our simulator predicts an MPS certified fidelity of 0.78, the discrepancy is explained later).  

Since certified fidelities are only lower bounds, it is natural to ask exactly where the state fidelities actually lie. We perform direct fidelity estimation (DFE) \cite{Flammia:2011, daSilva:2011} to determine the overlap between the 14-spin simulator at $t_{14}$ and $\ket{\Psi^3_c}$. 250 observables are measured, randomly-drawn from the set with support in $\ket{\Psi^3_c}$ (Supp. Mat.). The result is a fidelity of $0.74 \pm{0.05}$. 

Clearly MPS tomography provided an accurate estimate of the 14-spin simulator state, and the fidelity lower bound of $F^3_c=0.39\pm{0.08}$ is correct. However, the bound is rather conservative and even lies quite far from the lower bound expected from an idealised model of our system (using states $\ket{\Phi(t_{14})}$) of 0.78. Via analysis of the local measurements, we find that this discrepancy can be largely explained by errors in the initial state preparation and modelled by adding mixture to each spin separately (local noise), yielding a predicted 14-spin certified fidelity lower bound at $t_{14}$ of $0.49\pm{0.07}$ (see Supp. Mat.). These errors limit the ability of the certification step to guarantee the accuracy of the MPS estimate, although the estimate is still a good description. The local measurements also reveal that we made more errors per spin when preparing the initial state for 14-spins than for 8-spins: our current optical setup makes it more difficult to control ions at the ends of the string with lasers, as the number of ions increases. The increase in error-per-spin as our current simulator is scaled-up in size, is seen to limit the ability to accurately characterise its state. A new optical setup should allow for a constant and small error-per-spin up to several tens of spins. 

  \begin{figure}[t]
%\vspace{-5mm}
  \begin{center}
    \includegraphics[width=1\columnwidth]{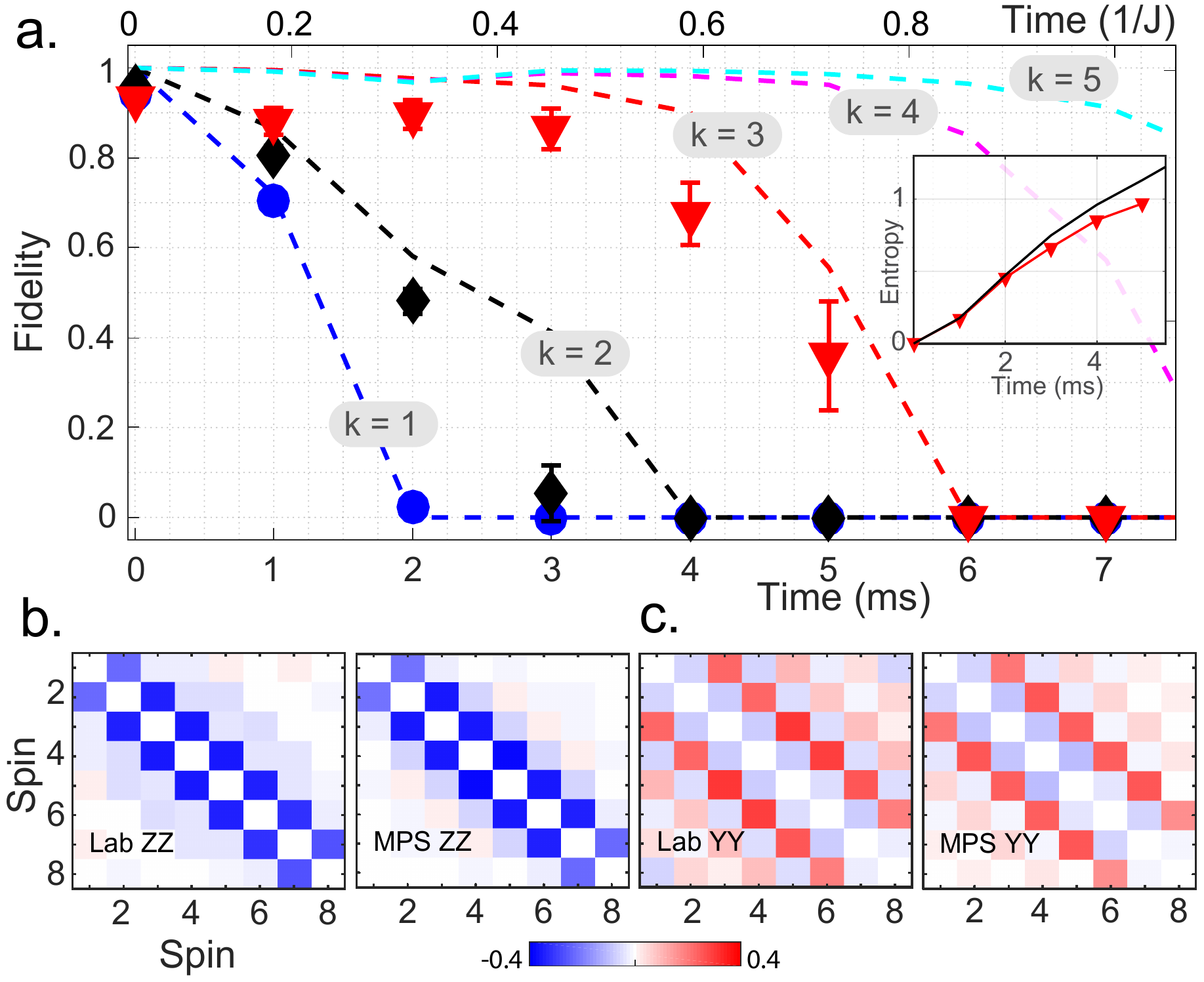}
  \vspace{-3mm}
   \caption{\textbf{MPS tomography results for an 8 spin quench.} 
\textbf{a.} 
Certified lower bounds $F^k_c$ on the fidelity between MPS $\ket{\Psi^k_{c}}$, reconstructed from measurements over $k$ sites, and the quantum simulator state $\rho_{lab}$. Shapes: data points with errors (uncertainty due to finite measurement number). Dashed lines: model, MPS tomography applied to idealised simulator dynamics ($\ket{\Phi(t)}$) with exact knowledge of local observables. Color: Blue, black, red, magenta and cyan represent local reductions of length k=1,2,3,4,5 sites, respectively. 
Insert: half-chain Von Neumann entropy of the pure global state. Red triangles: from data ($\ket{\Psi^3_{c}}$). Black line: from ideal model ($\ket{\Phi(t)}$).  
\textbf{b.}  Spin pair correlation matrices showing observable $\langle\text{Z(t)}_i\text{Z(t)}_j\rangle-\langle\text{Z(t)}_i\rangle\langle\text{Z(t)}_j\rangle$ at  $t=3$ms, for spins $i$ and $j$. Results directly measured on $\rho_{lab}$ (LHS) are compared with those derived from $\ket{\Psi^3_{c}}$ (RHS, see titles). \textbf{c.}  Same as \text{b.} but for observable $\langle\text{Y(t)}_i\text{Y(t)}_j\rangle-\langle\text{Y(t)}_i\rangle\langle\text{Y(t)}_j\rangle$. Correlation matrices from an idealised model ($\ket{\Phi(t)}$) are visually indistinguishable from those directly measured in the lab (not shown, see Supp. Mat.).  
}
  \label{figure 3}
 \vspace{-5mm}
  \end{center}
 \end{figure}

Comparison of the correlation matrices (Figure 4c), shows that the entangled 14-spin MPS estimate $\ket{\Psi^3_{c}}$ at $t_{14}$ captures many of the correlations between spins up to 4 sites apart (see Supp. Mat. for extended results). The weak correlations over greater distances in the laboratory state develop effectively instantly in quench dynamics, due to the long-range components of our interactions. 
The entanglement content and distribution in $\ket{\Psi^3_{c}}$ is consistent with the amount expected from an ideal model and the state has no separable partitions.

An appealing strategy is to use MPS tomography to acquire a state estimate and fidelity lower bound with minimal effort, then use DFE to find the exact fidelity. However, we find that the number of additional measurements for DFE becomes impractically large for more than 14-spins in our system. It is an open question as to whether DFE scales efficiently for MPS \cite{Flammia:2011, daSilva:2011}.

In conclusion, MPS tomography is guaranteed to provide an accurate state estimate with effort that scales efficiently in system size for a broad range of physically relevant states e.g. 2D cluster states, and the static and dynamic states found in 1D systems with finite-range interactions. Our experiments show that its scope of application is even broader, allowing characterisation of many-body entangled states and their dynamics even in systems without finite-range interactions. Since no prior knowledge of the state in the laboratory is required, MPS tomography provides a practical and efficient approach to obtaining a reliable state estimate and should therefore be a powerful addition to the toolbox for verifying and benchmarking engineered quantum systems.

\noindent {\bf Acknowledgments.} 
Work in Innsbruck was supported by the Austrian Science Fund (FWF) under the grant number P25354-N20,  by the European Commission via the integrated project SIQS, by the Institut f\"ur Quanteninformation GmbH and by the U.S. Army Research Office through grant W911NF-14-1-0103. 
All statements of fact, opinion or conclusions contained herein are those of the authors and should not be construed as representing the official views or policies of ARO, the ODNI, or the U.S. Government. 
Work in Ulm was supported by an Alexander von Humboldt Professorship, the ERC Synergy grant BioQ, the
EU projects QUCHIP and EQUAM, the US-Army Research Office Grant No. W91-1NF-14-1-0133 and the 
BMBF Verbundproject QuOReP. 
Numerical computations have been supported by the state of Baden-W\"urttemberg through bwHPC and the German Research Foundation (DFG) through grant no INST 40/467-1 FUGG.
I.D. acknowledges support from the Alexander von Humboldt Foundation. 
M. H. acknowledges contributions from Daniel Suess to jointly developed code used for data analysis. 
Work at Strathclyde is supported by the European Union Horizon 2020 collaborative project QuProCS (grant agreement 641277), and by AFOSR grant FA9550-12-1-0057
M.C. acknowledges: the ERC grant QFTCMPS and SIQS, the cluster of excellence EXC201 Quantum Engineering and Space-Time Research, and the DFG SFB 1227 (DQ-mat).
T. B. acknowledges: EPSRC (EP/K04057X/2) and the UK National Quantum Technologies Programme (EP/M01326X/1).

\noindent {\bf Author contributions.} 
B.P.L, C.F.R, M.B.P. and M.C. developed and supervised the project; 
C.M., C.H., B.P.L., P.J., R.B. and C.F.R. performed and contributed to the experiments; 
B.P.L., M.H., T.B., C.M, C.F.R., I.D., A.B. and A.D. performed data analysis and modelling; B.P.L. wrote the manuscript, with contributions from all authors. 

\vspace{1cm}

   \begin{figure}[t]
%\vspace{-5mm}
  \begin{center}
    \includegraphics[width=1\columnwidth]{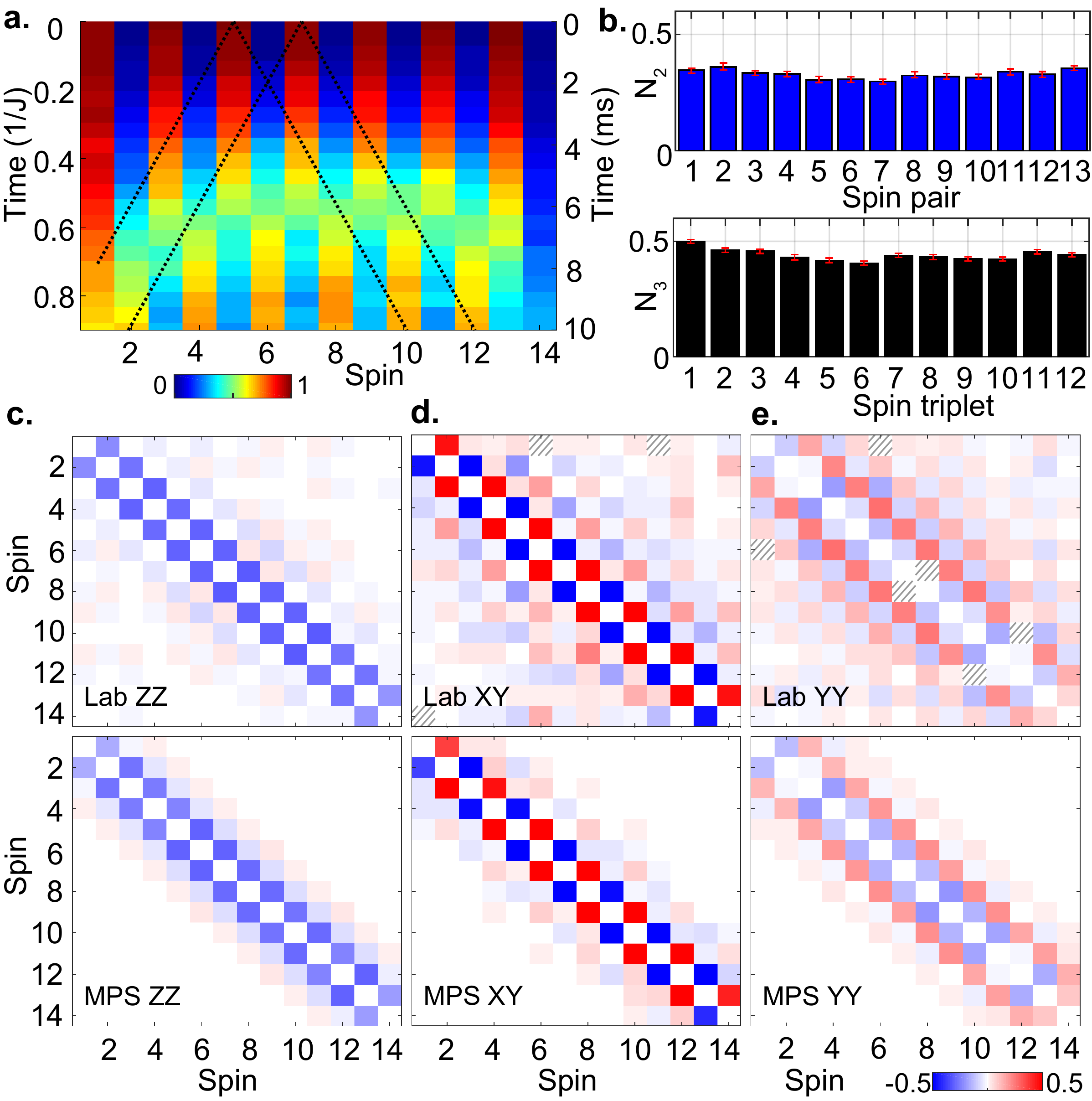}
\vspace{-5mm}
   \caption{\textbf{MPS tomography results for a 14 spin quench.} 
\textbf{a.-b}. Results from local measurements. \textbf{a}. Spin magnetisation $(1+\langle \sigma_i^z(t)\rangle)/2$ with two approximate light-like cones (Supp. Mat.). 
\textbf{b}. Entanglement in local reductions at $t=4$~ms, from density matrices reconstructed via QST. Upper: between neighbouring spin pairs (negativity). Lower: spin triplets (tripartite negativity), from QST of corresponding density matrices. 
\textbf{c.-e.} Comparison of correlation matrices directly measured in lab and from MPS estimates at $t=4$~ms, showing observable $\langle\text{A(t)}_i\text{B(t)}_j\rangle-\langle\text{A(t)}_i\rangle\langle\text{B(t)}_j\rangle$, for spins $i$ and $j$. A, B as labelled. Not all correlations were measured in the lab (hatched squares).
}
 \vspace{-7mm}
  \end{center}
 \end{figure}

%%%%%%%%%%
\clearpage
\newpage

% Does not work with revtex:
% %\includepdf[pages=-]{supplementary_material/MPS_paper_supplementary_FINAL.pdf}
%
% Alternative from http://tex.stackexchange.com/a/227416:
% 


\section*{Supplementary material}

# Efficient tomography of a quantum many-body system:
# Supplementary Material

B. P. Lanyon,[1,2,*] C. Maier,[1,2,*] M. Holzäpfel,[3] T. Baumgratz,[3,4,5]
C. Hempel,[2,6] P. Jurcevic,[1,2] I. Dhand,[3] A. S. Buyskikh,[7] A. J.
Daley,[7] M. Cramer,[3,8] M. B. Plenio,[3] R. Blatt,[1,2] and C. F. Roos[1,2]

*[1]Institut für Quantenoptik und Quanteninformation,*
*Österreichische Akademie der Wissenschaften,*
*Technikerstr. 21A, 6020 Innsbruck, Austria*
*[2] Institut für Experimentalphysik, Universität Innsbruck,*
*Technikerstr. 25, 6020 Innsbruck, Austria*
*[3]Institut für Theoretische Physik and IQST,*
*Albert-Einstein-Allee 11, Universität Ulm, 89069 Ulm, Germany*
*[4]Clarendon Laboratory, Department of Physics,*
*University of Oxford, Oxford OX1 3PU, United Kingdom*
*[5]Department of Physics, University of Warwick,*
*Coventry CV4 7AL, United Kingdom*
*[6]ARC Centre for Engineered Quantum Systems,*
*School of Physics, The University of Sydney,*
*Sydney, New South Wales 2006, Australia.*
*[7]Department of Physics and SUPA, University of Strathclyde, Glasgow G4 0NG, UK*
*[8]Institut für Theoretische Physik, Leibniz Universität Hannover, Hannover, Germany*
(Dated: December 23, 2016)

## Contents

---

*These authors contributed equally to this work.



# I. TRAPPED-ION QUANTUM SIMULATOR

## A. Ion trapping frequencies

We refer to the 'axial' direction along the ion string principle axis as $z$ and the two 'radial' directions, orthogonal to the string principle axis, as the $x$ and $y$ axes. Ion strings are loaded into a highly anisotropic trapping potential: the radial confinement is far stronger than the axial confinement. For all experiments presented, the frequencies of the centre of mass vibrational modes are: $\omega_z = 2\pi \times 0.214$ MHz, $\omega_y = 2\pi \times 2.69$ MHz, $\omega_x = 2\pi \times 2.71$ MHz.

## B. Simulator Hilbert space and laser beams

We identify the two electronic Zeeman states $|S_{1/2}, m = +1/2\rangle$ and $|D_{5/2}, m' = +5/2\rangle$ of trapped $^{40}$Ca$^+$ ions with the $|\downarrow\rangle$ and $|\uparrow\rangle$ states of spin-1/2 particles, respectively. These atomic states are coupled by an electric quadrupole transition at the optical wavelength of 729 nm. The quantum states of the spin states are coherently manipulated using an approximately 1Hz linewidth Ti:Sa CW laser. Two laser beam paths are employed for this: a global beam illuminating the ion string approximately equally from a direction perpendicular to the ion string axis and at an angle approximately half way between the two radial mode directions $x$ and $y$ (see previous subsection). Consider the standard Pauli spin operators on spin $j$: $\sigma_x^j$, $\sigma_y^j$ and $\sigma_z^j$. The global beam is used to perform global $\sigma_x$ and $\sigma_y$ rotations, approximately equally on all spins, e.g. $G_x(\theta) = \exp(-i\theta \sum_{j=1}^{N} \sigma_x^j)$ and $G_y(\theta) = \exp(-i\theta \sum_{j=1}^{N} \sigma_y^j)$. The global beam is also used to implement standard frequency-resolved sideband cooling, optical pumping on the quadrupole transition and the spin-spin interaction Hamiltonian (see later). The second 'addressed' beam path comes in parallel to the global beam (radial direction) but from the opposite direction. This second beam is focused to a single ion. The direction of the laser beam can be switched to have its focus pointing at different ions within 12 $\mu$s, using an acousto-optic deflector. This addressed beam is frequency-detuned by about 80 MHz from the spin transition and thereby performs an AC Stark rotation on addressed ion $j$ of the form: $R_z^j(\theta) = \exp(-i\theta\sigma_z^j)$. The combination of a global resonant beam and a focused detuned beam, inducing AC-Stark shifts for carrying out arbitrary single-spin rotations, has the advantage of not requiring laser beam paths whose optical path length difference is interferometrically stabilized. For an overview of the use of global and addressed beams to manipulate ionic spins (qubits) see [1].

## C. Simulator initialisation

*Cooling and optical pumping.* Each experimental sequence begins with Doppler cooling ($\sim 3$ ms) and optical pumping ($\sim 500\,\mu$s) to initialize all $N$ ions in the string into the $|\downarrow\rangle$ state. Next, all $2N$ radial motional modes, transverse to the string, are cooled to the ground state via $\sim 10$ frequency-resolved sideband cooling pulses taking about 10 ms in total, followed by a second frequency-resolved optical pumping step of $\sim 500\,\mu$s. The system is now prepared in a pure electronic and motional quantum state and is ready for preparation of the Néel state.

*Preparing the Néel state.* The Néel state is prepared using a combination of global and addressed pulses. The addressed operation $A_z(\theta) = \exp(-i\theta \sum_{[j=1,3,5...N-1]} \sigma_z^j)$ is employed, corresponding to (ideally) equal rotations around the $z$ axis of a subset of spins in the string performed sequentially. To create the Néel state from the initial state $|\downarrow\rangle^{\otimes N} = |\downarrow, \downarrow, \downarrow ...\rangle$ requires flipping every second spin to the $\uparrow$ state. This is done with the following composite pulse sequence $G_x(\pi/4)G_y(\pi/4)A_z(\pi/2)G_y(\pi/4)G_x(\pi/4)|\downarrow\rangle^{\otimes N} = |\uparrow, \downarrow, \uparrow ...\rangle$. To first order, the state created by this sequence is insensitive to errors in the rotation angles of the $x$ and $y$ rotations. These errors come from the unequal coupling strength of the global beam across the string, due to its Gaussian beam shape. We prepare the Néel state for 8 and 14 ion strings. Figure 5 shows the measured probabilities of prepar-

ing each spin in $|\uparrow\rangle$. The fidelities of these states, with the ideal Néel state is obtained by directly measuring in the $z$-basis on all spins. For 8 spins, the direct fidelity with the Néel state is $0.967 \pm 0.006$, corresponding to an average error per spin of $-\log_2(0.967)/8 = 0.006 \pm 0.001$. For 14 spins, the highest radial direct fidelity with the Néel state is $0.89 \pm 0.01$, corresponding to an average error per spin of $-\log_2(0.89)/14 = 0.012 \pm 0.001$. Clearly the error-per-particle is significantly larger for the 14 spin initial state, than for 8. Note that the aforementioned direct fidelities of the initial state agree well with the certified lower bounds obtained for measurements on single sites, via MPS tomography (see main text).

### D. Spin-spin interactions

The experimental implementation of the $XY$ spin-spin Hamiltonian is described in [2] and [3]. In summary, the model is realised via the global laser path via a beam containing three frequencies (trichromat) two of which off-resonantly drive all $2N$ radial vibrational modes of the string and are symmetrically detuned by $\pm\Delta$ from the spin flip transition. The magnitude of the detuning $|\Delta|$ is larger than the highest radial COM mode $\omega_x/(2\pi) = 2.71$ MHz by $2\pi \cdot 79$ kHz (8 spins) or $2\pi \cdot 76$ kHz (14 spins). The third frequency lies around 1MHz detuned from the spin flip transition and compensates for AC Stark shifts. The XY model is obtained (is a good description) by the addition of an overall detuning of all three frequencies by $2\pi \cdot 3$ kHz (8 spins) or $2\pi \cdot 5$ kHz (14 spins) (ensure that the detuning is much larger than the absolute value of the spin-spin coupling terms of a few tens to hundreds of Hz, see timescales of simulator dynamics) and careful attention to the minimisation of any process that causes the energy splitting of spin states to be different across the ion string (e.g. by minimising magnetic field gradients across the string). The coupling matrix element $J_{ij}$ (see main paper) determines the rate at which a quasiparticle excitation (spin up) at site $i$ can hop to an unoccupied site $j$ (spin down), and vice versa. While a single excitation in the system will simply disperse (like a single particle quantum random walk in 1D), multiple excitations interact and scatter [3]. In [2] we showed, from direct measurements, that the $J_{ij}$ achieved in the experiment is very well described by the theoretical $XY$ model of our system. More about the modelling of our spin-spin interactions is given in a later section.

### E. Electron shelving

At the end of every measurement, the spin state is determined via the standard electron shelving technique: light at 397 nm (and 866 nm, repumping) is sent to the ion string, coupling to the $S_{1/2} \rightarrow P_{1/2}$ and $D_{3/2} \rightarrow P_{1/2}$ transitions respectively. 397 nm light is only scattered if the electron is in the $|\downarrow\rangle$ state. This scattered light is detected using a single-ion resolving CCD camera. Figure 6 shows examples of CCD camera pictures of an 8 ion string, with all ions in the ground state and (separately) in the Néel-ordered state. Both single images and averaged images are presented. The colorbar ranges from dark blue to dark red, where red denotes the qubit being in the (fluorescing) ground state $|\downarrow\rangle$, while blue denotes the spin being in the excited state $|\uparrow\rangle$.

## II. MODELLING THE SIMULATOR

### A. Ideal simulator model

In the main paper we give approximate spin-spin interaction power-law ranges ($\alpha$ values), light-like cones for information spreading (Figures 2 and 4) and, in this supplementary material, we compare data with a theoretical model for our simulator. In this section we explain how we do this modelling.

Our model for the simulator dynamics is the XY Hamiltonian, as described in the main text. In previous work we have shown that the simulator dynamics is well described by this model in low-energy regimes, that is, when the initial state is close to the ground state (containing one [2] or a few [3] quasiparticle excitations). In

this work we explore the dynamics of highly excited initial states: the Néel state contains $N/2$ excitations. Note that the XY Hamiltonian preserves the excitation number throughout dynamical evolutions (subspaces with different excitation number are decoupled).

The XY model is parameterised by the spin-spin coupling matrix $J_{ij}$. The model of $J_{ij}$ in terms of experimental parameters is described in the supplementary material of [2]. In summary, $J_{ij}$ depends on the ion string vibrational mode frequencies, eigenvectors, detuning from the laser fields, ionic mass and laser-ion coupling strength. All else held constant, the interaction range (modelled by a power law, see main text) can be changed by a single experimental parameter: the detuning of laser fields from the motional resonances. We independently measure all the aforementioned parameters in our experimental system and thereby derive the spin-spin coupling matrix (figure 7).

The XY model is derived from a transverse Ising model with large transverse field [2]. Deviations from the XY model are due to e.g. small inhomogeneities in the transverse fields (different for each spin) which act like local potential barriers to spin excitations hopping around the string. The inhomogenuities comes from e.g. electric quadrupole shifts which differ along the string, magnetic field gradients across the string (in addition to our standard constant 4 Gauss field), and AC Stark shifts of the spin transitions due to the presence of laser fields with intensity gradients across the string (Gaussian beam profiles). These inhomogeneities are measured and included as an additional transverse field in the model ($b_k$). That is $\tilde{H}_{XY} = 0.5\hbar \sum_{i \neq j} J_{ij} (\sigma_i^+ \sigma_j^- + \sigma_i^- \sigma_j^+) + \hbar \sum_k (B + B_k) \sigma_k^z$. Here $J_{ij}$ is an $N \times N$ spin-spin coupling matrix, $\sigma_i^+$ ($\sigma_i^-$) is the spin raising (lowering) operator for spin $i$ and $\sigma_z^k$ is the Pauli $z$ operator for spin $k$. The transverse field consists of an overall constant $B$ and site-dependent perturbations $B_k$. After careful attention to their minimisation, the inhomogeneities are small (compared to the spin-spin coupling strength) and play little role in obtaining an accurate description of our system dynamics at the evolution times considered in the main paper.

Time-evolved model simulator states are calculated by brute force matrix exponentiation for up to 8 spins e.g. $|\Phi(t)\rangle = \exp(-i\tilde{H}_{XZt}t) |\Phi(0)\rangle$. For 14 spins, this approach takes hours and hours to run, using the computers that we have readily available. Therefore, it was time efficient for 14 spins to use the Krylov subspace projection methods (Arnoldi and Lanczos processes) which, in the case of sparse Hamiltonians, give a substantial speed up and well controlled error bounds [4].

### B. Interaction range in experiments

The realised spin-spin interactions are approximated by a power law dependence on the distance $|i - j|$: $J_{ij} \propto |i - j|^{-\alpha}$ with decay parameter $\alpha$. In the experiment there are two ways to tune the interaction range: either by varying the laser detuning from the motional resonances or by bunching up or fanning out the transverse modes in frequency space. Here the detuning is directly chosen whereas the mode-bunching depends on the effective trapping parameters (therewith also on the number of ions). Figure 8 compares the experimental coupling strengths as a function of the distance, with ideal power-law decay lines for $\alpha = 1, 2, 3$. It shows that it is not possible to extract an unambiguous decay parameter $\alpha$ by a direct fit in real space. However, an effective value for $\alpha$ can be estimated by fitting the eigenmode spectrum (or quasiparticle dispersion relation) of our system with the eigenmode spectrum for power-law interactions [2]. The power-law exponent $\alpha$ yielding the best fit gives an estimate for the interaction range. Figure 9 shows the best fit, providing $\alpha = 1.58$ (8 spins) and $\alpha = 1.27$ (14 spins).

### C. Light-like cones

The velocity at which excitation and quantum correlations spread in our system can not be arbitrarily fast. It rather happens in a light-like fashion, where information propagation outside the light-cone (that is, with velocities faster than a certain group velocity) is suppressed [2]. To visualise how fast this spread $v$ is in our system, we insert lines $t = d/v$ in figures 10 and 11, delineating the light cones of a system with

nearest-neighbour interactions only. For this we assume a nearst-neighbour model with a constant coupling strength, corresponding to the averaged nearest-neighbour coupling of the original matrix. Next we calculate the Eigenmode spectrum and determine the gradient between every pair of consecutive eigenvalues (depicted in figure 9). The largest of these gradients corresponds to the maximum velocity $v$ at which energy and correlations disperse in the system. Finally we renormalise the quantity by the algebraic tail of the original coupling matrix $J_{ij}$. Therefore we choose the central ion $i_c = 4$ (7) respectively for 8 (14) spins and average between the left and right algebraic tail:

$$N = \frac{1}{2} \sum_{j=1}^{N} J_{ji_c} \quad , \quad i = 1 ... N - 1 \tag{1}$$

### D. Normalised time unit (1/J)

We use two ways to label time axes in our plots: one way indicates the real laboratory time (in ms) passed during the evolution (e.g. figure 11 right y-axes), while the other way shows the time normalized by the averaged nearest-neighbour interaction strength of the original matrix $J_{ij}$ (e.g. figure 11 left side):

$$J = \sum_{i}^{N-1} \frac{J_{i,i+1}}{N-1} \quad . \tag{2}$$

## III. MEASURING AND RECONSTRUCTING LOCAL REDUCTIONS

In this section, we describe the measurements performed in the experiment and how these measurements are employed in the analysis. MPS tomography requires the ability to estimate the local reduced density matrices of all blocks of $k$ neighbouring spins. On a linear chain of $N$ spins, there are $N - k + 1$ such blocks. A straightforward method of reconstructing all these blocks requires a total of $(N - k + 1)4^k$ measurements, each performed on one of the $N - k + 1$ local blocks of $k$ spins. Instead, we perform $3^k$ measurements, each on the entire system of $N$ spins and use these measurements to infer the local reductions. First, we describe these $3^k$ measurements and show that these suffice.

### A. Chosen measurement setting

Here we recall the straightforward method of reconstructing the local reduced density matrices. One can obtain the density matrix of $k$ spins from the expectation values of a linearly independent set of $4^k$ observables. The set of all $k$-fold tensor products of the three Pauli operators $X = \sigma_x$, $Y = \sigma_y$, $Z = \sigma_z$ and the identity operator $\mathbb{1}$ provides one such set. For example, for $k = 2$ spins, the density matrix can be obtained from the following 16 expectation values

$$\begin{array}{cccccccc}
\langle Z_1 Z_2 \rangle & \langle Z_1 X_2 \rangle & \langle Z_1 Y_2 \rangle & \langle Z_1 \mathbb{1}_2 \rangle & \langle X_1 Z_2 \rangle & \langle X_1 X_2 \rangle & \langle X_1 Y_2 \rangle & \langle X_1 \mathbb{1}_2 \rangle \\
\langle Y_1 Z_2 \rangle & \langle Y_1 X_2 \rangle & \langle Y_1 Y_2 \rangle & \langle Y_1 \mathbb{1}_2 \rangle & \langle \mathbb{1}_1 Z_2 \rangle & \langle \mathbb{1}_1 X_2 \rangle & \langle \mathbb{1}_1 Y_2 \rangle & \langle \mathbb{1}_1 \mathbb{1}_2 \rangle .
\end{array} \tag{3}$$

In order to obtain the expectation value of say $\langle Z_1 X_2 \rangle$, the following measurement would be performed in the experiment: Spin-I is measured in the eigenbasis of $Z$ while Spin-II is measured in the eigenbasis of $X$. We refer to this measurement setting as $[Z, X]$; the eigenbasis of the $i$-th vector element provides the measurement basis for the $i$-th spin. The measurement setting $[Z, X]$ has four distinguishable outcomes (resolved on the CCD camera in our experiment). We obtain spin up ($\uparrow$) or spin down ($\downarrow$) in the $Z$ basis for Spin-I and spin up ($\uparrow$) or spin down ($\downarrow$) in the $X$ basis for Spin-II. By repeating the measurement $[Z, X]$ many times, we can estimate the four outcome probabilities $p_{\uparrow\uparrow}, p_{\uparrow\downarrow}, p_{\downarrow\uparrow}, p_{\downarrow\downarrow}$.

These probabilities can be used not only for extracting the expectation value $\langle Z_1 X_2 \rangle$, but also for extracting the expectation values $\langle Z_1 \mathbb{1}_2 \rangle$ and $\langle \mathbb{1}_1 X_2 \rangle$. This insight generalises to $k \geq 2$ spins and to more general measurement settings, and it

enables us to estimate the expectation values of the $4^k$ measurement observables (3) from only $3^k$ measurements.

For each local block of $k$ spins, the following $3^k$ measurement settings suffice. Each of the $k$ spins requires measurement in the basis of the three Pauli operators, thus leading to $3^k$ measurement settings. For instance, consider $k = 2$. In this case, the $3^k = 9$ measurement settings are given by

$$[Z, Z] \quad [Z, X] \quad [Z, Y] \quad [X, Z] \quad [X, X] \quad [X, Y] \quad [Y, Z] \quad [Y, X] \quad [Y, Y]. \quad (4)$$

Each of the $3^k$ measurement settings has $2^k$ distinguishable outcomes. In total, we estimate $3^k \times 2^k = 6^k$ outcome probabilities. Each of the $4^k$ expectation values required for obtaining the local reduced density matrices can be estimated from this set of $6^k$ outcome probabilities. Therefore, the set of $6^k$ outcome probabilities is sufficient to estimate a density matrix on $k$ spins [23].

The $3^k$ measurement settings described above are to be repeated for each of the $N - k + 1$ local blocks on the chain. Performing measurement independently for the local blocks would require $(N - k + 1)3^k$ measurement settings, where measurements are performed on the local blocks and remaining spins are ignored. However, a more judicious choice can provide the required information with fewer measurement settings.

We choose a set of $3^k$ total measurement settings such that measurements are performed on the entire spin chain rather than just the local blocks. We repeat each of the $3^k$ measurement settings on $k$ spins along the chain. Specifically, for each of the $3^k$ measurement settings, we split the system into $\lfloor N/k \rfloor + 1$ blocks and replicate the same measurement settings on each of the blocks. For instance, the case of $k = 2$ requires the measurement settings

$$
\begin{array}{lll}
[X, X, X, X, \dots] & [X, Y, X, Y, \dots] & [X, Z, X, Z, \dots] \\
[Y, X, Y, X, \dots] & [Y, Y, Y, Y, \dots] & [Y, Z, Y, Z, \dots] \\
[Z, X, Z, X, \dots] & [Z, Y, Z, Y, \dots] & [Z, Z, Z, Z, \dots].
\end{array} \quad (5)
$$

In our experiment, we set $k = 3$, i.e., we perform measurements on $3^3 = 27$ settings. Formally, we perform measurements in the $3^k$ with $k = 3$ different measurement bases

$$[A_1, \dots, A_k, A_1, \dots, A_k, \dots]: \quad A_i \in \{X, Y, Z\}, \quad i \in \{1, \dots, k\} \quad (6)$$

on $N$ spins. Each of the chosen $3^k$ measurement settings has $2^N$ distinguishable outcomes. $m = 1000$ outcomes (which could take any of the $2^N$ unique values) were recorded for each of the 27 settings.

To summarize, we choose $3^k$ measurement settings comprising repetitions of local $3^k$ non-trivial Pauli measurements. This brings the total measurement setting requirement down from $(N - k + 1)4^k$ local measurements to $3^k$ measurements on the entire chain.

## B. Using measurement outcomes

Here we describe how the outcomes obtained from measurement settings (6) are used in the subsequent analysis. The measurement data are used either (i) to reconstruct the state via certified MPS tomography or (ii) to estimate local reduced density matrices on $k$ spins, for instance to estimate 3-spin entanglement.

The measurement data are input to the certified MPS tomography algorithms (Section IV) after converting to one out of the following two forms. The first form is that of $(N - k + 1)4^k$ local expectation values

$$\langle A_{s+1} \otimes \cdots \otimes A_{s+k} \rangle: \quad A_{s+i} \in \{\mathbb{1}, X, Y, Z\}, \quad (7)$$

where $s \in \{0, \dots, N - k\}$ is the first site of the local block and $i \in \{1, \dots, k\}$ labels sites within the respective local block. An alternate but equivalent form of the input to certified tomography is that of the outcome probabilities of the $6^k$ non-identity Pauli measurements performed on each of the $N - k + 1$ local blocks. Formally, the

following $(N - k + 1)6^k$ local outcome probabilities are estimated:

$$\langle P_{s+1,a_1,b_1} \otimes \cdots \otimes P_{s+k,a_k,b_k} \rangle: \quad a_i \in \{X, Y, Z\}, \quad b_i \in \{-1, +1\},$$
$$s \in \{0, \ldots, N - k\}, \quad i \in \{1, \ldots, k\}, \tag{8}$$

where $P_{j,a_i,b_i} = |a_i b_i\rangle_j \langle a_i b_i|$ projects spin $j$ onto the eigenvector $|a_i b_i\rangle$ of the Pauli operator $a_i \ (= X, Y \text{ or } Z)$ with eigenvalue $b_i$. The methods for estimating quantities (7) or 8 from the measurement data are detailed in Section IV.

The second use of the measurement data is to reconstruct local reduced density matrices of $k$ neighbouring spins, or in other words, to perform full quantum state tomography of the local blocks. We use maximum likelihood estimation [5] to obtain density matrix estimates $\rho_{\mathrm{qst}}^k$ from the local $k$-spin outcome probabilities. Quantities of interest are computed from the density matrix estimates, e.g., the entanglement measures (Section V) in Figures 2 and 4 of the main text. Error bars in quantities derived from local reconstructions are obtained from standard Monte-Carlo simulations of quantum projection noise.

## IV. CERTIFIED MPS TOMOGRAPHY

In this section, we describe our method for certified MPS tomography, which is based on the results of Refs. [6] and [7]. We use the modified SVT algorithm from [6] and the scalable maximum likelihood estimation method for quantum state tomography from [7] to obtain an estimate of the unknown lab state from experimental measurement data. Because the experimentally measured observables do not contain complete information on the unknown state, an additional step is necessary to verify the correctness of the result. We use the assumption-free lower bound on the fidelity between our estimate and the unknown lab state from Ref. [6] for this purpose; we call such a lower bound on the fidelity a certificate. We refer to the combined procedure of MPS tomography and certification as certified MPS tomography.

### A. Details of procedure

We discuss how to reconstruct and certify a pure quantum state on $N$ qubits from local informationally-complete measurements, i.e., measurements whose positive-operator valued measure (POVM) [8] elements span the complete operator space on all blocks of $k$ neighbouring spins . The information completeness of local POVMs ensures that the corresponding (reduced) density matrix can be reconstructed from the measurement outcomes. The measurement settings described in Sec. III satisfy this property.

As there are $N - k + 1$ such contiguous blocks of size $k$ on a linear chain, the total measurement effort scales at most linearly with the number $N$ of qubits. Our discussion is formulated for $N$ qubits (spin-$\frac{1}{2}$ particles), but it equally applies to $N$ qudits. Figure 14 on page 35 provides a schematic overview over the following subsections.

We will proceed as follows: Measurement data is split into two parts; the first part is used for MPS tomography while the second part is used for certification (Sec. IV A 1). We apply existing MPS tomography algorithms to obtain an initial estimate $|\psi_{\mathrm{est}}\rangle$ of the unknown lab state (Sec. IV A 2). From the initial estimate, a family of candidates for a so-called parent Hamiltonian is constructed and one of them is selected, denoted by $H$. The parent Hamiltonian $H$ provides the certificate and its ground state $|\psi_{\mathrm{GS}}\rangle$ is the certified estimate of the unknown lab state (Sec. IV A 3). The general approach to obtain the measurement uncertainty of the fidelity lower bound is derived (Sec. IV A 4). The fact that local probabilities have been obtained from global measurements in the experiment complicates obtaining the measurement uncertainty of the fidelity lower bound; remaining technical details related to this issue are covered at the end (Sec. IV A 5).

#### 1. Measurements

Our method begins with measurements on the unknown lab state $\rho_{\mathrm{lab}}$. We use the data from the measurements described in Sec. III. The data comprise $m = 1000$ out-

comes for each of the $q = 3^k$ different measurement settings 6 on $N$ qubits. The samples are split into two parts of 500 samples each. The first part is used to obtain an estimate of the unknown lab state, while the second part is used to obtain the certificate, i.e. the lower bound on the fidelity between the unknown lab state and our estimate of the lab state. This splitting is performed to avoid any risk of overestimating fidelity by constructing or selecting a parent Hamiltonian (see below) which is tuned to the particular set of statistical fluctuations in a single set of measurement data. Future work could study whether one can use measurement data in a more efficient way.

### 2. Uncertified MPS tomography

We obtain an estimate of the unknown lab state by combining two efficient MPS tomography algorithms [9]. We use the modified SVT algorithm from Ref. [6] to obtain a pure state. This pure state is used as start vector for the iterative likelihood maximization scheme over pure states from Ref. [7]. The computation time required for both algorithms scales polynomially with the number of qubits $N$.

The input for the modified SVT algorithm consists of the local expectation values from Eq. (7) (Sec. III). Alternatively, one can specify the input as estimates of the local reduced states on $k$ neigbouring qubits (the difference is only an operator basis change in Hilbert-Schmidt space). In our implementation, we choose the latter and convert the local outcome probabilities from Eq. (8) into local reduced states using linear inversion. This is accomplished using the Moore-Penrose pseudoinverse of the map $\mathcal{M}_s$, which we describe in Eq. (34) in Section IV A 5.

The input for the scalable maximum likelihood estimation (MLE) scheme consists of the local outcome probabilities from Eq. (8). In principle, one could perform MLE with more information than only the local outcome probabilities. For example, one could extract all pairwise correlations available from the existing measurement data and provide them to the MLE algorithm. This is another avenue for further work. The scalable MLE algorithm returns a initial estimate of the unknown lab state, denoted by $|\psi_{\text{est}}\rangle$.

In both methods mentioned above, the pure state is represented as a matrix product state [10] with limited bond dimension $\Delta$. In some cases, we observe that the fidelity lower bound obtained at the end of the scheme decreases if we allow for a larger bond dimension $\Delta$. Presumably, this is a result of statistical noise adding spurious correlations to our state estimate, which is prevented by lowering the bond dimension. For 8 qubits, we use $\Delta = 2$ for $t \leq 2\,\text{ms}$ and $\Delta = 4$ for all remaining times. For 14 qubits, we use $\Delta = 16$ for all times.

### 3. Fidelity lower bound and selection of a parent Hamiltonian

In the last subsection, we have obtained the initial estimate $|\psi_{\text{est}}\rangle$ of the unknown lab state $\rho_{\text{lab}}$. At this point, we do not know whether $|\psi_{\text{est}}\rangle$ is close to the lab state $\rho_{\text{lab}}$. We continue by finding a certified estimate $|\psi_{\text{GS}}\rangle$ and a lower bound to the fidelity between $|\psi_{\text{GS}}\rangle$ and the lab state $\rho_{\text{lab}}$. In the remainder of this section, we define $|\psi_{\text{GS}}\rangle$ and its parent Hamiltonian $H$, and we describe how these are constructed based on estimated state $|\psi_{\text{est}}\rangle$.

The fidelity lower bound is obtained using a so-called parent Hamiltonian. A parent Hamiltonian of a pure state $|\psi_{\text{GS}}\rangle$ is any Hermitian linear operator $H$ such that $|\psi_{\text{GS}}\rangle$ is the non-degenerate ground state of $H$. Let $E_0$ and $E_1$ be the smallest and second smallest eigenvalues of $H$. Then, a lower bound to the fidelity between the ground state $|\psi_{\text{GS}}\rangle$ and any other pure or mixed state $\rho_{\text{lab}}$ is given by [6]

$$\langle \psi_{\text{GS}} | \rho_{\text{lab}} | \psi_{\text{GS}} \rangle \geq 1 - \frac{E - E_0}{E_1 - E_0} \tag{9}$$

with the energy $E = \text{tr}(H \rho_{\text{lab}})$ of the unknown state $\rho_{\text{lab}}$ in terms of the parent Hamiltonian $H$. Note that $H$ is usually completely artificial and unrelated to any energy in the physical system in the lab. If $H$ is a sum of local terms—that is, terms acting nontrivially only on $k$ neighbouring spins—the measurements described above suffice to

obtain the energy $E$ and the fidelity lower bound. It remains to find such a parent Hamiltonian, given the initial estimate $|\psi_\mathrm{est}\rangle$.

In order to find a larger-than-zero fidelity lower bound, we have to find a parent Hamiltonian which must satisfy two conditions: (i) the ground state $|\psi_\mathrm{GS}\rangle$ must be close to the initial estimate $|\psi_\mathrm{est}\rangle$ and (ii) the gap $E_1 - E_0$ between the two smallest eigenvalues must be much bigger than the measurement uncertainty about the value of $E$. If condition (ii) is not satisfied, we will not learn anything about the fidelity. The following qualitative argument illustrates that condition (i) is necessary as well: If the initial estimate $|\psi_\mathrm{est}\rangle$ is far away from the lab state $\rho_\mathrm{lab}$, we do not try to find a useful parent Hamiltonian because it would be unlikely to succeed. Therefore, we only consider the case where the fidelity of the initial estimate $|\psi_\mathrm{est}\rangle$ and the lab state $\rho_\mathrm{lab}$ is high: In this case, a high fidelity between the ground state $|\psi_\mathrm{GS}\rangle$ of the Hamiltonian and lab state $\rho_\mathrm{lab}$ is possible only if the fidelity of $|\psi_\mathrm{est}\rangle$ and $|\psi_\mathrm{GS}\rangle$ is high as well.

First, we attempt to construct a parent Hamiltonian whose ground state $|\psi_\mathrm{GS}\rangle$ is the same as $|\psi_\mathrm{est}\rangle$. If the matrix product state $|\psi_\mathrm{est}\rangle$ belongs to the class of *injective MPS* for a certain number $k$ of neighbouring spins [11, 12], then the operator

$$H = \sum_{s=1}^{N-k+1} \mathbb{1}_{1,\dots,s-1} \otimes P_{\mathrm{ker}(\rho_s)} \otimes \mathbb{1}_{s+k,\dots,N}, \qquad \rho_s = \mathrm{tr}_{1,\dots,s-1,s+k,\dots,n}(|\psi_\mathrm{est}\rangle\langle\psi_\mathrm{est}|) \quad (10)$$

has $|\psi_\mathrm{est}\rangle$ as its unique ground state [11, 12]; $P_{\mathrm{ker}(\rho_s)}$ is the orthogonal projection onto the kernel of the reduced density matrix $\rho_s$ on the $k$ neighbouring spins from $s$ to $s+k-1$. The injectivity property implies certain restrictions on possible combinations between the bond dimension $D$ of $|\psi_\mathrm{est}\rangle$ and the number of neighbours $k$. However, the initial estimate $|\psi_\mathrm{est}\rangle$ generally is not an injective MPS for our given $k$ and Eq. (10) will not provide a parent Hamiltonian; what it will provide is a Hamiltonian which has $|\psi_\mathrm{est}\rangle$ as one of its degenerate ground states. This violates condition (ii) and leads to a zero fidelity lower bound.

To mitigate the problem of ground-state degeneracy of $H$ (10), we relax the requirement that $|\psi_\mathrm{est}\rangle$ is a ground state. Specifically, we introduce a threshold $\tau \geq 0$ and obtain candidates for parent Hamiltonians from

$$H = \sum_{s=1}^{N-k+1} \mathbb{1}_{1,\dots,s-1} \otimes P_{\mathrm{ker}(T_\tau(\rho_s))} \otimes \mathbb{1}_{s+k,\dots,N} \tag{11}$$

where the thresholding function $T_\tau$ replaces eigenvalues of $\rho_s$ smaller than or equal to $\tau$ by zero. We construct a set of candidates $H_1, H_2, \dots$ for parent Hamiltonians by considering all possible values of $\tau \geq 0$. We then try to find a compromise between conditions (i) and (ii) from above, $|\psi_\mathrm{est}\rangle$ and $|\psi_\mathrm{GS}\rangle$ being similar and a large gap, by choosing the operator $H$ which minimizes

$$c\mathcal{D}(|\psi_\mathrm{est}\rangle, |\psi_\mathrm{GS}\rangle) - (E_1 - E_0) \tag{12}$$

where $c > 0$ is some constant and

$$\mathcal{D}(|\psi\rangle, |\phi\rangle) \stackrel{\mathrm{def}}{=} \| |\psi\rangle\langle\psi| - |\phi\rangle\langle\phi| \|_1 / 2$$
$$= \sqrt{1 - |\langle\psi|\phi\rangle|^2} \tag{13}$$

is the trace distance [8]. We obtain a valid fidelity lower bound for any value of the constant $c$. However, we may obtain a very small lower bound or a lower bound associated with a large measurement uncertainty for some choices of this constant. We have used the value $c = 5$ and we do not observed significantly higher fidelity lower bounds for other values of this constant. Modifying Eq. (12) or choosing a more optimal value for $c$ in connection with the discussion in Corollary 7 on Page 28 has the potential to provide improved fidelity lower bounds.

We use the following numerical tools to compute parent Hamiltonians. For 8 qubits, the full spectrum of the parent Hamiltonian candidates has been computed with the library function `numpy.linalg.eigh()` from SciPy [13]. For 14 qubits, a DMRG-like iterative MPS ground state search with local optimization on two neighbouring qubits [10, Section 6.3] has been used to obtain two eigenvectors of the smallest eigenvalue(s). We have used the functions `mineig()` and `mineig_sum()`

available from the Python library mpnum [14]. The second eigenvector has been obtained as ground state of $H' = H + 5\,|\psi_{\text{GS}}\rangle\langle\psi_{\text{GS}}|$. For both eigenvectors, the quantity $\langle\psi|\,H^2\,|\psi\rangle - (\langle\psi|\,H\,|\psi\rangle)^2$ has been monitored to ensure sufficient convergence of the iterative search. It should be noted that, strictly speaking, DMRG-like algorithms only provide upper bounds on smallest eigenvalues, but in practice they have been observed to be very reliable [10]. The results from a first low-precision eigenvalue computation with MPS bond dimension 8 have been used to select a parent Hamiltonian. The eigenvalues from a second high-precision eigenvalue computation with MPS bond dimension 24 have been used to obtain the certificate.

As the certified estimate $|\psi_{\text{GS}}\rangle$ is the result of an eigenvector computation, the eigenvector computation determines the maximal bond dimension of $|\psi_{\text{GS}}\rangle$: For 8 qubits, its bond dimension may reach the maximal value of 16 and for 14 qubits, its bond dimension can be up to 24. The case $k = 1$ is an exception; there, it is easy to see that a non-degenerate ground state must be a product state (bond dimension 1).

### 4. Statistical analysis of the fidelity lower bound

The fidelity lower bound (9) relies on using the parent Hamiltonian and information about the lab state to estimate the energy $E = \text{tr}(H\rho_{\text{lab}})$. The information about the lab state is obtained from a finite number of measurement outcomes distributed according to an unknown probability distribution. This leads to uncertainty in our estimate of the energy $E$. To determine this uncertainty in $E$, we construct an estimator $\epsilon(D)$ for $E$, where $D$ represents the measurement data. The term *estimator* refers to a function that uses values of random variables—in our case, the measurement data $D$—to obtain an estimate $\epsilon(D)$ of the true value $E$ [15]. In this section, we present the estimator and will determine its variance and mean squared error.

In order to introduce the estimator $\epsilon(D)$, we have to define the measurement data $D$. The measurement settings used in the experiment were given by Eq. (6) (Sec. III). Here, we describe each measurement setting as one POVM $\Pi_j$ and we collect the POVMs for all measurement settings in the POVM set $\Pi_M = \{\Pi_j : j\}$. We describe the $3^k$ measurement settings from Sec. III with $3^k$ POVMs:

$$\Pi_M = \Big\{\, \Pi_j : \quad j = (j_1, \ldots, j_k), \quad j_a \in \{X, Y, Z\}, \quad a \in \{1, 2, \ldots, k\} \,\Big\}. \qquad (14)$$

As there are exactly $3^k$ different values of the POVM label $j$, we can equivalently treat $j$ as integer $j \in \{1, \ldots, q\}$, $q = 3^k$.

Each POVM has $2^N$ distinguishable outcomes, i.e. $2^N$ POVM elements:

$$\Pi_j = \Big\{\, \Pi_{jl} : \quad l = (l_1, \ldots, l_N), \quad l_c \in \{-1, 1\}, \quad c \in \{1, \ldots, N\} \,\Big\}. \qquad (15)$$

The POVM elements are given by

$$\Pi_{jl} = P_{1, j_1, l_1} \otimes \ldots \otimes P_{k, j_k, l_k} \otimes P_{k+1, j_1, l_{k+1}} \otimes \ldots \otimes P_{2k, j_k, l_{2k}} \otimes \ldots, \qquad (16)$$

where $P_{c, a_i, b_i} = |a_i b_i\rangle_c \langle a_i b_i|$ projects spin $c$ onto the eigenvector $|a_i b_i\rangle$ of the Pauli operator $a_i$ ($= X$, $Y$ or $Z$) with eigenvalue $b_i$. Note that the single-qubit measurement basis, as indicated by $j_1, \ldots j_k$, repeats after $k$ qubits. These POVM elements describe exactly the measurement settings mentioned in Eq. (6) (Sec. III). A single measurement of one of the POVMs produces a single outcome $l = (l_1, \ldots, l_N)$ (Eq. (15)). We will refer to an outcome from a measurement of $\Pi_j$ as $y_j = l = (l_1, \ldots, l_N)$.

For simplicity, we consider the case where each $\Pi_j \in \Pi_M$ has been measured exactly $m$ times. This allows us to take one single outcome $y_j$ from each $\Pi_j \in \Pi_M$ and store them into a vector $x = (y_1, \ldots, y_q)$. From now on, we will refer to $x$ as a "single outcome" or as a "(single) sample". The complete dataset of $m$ outcomes from $q$ POVMs is then structured as $D = (x_1, \ldots, x_m)$. A single element $x_i$ of the dataset $D$ is distributed according a probability density $p(x)$. The sampling distribution $p_m$ describes the distribution of the complete dataset $D$ and is given by $p_m(D) = p(x_1) \ldots p(x_m)$. (The explicit form of $p(x)$ and $p_m(D)$ will be provided in Sec. IV A 5.)

Our estimator will be given in terms of a real-valued function $f(x)$ of a single outcome $x$. To describe its properties, we will need the expectation value (often

referred to as population average)

$$\mathbb{E}_p(f) = \int f(x)p(x)\mathrm{d}x. \tag{17}$$

An expectation value corresponds to an exact value which we want to obtain but cannot access directly because we do not know $p(x)$. We only have access to $m$ samples from $p(x)$, given by $D = (x_1, \ldots, x_m)$ (which is the measurement data from the experiment). Using this data, we define the data expectation value (often referred to as sample average)

$$\mathbb{E}_D(f) = \frac{1}{m} \sum_{i=1}^{m} f(x_i). \tag{18}$$

The data expectation value is a quantity which we can compute from the samples we have, and we will use it to estimate the expectation value. The covariance and data covariance are defined as usual by

$$\mathbb{V}_p(f, g) = \mathbb{E}_p(fg) - \mathbb{E}_p(f)\mathbb{E}_p(g), \qquad \mathbb{V}_D(f, g) = \mathbb{E}_D(fg) - \mathbb{E}_D(f)\mathbb{E}_D(g), \tag{19}$$

and the variance is given by $\mathbb{V}_p(f) = \mathbb{V}_p(f, f)$. We use the following textbook relations (see Sec. IV A 6 below for a proof).

**Lemma 1.** *For two functions $f$ and $g$ of a random variable, we have*

$$\mathbb{E}_{p_m}[\mathbb{E}_D(f)] = \mathbb{E}_p(f), \tag{20}$$

$$\mathbb{V}_{p_m}[\mathbb{E}_D(f), \mathbb{E}_D(g)] = \frac{1}{m}\mathbb{V}_p(f, g), \tag{21}$$

$$\mathbb{V}_p(f, g) = \frac{m}{m-1}\mathbb{E}_{p_m}[\mathbb{V}_D(f, g)]. \tag{22}$$

Our strategy is now to define a function $f_\epsilon(x)$ which provides an estimator for $E = \mathrm{tr}(H\rho_{\mathrm{lab}})$ via $\epsilon(D) = \mathbb{E}_D(f_\epsilon)$. In particular, we will define $f_\epsilon$ such that

$$\mathbb{E}_{p_m}(\epsilon(D)) = \mathbb{E}_p(f_\epsilon) = E. \tag{23}$$

In other words, $\epsilon(D)$ will be an unbiased estimator of $E$. This provides the equality

$$\mathbb{E}_{p_m}[(\epsilon(D) - E)^2] = \mathbb{V}_{p_m}(\epsilon(D)), \tag{24}$$

i.e., the mean squared error (left-hand side) of the estimator is equal its variance (right-hand side). We can estimate the estimator's variance using

$$V_\epsilon(D) = \frac{1}{m}\frac{m}{m-1}\mathbb{V}_D(f_\epsilon). \tag{25}$$

Combining Eqs. (21) and (22) shows

$$\mathbb{E}_{p_m}(V_\epsilon(D)) = \mathbb{V}_{p_m}(\mathbb{E}_D(f_\epsilon)) = \mathbb{V}_{p_m}(\epsilon(D)), \tag{26}$$

i.e., $V_\epsilon(D)$ is an unbiased estimator of the variance of the estimator $\epsilon(D)$ under the sampling distribution $p_m$. We still have to define a function $f_\epsilon(x)$ which satisfies Eq. (23). The definition of $f_\epsilon(x)$ is rather technical and we defer it to the next subsection.

We summarize the results from this section, using the notation from the main text. Using Eq. (9), a lower bound to the fidelity between the certified estimate $|\psi_c^k\rangle = |\psi_{\mathrm{GS}}\rangle$ and the unknown lab state $\rho_{\mathrm{lab}}$ was obtained:

$$\langle\psi_c^k|\rho_{\mathrm{lab}}|\psi_c^k\rangle \geq F_c^k \pm \Delta F_c^k. \tag{27}$$

The value of the fidelity lower bound $F_c^k$ and its measurement uncertainty $\Delta F_c^k$ are given by

$$F_c^k = 1 - \frac{\epsilon(D) - E_0}{E_1 - E_0}, \qquad\qquad \Delta F_c^k = \frac{\sqrt{V_\epsilon(D)}}{E_1 - E_0}, \tag{28}$$

with $\epsilon(D) = \mathbb{E}_D(f_\epsilon)$, $V_\epsilon(D)$ from Eq. (25) and the function $f_\epsilon(x)$ given in the next subsection. Values of the fidelity lower bound $F_c^k$ are mentioned in the main text and shown in Fig. 3 of the main text.

The estimator $\epsilon(D)$ will be seen to be a weighted sum of functions of many independent random variables, in our case we have 27000 random variables from 27 measurement bases and 1000 measurements per measurement basis. While not all 27 measurement bases contribute equally to the weighted sum, all 1000 measurements do contribute equally and it is reasonable to expect that $\epsilon(D)$ will be distributed according to a normal distribution.

### 5. Estimator for the parent Hamiltonian energy

In the last section, we have replaced the uncertified initial estimate by a certified estimate, and we have provided the functional form of the fidelity lower bound which provides the certificate. The function $f_\epsilon$ introduced in the last section still needs to be defined; it is required to obtain values of the fidelity lower bound and its uncertainty. While the value of the fidelity lower bound $F_c^k$ could be obtained from an easier computation than presented below, determining its measurement uncertainty $\Delta F_c^k$ requires the full discussion of this subsection. Incorporating the fact that the local probabilities of Eq. (8) (Sec. III) have been obtained from the global measurement bases of Eq. (6) complicates the computation of the measurement uncertainty $\Delta F_c^k$.

The parent Hamiltonian $H$ from Eq. (11) takes the form

$$H = \sum_{s=1}^{N-k+1} \mathbb{1}_{1,\ldots,s-1} \otimes h_s \otimes \mathbb{1}_{s+k,\ldots,N}, \tag{29}$$

where each term $h_s$ acts only on $k$ out of the total $N$ qubits. We need to estimate the energy $E$, given by

$$E = \sum_{s=1}^{N-k+1} \operatorname{tr}(h_s \rho_s), \tag{30}$$

where $\rho_s$ is the reduced state of $\rho_{\text{lab}}$ on sites $s, s+1, \ldots, s+k-1$. Our first step is providing an expression for $E$ in terms of the local probabilities from Eq. (8) (Sec. III). For this purpose, we define a POVM set $\Pi_L$ whose outcome probabilities correspond to the named local probabilities:

$$\Pi_L = \{\, Q_s: \quad s = 1, \ldots, N-k+1 \,\}. \tag{31}$$

The individual POVMs $Q_s$ are given by

$$Q_s = \{\, Q_{si}: \quad i = (a_1, \ldots, a_k, b_1, \ldots, b_k), \quad a_c \in \{X, Y, Z\}, \quad b_c \in \{-1, 1\} \,\} \tag{32}$$

with $c \in \{1, \ldots, k\}$. Their $6^k$ elements are given by

$$Q_{si} = P_{s, a_1, b_1} \otimes \ldots \otimes P_{s+k-1, a_k, b_k}. \tag{33}$$

As above, $P_{c, a_i, b_i} = |a_i b_i\rangle_c \langle a_i b_i|$ projects spin $c$ onto the eigenvector $|a_i b_i\rangle$ of the Pauli operator $a_i$ ($= X$, $Y$ or $Z$) with eigenvalue $b_i$. It is understood that the elements of $Q_s$ act only on the sites $s, \ldots, s+k-1$ of the $N$-qubit lab state $\rho_{\text{lab}}$.

We will use the linear map

$$\mathcal{M}_s(\rho_s) = [\operatorname{tr}(Q_{si}\rho_s)]_{Q_{si} \in Q_s} \tag{34}$$

which maps a given $k$-qubit state $\rho_s$ on the vector of POVM probabilities $p_{si} = \operatorname{tr}(Q_{si}\rho_s)$. Because the POVMs $Q_s \in \Pi_L$ are informationally-complete, the identity $\overline{\mathcal{M}_s}\mathcal{M}_s(\rho_s) = \rho_s$ holds; here, $\overline{\mathcal{M}_s}$ is the Moore–Penrose pseudoinverse of $\mathcal{M}_s$. This relation provides

$$E = \sum_{s=1}^{N-k+1} \operatorname{tr}(h_s \overline{\mathcal{M}_s}(\mathcal{M}_s(\rho_s))) = \sum_{s=1}^{N-k+1} \sum_i c_{si} p_{si},$$

$$c_{si} = \operatorname{tr}(h_s \overline{\mathcal{M}_s}(e_i)), \quad p_{si} = \operatorname{tr}(Q_{si}\rho_s), \tag{35}$$

where $e_i$ is the $i$-th standard basis vector. We have accomplished the goal of expressing the energy $E$ in terms of the local probabilities $p_{si}$.

If we had measurement data for the POVM set $\Pi_L$ available, we could use simple counting functions

$$\theta_{si}(x) = \begin{cases} 1, & \text{if } y_s = i \text{ with } x = (y_1, \ldots, y_{N-k+1}) \\ 0, & \text{otherwise} \end{cases}, \tag{36}$$

where we have combined single outcomes $y_s$ from $Q_s \in \Pi_L$ into vectors $x = (y_s)_s$ (as has been done in the last subsection for $\Pi_j \in \Pi_M$). It is simple to see that such counting functions estimate probabilities (see below for an explicit example):

$$\mathbb{E}_{p_m}[\mathbb{E}_D(\theta_{si})] = \mathbb{E}_p(\theta_{si}) = p_{si}.$$

In this case, we could obtain the function $f_\epsilon$ by replacing $p_{si}$ with $\theta_{si}$ in Eq. (35).

However, our measurement data is data for the POVM set $\Pi_M$ defined above in Eqs. (14)–(16) and we must estimate the local probabilities $p_{si}$ from that data. We need to establish a relation between the two POVM sets $\Pi_M$ and $\Pi_L$. Because of the particular choice we made for these two sets, it is easy to find non-negative coefficients $c_{si,jl'}$ such that [24]

$$Q_{si} = \sum_{jl'} c_{si,jl'} \Pi_{jl'}^{(s)}, \quad Q_{si} \in Q_s, \quad Q_s \in \Pi_L, \tag{37}$$

where the sum over $l'$ is over the $2^k$ different partial traces

$$\Pi_{jl'}^{(s)} = \text{tr}_{1,\ldots,s-1,s+1,\ldots,N}(\Pi_{jl}) \tag{38}$$

of the $2^N$ elements $\Pi_{jl} \in \Pi_j$. (As before, $\Pi_j \in \Pi_M$.) The partial traces $\Pi_{jl'}^{(s)}$ are rank-1 projectors onto a particular $k$-fold tensor product of eigenvectors of Pauli matrices. Therefore, we enumerate them with an index vector $l' = (l'_1, \ldots, l'_k)$, $l'_a \in \{+1, -1\}$, $a \in \{1, \ldots, k\}$ where $l'_a$ specifies whether the $a$-th eigenvector is spin up or spin down is some direction ($X$, $Y$ or $Z$).

In order to estimate the local probabilities $p_{si}$ from data of the global POVM set $\Pi_M$, we define counting functions for occurences of a local part $(l_s, \ldots, l_{s+k-1})$ of a global outcome $(l_1, \ldots, l_N)$:

$$\theta_{jsl'}(x) = \begin{cases} 1, & \text{if } (l_s, \ldots, l_{s+k-1}) = (l'_1, \ldots, l'_k) \\ & \text{with } x = (y_1, \ldots, y_q) \text{ and } y_j = (l_1, \ldots, l_N), \\ 0, & \text{otherwise}. \end{cases} \tag{39}$$

To show that $\theta_{jsl'}$ can be used to estimate $\text{tr}(\Pi_{jl'}^{(s)}\rho_s)$, we have to complete some definitions. As has been mentioned above, the sampling distribution $p_m(D) = p(x_1) \ldots p(x_m)$ describes the probability distribution of the complete dataset $D = (x_1, \ldots, x_m)$. Single outcomes $x = (y_1, \ldots, y_q)$ contain one outcome $y_j$ for each POVM $\Pi_j \in \Pi_M$. The single-outcome probability density therefore is $p(x) = p_1(y_1) \ldots p_q(y_q)$. We embed the discrete probability distribution with probabilities $p_l^{(j)} = \text{tr}(\Pi_{jl}\rho_{\text{lab}})$, $\Pi_{jl} \in \Pi_j$ of the POVM $\Pi_j$ into a probability density via

$$p_j(y_j) = \sum_l \delta(y_j - l) \, \text{tr}(\Pi_{jl}\rho_{\text{lab}})$$

where we have used $l$ as an integer from $\{1, 2, \ldots, 2^N\}$. The counting functions defined above then have the property

$$\mathbb{E}_p(\theta_{jsl'}) = \text{tr}(\Pi_{jl'}^{(s)}\rho_{\text{lab}}) = \text{tr}(\Pi_{jl'}^{(s)}\rho_s). \tag{40}$$

Finally, we can put everything together to obtain the final function $f_\epsilon$, which will provide the estimator $\epsilon(D) = \mathbb{E}_D(f_\epsilon)$ of $E$. First, we define

$$f_{si}(x) = \sum_{jl'} c_{si,jl'} \theta_{jsl'}(x) \tag{41}$$

and observe

$$\mathbb{E}_p(f_{si}) = \sum_{jl'} c_{si,jl'} \operatorname{tr}(\Pi_{jl'}^{(s)} \rho_s) = \operatorname{tr}(Q_{si}\rho_s) = p_{si}. \qquad (42)$$

We define $f_\epsilon$ by

$$f_\epsilon(x) = \sum_s \sum_i c_{si} f_{si}(x), \quad c_{si} = \operatorname{tr}(h_s \overline{\mathcal{M}_s}(e_i)). \qquad (43)$$

Using what we have learned so far, we obtain

$$\mathbb{E}_{p_m}(\epsilon(D)) = \mathbb{E}_{p_m}(\mathbb{E}_D(f_\epsilon)) = \mathbb{E}_p(f_\epsilon) = \sum_{si} c_{si} \mathbb{E}_p(f_{si}) = E. \qquad (44)$$

This shows that the function $\epsilon(D)$ is an unbiased estimator of the energy $E = \operatorname{tr}(H\rho_{\text{lab}})$. This scheme has been implemented as part of the Python library mpnum [14, function `mpnum.povm.MPPovmList.est_lfun_from`].

### 6. Proof for basic variance relations

In this section, we proof three basic equalities used in Sec. IV A 4. Their proof is included for completeness.

Let $p(x)$ be some probability, $D = (x_1, \ldots, x_m)$ a dataset of $m$ samples from $p$, and let $p_m(D) = p(x_1) \ldots p(x_m)$ the sampling distribution which describes the probability density of the complete dataset $D$. We will also use the definitions from Eqs. (17)–(19) on page 12. The equalities which we proof here provide relations between expectation values, sampling distribution expectation values and data expectation values. Above, they have been used to estimate the variance of an estimator from data. Lemma 1 states: For two functions $f$ and $g$ of a random variable, we have

$$\mathbb{E}_{p_m}[\mathbb{E}_D(f)] = \mathbb{E}_p(f),$$

$$\mathbb{V}_{p_m}[\mathbb{E}_D(f), \mathbb{E}_D(g)] = \frac{1}{m}\mathbb{V}_p(f, g),$$

$$\mathbb{V}_p(f, g) = \frac{m}{m-1}\mathbb{E}_{p_m}[\mathbb{V}_D(f, g)]$$

*Proof.* First equation:

$$\mathbb{E}_{p_m}[\mathbb{E}_D(f)] = \int \frac{1}{m}\sum_{i=1}^m f(x_i)p(x_1) \ldots p(x_m)\mathrm{d}x_1 \ldots \mathrm{d}x_m = \frac{m}{m}\mathbb{E}_p(f). \qquad (45)$$

For the second and third equation, we first compute

$$\begin{aligned}
\mathbb{E}_{p_m}[\mathbb{E}_D(f), \mathbb{E}_D(g)] &= \int \frac{1}{m^2}\sum_{i=1}^m\sum_{j=1}^m f(x_i)g(x_j)p(x_1) \ldots p(x_m)\mathrm{d}x_1 \ldots \mathrm{d}x_m \\
&= \frac{m}{m^2}\mathbb{E}_p(fg) + \frac{m^2-m}{m^2}\mathbb{E}_p(f)\mathbb{E}_p(g) \\
&= \frac{1}{m}\mathbb{V}_p(f, g) + \mathbb{E}_p(f)\mathbb{E}_p(g)
\end{aligned} \qquad (46)$$

This provides

$$\begin{aligned}
\mathbb{V}_{p_m}[\mathbb{E}_D(f), \mathbb{E}_D(g)] &= \mathbb{E}_{p_m}[\mathbb{E}_D(f), \mathbb{E}_D(g)] - \mathbb{E}_{p_m}[\mathbb{E}_D(f)]\,\mathbb{E}_{p_m}[\mathbb{E}_D(g)] \\
&= \frac{1}{m}\mathbb{V}_p(f, g)
\end{aligned} \qquad (47)$$

and

$$\mathbb{E}_{p_m}[\mathbb{V}_D(f, g)] = \mathbb{E}_p(fg) - \mathbb{E}_{p_m}[\mathbb{E}_D(f), \mathbb{E}_D(g)] = (1 - \frac{1}{m})\mathbb{V}_p(f, g), \qquad (48)$$

which is the required relation. □

## B. Simulations of MPS tomography and certification

In Figure 3a of the main text, fidelity lower bounds $F_c^k$ based on an ideal model of the tomographic process are presented. Here we explain how they were obtained.

The ideal model of the tomographic process assumes a perfect initial state $|\Phi(0)\rangle = |\uparrow\downarrow\uparrow\downarrow\ldots\rangle$ in the $\sigma^z$ basis and an idealised time evolution of the quantum simulator, described by the Hamiltonian

$$H = \sum_{i=1}^{N}(B + B_i)\sigma_i^z + \sum_{i,j=1}^{N} J_{ij}\sigma_i^x\sigma_j^x. \tag{49}$$

The transverse fields $B$, $B_i$ and coupling matrix elements $J_{ij}$ have been described above. Note that in the limit $B \gg |J_{ij}|$, which is upheld in the experiments, the above Hamiltonian is equivalent to an XY model in a transverse field, as described in II. For more details see the supplementary material of [3]. The ideal time evolution $|\Phi(t)\rangle = \exp(-iHt)|\Phi(0)\rangle$ is computed using the library function `scipy.sparse.linalg.expm_multiply` [13]. For the simulation with a mixed initial state discussed in Sec. IV C below, the same library function has been used to propagate 15 different pure initial states in time. It was convenient to convert the resulting state to a purified MPS representation with a single ancilla site of dimension 15.

The simulated MPS tomography and certification can be performed with exact knowledge of local probabilities or data from a finite number of simulated measurements (simulation mentioned in Sec. IV C below).

*Exact knowledge of local observables.* Assuming exact knowledge of local observables simplifies the simulation. The MPS tomography algorithms are run with the exact values of the $6^k$ probabilities describing the measurement outcomes of the $3^k$ $k$-fold tensor products of the Pauli X, Y and Z matrices for each of the $N-k+1$ local blocks. Computing the energy $E = \mathrm{tr}(H\rho_{\mathrm{lab}})$ of the (now known) ideal lab state $|\Phi(t)\rangle$ in terms of the parent Hamiltonian $H$ is simplified considerably as we can compute the exact local reductions of $\rho_{\mathrm{lab}}$. As a consequence, the resulting fidelity lower bound is known without uncertainty as well.

This numerical simulation represents the expected performance of certified MPS tomography in a case where the perfect initial state is prepared, the simulator produces ideal unitary dynamics and an infinite number of perfect measurements are performed. This is shown in Fig. 3a in the main text.

*Finite number of simulated measurements of global observables.* While the measurement of a global observable such as $X^{\otimes N}$ can yield one of $2^N$ outcomes, it is simple to draw a sample from the probability distribution of measurement outcomes if a matrix product description of the state on which measurements shall be simulated is available: One can simply compute the local reduced state on the first qubit, simulate a single-qubit measurement there and continue by computing the reduced state state of the second qubit conditioned on the outcome of the first measurement, etc. A more direct way to implement the same procedure uses a matrix product description of the POVM in question to obtain a matrix product representation of the measurement outcome probabilities. Marginal and conditional distributions of the full outcome probability distribution can be efficiently obtained. This procedure has been implemented as part of the mpnum library [14] (function `mpnum.povm.MPPovm.sample(method='cond')`). However, for 14 qubits, it was still more convenient to convert the matrix product representation of the outcome probability distribution to a full array with $2^{14}$ elements and sample from the full description.

## C. Modelling initial Néel state errors for the 14-spin experiments

In the main text we state that the differences, between the experimentally-obtained and ideal-simulator model fidelity bounds $F_c^3$ at $t = 4\ ms = t_{14}$ are largely explained by errors in the initial Néel state preparation for 14 spins. Here we aim to convince the reader of that.

In a previous section of this supplementary material entitled 'Simulator Initialisation' we showed that the initial 14-spin Néel state was prepared with a (directly-measured) fidelity of $0.89 \pm 0.01$, compared to $0.967 \pm 0.006$ for the 8-spin case. This

corresponds to a significantly larger error-per-particle for the 14-spin case. We performed numerical simulations to determine the effect of the errors in the 14-spin initial state have on the MPS reconstruction of the 14-spin quench experiment. Specifically, we asked how detrimental the initial state error are expected to be to the performance of the MPS tomography for 3-site measurements after $t = 4$ ms of evolution (the time at which direct state fidelity estimation was carried out).

To do this, we first modelled the initial state error in the following way. Analysis of the direct measurement results for the 14-spin initial state in the lab show that, out of 1000 times that we prepared and measured the state in the $z$-basis, 893 times we observed the Néel state (hence the fidelity of 0.893). 93 times we observe a state with one spin flip error. For the remaining 12 times, we observed two spin flip errors. The errors are most likely caused by fluctuations in the intensity of our addressed laser beam, meaning that these erroneous states are added in mixture with the ideal Neel state. We built a noisy model for the initial lab state as an appropriately weighted mixture of the ideal Neel state and single spin flip errors. We call this model, of the noisy initial 14 spin state, $\rho^{14}_{noisysim}$.

In the next step, we numerically simulate obtaining $k = 3$-site estimates of the local reductions of $\rho^{14}_{noisysim}$, using 1000 measurements per measurement basis (the number of measurements that we made in the actual experiments in the lab). We insert these noisy local reductions into the MPS tomography search algorithm. The MPS reconstruction then proceeds as usual, as described in a previous section. The output is a certified estimate for the fidelity lower bound $F^3_{c,noisy}$, that we would expect to achieve when measuring such a noisy state in the lab. The result, for $t = 0$ ms, for the 14 spin noisy model, is the lower bound $F^3_{c,noisysim} = 0.90 \pm 0.03$. This is to be compared with the direct (exact) fidelity measurement of $0.89 \pm 0.01$ in the lab and the MPS-tomography lower fidelity bound, for measurements on $k = 1$-site, of $0.90 \pm 0.04$. Clearly all agree well and the lower bounds are tight.

The result, for the 14 spin noisy model at $t = 4$ ms, is $F^3_{c,noisysim} = 0.49 \pm 0.07$. This is to be compared with the lower bound obtained in the experiment $F^3_c = 0.39 \pm 0.08$. Of course, the described errors in the initial state are not the only errors in the experiment, however, we conclude that they are largely responsible for the difference between experimentally obtained lower bound via MPS tomography, and idealised model of a perfect simulator. So, noise adding mixture to the initial state preparation explains why the experimentally obtained fidelity lower bounds for 14 spins are lower than the ideal case. In the future, we should work on keeping the error-per-particle constant when scaling up our system. This should be relatively straightforward for up to several tens of spins, by rebuilding the optical setup used to deliver the addressed laser beam. Other sources of error that we considered, and found to play minor roles by equivalent numerical modelling, are: the finite lifetime of the excited spin (atomic) state; correlated dephasing due to correlated fast fluctuations in real magnetic fields around the ion string; small errors in the analysis pulses that determine the measurement bases.

## V. BIPARTITE AND TRIPARTITE NEGATIVITY:

In Figures 2 and 4 of the main text we present the entanglement observed in local reductions, quantified by two forms of negativity. In this section, we recall the definitions of these quantities.

Negativity is an entanglement measure that can be computed effectively and easily for a generic bipartite mixed state $\rho$, based on the trace norm of the partial transpose $\rho^{T_A}$ [16]:

$$\mathcal{N}(\rho) = \frac{\|\rho^{T_A}\|_1 - 1}{2} \qquad .$$

This expression corresponds to the absolute value of the sum of negative eigenvalues of $\rho^{T_A}$ and vanishes for unentangled states. We rescale the negativity such that maximum entanglement corresponds to $\mathcal{N} = 1$ and use this to quantify the degree of entanglement in the reduced 2-qubit density matrices $\rho^{(2)}$ of neighbouring spin pairs:

$$N_2(\rho^{(2)}) = \|\rho^{T_A}\|_1 - 1 = 2 \cdot |\sum_j \mu_j| \qquad ,$$

where $\mu_j$ are the negative eigenvalues of $\rho^{T_A}$. For estimating the degree of entanglement in neighbouring spin triplets $\rho^{(3)}$ we use the tripartite negativity $N_3$ [17]. It is defined as the geometric mean of all three bipartite negativity splittings:

$$N_3(\rho^{(3)}) = \sqrt[3]{N_2(\rho_{1,2}) \cdot N_2(\rho_{2,3}) \cdot N_2(\rho_{1,3})}\qquad,$$

with the reduced 2-qubit density matrices of spin 1 and 2 ($\rho_{1,2}$), spin 2 and 3 ($\rho_{2,3}$), spin 1 and 3 ($\rho_{1,3}$).

## VI.    EXTENDED EXPERIMENTAL ANALYSIS AND RESULTS

### A.    Single site magnetisation dynamics for 8 and 14 spin quenches

Figures 2 and 4 in the main text present the measured single-site magnetisation dynamics for 8 and 14-spin quenches, for the Néel initial state. In order to calibrate our experimental system, and compare the model dynamics with the results in the lab, we run quenches starting with a spin state that contains a single spin excitation (local quench [2]). The subsequent dynamics reveal the spreading of correlations from a single site and provide a useful visualisation of the approximate light-like cones (approximate maximum group velocity for the spread of information). Figure 10 compares the measured and model single-site magnetisation dynamics for a local quench, showing how the single excitation spreads out in the system. Two kinds of light-like cones are presented, as described in the caption. The faster of which is the same as those presented in the main text. The maximum rate of information spreading should not depend on the initial state, but only the spin-spin interaction Hamiltonian. Figure 11 presents the single site magnetisation dynamics for the Néel initial state, for 8 and 14 quenches, with the same light-like cones. The experimental results are the same as those in figures 2 and 4 in the main text. In all cases, the match between data and model is excellent.

### B.    Local reductions and correlation matrices for the 8-spin quench

We now present dynamical properties of the experimentally reconstructed local reductions of single spins, neighbouring spin pairs and neighbouring spin triplets, during the 8-spin quench experiments. These local reductions are reconstructed directly from the local measurements, using full quantum state tomography (not MPS tomography), which searches over all possible physical states (pure and mixed) to find an optimum fit with the data. The fidelities of the reconstructed local reductions with the ideal simulator model, are presented in figure 12. For fidelity $F$ of an $N$-spin state, a measure of error-per particle is $E_N = \log_2(F)/N$. For the initial states (time $t = 0$) in figure 12, we find $E_1$, $E_2$ and $E_3$ all agree to within measurement uncertainty. That is, the error in the initial states of single spins, pairs and triplets is statistically indistinguishable. The single spin fidelities reach unity after a few ms of evolution. This can be understood by considering that single spin states rapidly become fully mixed due to quantum correlations. The fully mixed state is unchanged under any unitary rotations and many physical noise sources. The Von Neumann entropy ('quantum entropy') of example local reductions during the 8 spin quench are presented in figure 13. The maximal value for Von Neumann entropy of an $N$ spin (qubit) state is $N$, corresponding to a maximally mixed state (shown as horizontal dashed lines in the figure).

Figure 15 presents the dynamics of entanglement, quantified by the negativity, in the experimentally-reconstructed local reductions. Spin pair entanglement maximises at 2 ms and spin triplet entanglement between 3 and 4 ms. As the simulator evolves further, entanglement reduces, first in pairs then in triplets agreeing with the spread of correlations in the system. The measured results closely fit an ideal model of the simulator (not shown for clarity).

16 presents the fidelity between the experimentally reconstructed neighbouring spin-pair states and a maximally entangled two-spin state. The entanglement between neighbours reaches a maximum between 2 and 3 ms. Quantifying entanglement in terms of the fidelity with a maximally entangled two-qubit state has operational meaning: states with fidelities above 50% are distillable. That is, given many

copies of states with fidelities above this threshold, fewer states with higher quality entanglement can be distilled [18].

In Figure 3 of the main text, correlation matrices are presented showing correlations in various bases between spin pairs across the 8-spin system. Figure 17 presents correlations measured in additional bases and compares them with those captured in the measured MPS reconstructions and those from an ideal model of the simulator.

### C. Correlation matrices for the 14-spin quench

In Figure 4 of the main text, correlation matrices are presented showing correlations in various bases between spin pairs across the 14-spin system. Figure 18 presents correlations measured in additional bases and compares them with those captured in the measured MPS reconstructions and those from an ideal model of the simulator.

### D. Certified MPS reconstructions for 8 spin quench

In Figure 3a of the main text, the fidelity bounds for the 8-spin quench experiment are presented. In that figure, the data are compared with a theoretical model (numerical simulation) which shows the MPS tomography fidelity bounds that would be obtained for a idealised model of the simulator. Specifically, the exact local reductions of the model state $|\Phi(t)\rangle = \exp(-iH_{XY}t)|\Phi(0)\rangle$ are used as input to the MPS tomography algorithm and certification process. This idealised simulation is described in section IV B. We would only expect to achieve these result in the laboratory if first, our system exactly implemented the XY model Hamiltonian and, second, we carried out an infinite number of perfect measurements to determine the local reductions. Clearly we do neither of these things. Figure 19 compares the same experimental data with a more realistic model (Shaded areas). This model again uses the ideal simulator states but considers the effect of carrying out a 1000 (perfect) measurements per basis to identify local reductions, as done in the experiments. This model is described in more detail in section IV B. From the figure 19 we conclude that: 1. the differences between data and the original perfect model are largely explained by the finite number of measurements used in experiments and: 2. there is not much to be gained from doing more measurements in the lab.

### E. Von Neumann Entropy over all bipartitions

Figure 20 presents the Von Neumann entropies $S$ for 8 and 14 spins for the reconstructed MPS from both experimental data and theoretical simulations. The entropy is plotted as a function of all bipartitions over the string.

$$S = -tr(\rho \log_2 \rho) \text{ or also } S = -\sum_{j=1}^{N} \eta_j \log_2 \eta_j \qquad,$$

with the state $\rho$ and the eigenvalues $\eta_j$. For the 8-spin system the time evolution of the entropy appears in different color-coding. It can be seen that the entropy increases with time and reaches half of its maximum possible value ($S_{max} = log_2(N) \approx 2$) at 5 ms. The entropy of the reconstructed MPS state agrees with the expected values derived from theoretical simulations.

## VII. DIRECT FIDELITY ESTIMATION

The fidelity lower bounds returned by the certification procedure described in the main text are merely lower bounds. That is, the actual overlap (fidelity) between the two states could take any value between the certificate and unity. Which value does the fidelity actually take is a natural question to ask. To estimate the overlap, we implement the method of direct fidelity estimation (DFE) [19, 20]. The DFE method uses measurements on a lab state to determine an estimate of the fidelity between the

lab state (generally mixed) and a given pure state, which we set as the output MPS from the efficient tomography procedure. In this section, we provide an overview of DFE with emphasis on the current experiment.

### A. Overview of direct fidelity estimation procedure

Before describing the employed DFE method, we recollect relevant notation. Consider state $\rho_{\text{lab}}$ implemented in the laboratory and let $\left|\Psi_c^3\right\rangle$ be the pure-state estimate obtained from MPS tomography on $\rho_{\text{lab}}$. The fidelity of the estimate with respect to the lab state is defined as

$$F\left(\left|\Psi_c^3\right\rangle, \rho_{\text{lab}}\right) \stackrel{\text{def}}{=} \left\langle\Psi_c^3\left|\rho_{\text{lab}}\right|\Psi_c^3\right\rangle. \tag{50}$$

Define $\{P^k : k = 1, 2, \ldots 4^N\}$ as the orthonormal Pauli operators $\text{tr}\left(P^\ell P^k\right) \stackrel{\text{def}}{=} \delta_{\ell,k}$. The Pauli operators form a basis for Hermitian operators acting on the system Hilbert space. In this basis, the lab state and the MPS estimate can be represented by their characteristic functions

$$\rho_{\text{lab}}^k = \text{tr}\left(P^k \rho_{\text{lab}}\right), \quad \sigma^k = \text{tr}\left(P^k\left|\Psi_c^3\right\rangle\left\langle\Psi_c^3\right|\right) = \left\langle\Psi_c^3\left|P^k\right|\Psi_c^3\right\rangle \tag{51}$$

respectively. The fidelity (50) is expressed in terms of the characteristic functions as

$$F\left(\left|\Psi_c^3\right\rangle, \rho_{\text{lab}}\right) = \sum_{k=1}^{4^N} \rho_{\text{lab}}^k \sigma^k. \tag{52}$$

Estimating the fidelity by a straightforward application of Equation (52) requires measuring $4^N$ observables, each of which requires exponentially many (in $N$) measurements, and is thus infeasible. This implies that over two hundred million observables need to be measured in our setting of fourteen spins, which is clearly infeasible.

The DFE method leverages the knowledge of the MPS estimate to overcome this infeasibility. Specifically, the full summation of Equation (52) is replaced by a preferential summation over those values of $k$ for which MPS-estimate components $\sigma^k$ are likely to be large. In other words, more measurements are made in those basis elements $P^k$ for which the MPS estimate is known to have a large expectation value.

The preferential summation to obtain the fidelity estimate is performed as follows. First, the fidelity is expressed as the expectation value

$$F\left(\left|\Psi_c^3\right\rangle, \rho\right) = \sum_{k=1}^{4^N} \rho_{\text{lab}}^k \sigma^k = \sum_{k=1}^{4^N} q^k \frac{\rho_{\text{lab}}^k}{\sigma^k}, \tag{53}$$

of a random variable $\rho_{\text{lab}}^k / \sigma^k$ over probability distribution $\left\{q^k \stackrel{\text{def}}{=} (\sigma^k)^2 : k = 1, 2, \ldots, 4^N\right\}$. Next, this expectation value is estimated using a Monte Carlo approach. That is, $M$ random indices $\left\{k_1, k_2, \ldots, k_M; k_m \in \{1, 2, \ldots, 4^N\}\right\}$ are generated according to the probability distribution $q^k$ where $M$ is chosen based on a desired threshold of error. In the experiment we set $M = 250$. In other words, the number of observables required to estimate the fidelity are reduced by six orders of magnitude.

The fidelity is obtained from the estimator

$$\overline{F} \stackrel{\text{def}}{=} \frac{1}{M} \sum_{i=1}^{M} \frac{\overline{\rho}_{\text{lab}}^{k_i}}{\sigma^{k_i}} \approx F\left(\left|\Psi_c^3\right\rangle, \rho_{\text{lab}}\right), \tag{54}$$

where $\overline{\rho}_{\text{lab}}^{k_m}$ estimates the lab-state expectation value (51). These estimates are obtained from measuring $M = 250$ observables using many copies of the state for each observable, where a total of $5 \times 10^5$ copies of the state were used. The number of copies $N_k$ spent to measure a particular Pauli operator $P_k$ was chosen to be proportional to the inverse square of its calculated expectation value $\sigma^{k_i}$ for $\Psi_c^3$ in order to prevent the error in the estimator $\overline{F}$ to be dominated by those terms of the sum in Eq. 54 for which $\sigma^{k_i}$ is very small.

The fidelity estimate (54) requires sampling the indices $\{k_1, k_2, \ldots, k_M\}$; this sampling is efficiently performed using classical algorithms outlined in the supplementary material of Reference [20]. Finally, the values $\{\sigma^{k_1}, \sigma^{k_2}, \ldots, \sigma^{k_M}\}$ are determined by efficient MPS-based classical algorithms. This completes a summary of the direct fidelity estimation procedure.

The values of $\overline{\rho}_{\text{lab}}^{k_m}$ obtained in the experiment and the corresponding calculated $\sigma^{k_m}$ are depicted in Figure 21(a) for the $M = 250$ different observables, which are indexed by $m$. The distribution of $\overline{\rho}_{\text{lab}}^{k_m} / \sigma^{k_m}$ for different $m$ is presented in Figure 21(b). Based on this distribution, we infer a fidelity estimate of 0.74. We present the procedure for calculating the error bars on this estimate in the next section.

### B. Mean-square error and bias of DFE estimates

The fidelity estimate (54) is amenable to random error, which arises from (i) choosing a smaller number $M$ of indices than the maximum possible $4^N$ and from (ii) using a finite number of measurements to estimate the expectation values $\{\rho_{\text{lab}}^{k_1}, \rho_{\text{lab}}^{k_2}, \ldots, \rho_{\text{lab}}^{k_M}\}$. This random error is quantified by the variance estimator

$$\overline{\text{var}}[F] \overset{\text{def}}{=} \frac{1}{M(M-1)} \sum_{m=1}^{M} \left( \frac{\overline{\rho}_{\text{lab}}^{k_m}}{\sigma^{k_m}} - \overline{F} \right)^2, \tag{55}$$

where $\overline{F}$ is determined using Equation 54. In the remainder of this section, we justify that $\overline{F}\left( \left| \Psi_c^3 \right\rangle, \rho_{\text{lab}} \right)$ (54) and $\overline{\text{var}}[F]$ (55) are unbiased estimators of the fidelity and the variance of the fidelity. In other words, the expectation value of the random variables $\overline{F}$ and $\overline{\text{var}}[F]$ are respectively equal to the true value of the fidelity and the fidelity variance.

In order to account for random error from the experiment, we connect the fidelity estimator (54) and variance estimator (55) with the measurement outcomes from the experiment. We account for random error in estimates $\overline{\rho}_{\text{lab}}^{k_m}$ by expressing $\rho_{\text{lab}}^{k_m}$ in terms of measurement outcomes:

$$\rho_{\text{lab}}^k = \text{tr}\left( P^k \rho_{\text{lab}} \right) = \sum_{i=1}^{2^N} \lambda_i^k \, \text{tr}\left( \left| \psi_i^k \right\rangle \left\langle \psi_i^k \right| \rho_{\text{lab}} \right) = \sum_{i=1}^{2^N} \lambda_i^k \left\langle \psi_i^k \left| \rho_{\text{lab}} \right| \psi_i^k \right\rangle = \sum_{i=1}^{2^N} \lambda_i^k p_i^k, \tag{56}$$

where $\{\lambda_i^k : k = 1, 2, \ldots, 2^N\}$ are the eigenvalues of Pauli operators

$$P^k = \sum_{i=1}^{2^N} \lambda_i^k \left| \psi_i^k \right\rangle \left\langle \psi_i^k \right|, \quad \lambda_i^k = \pm \frac{1}{\sqrt{2^N}}, \tag{57}$$

and $\left\{ p_i^k \overset{\text{def}}{=} \left\langle \psi_i^k \left| \rho_{\text{lab}} \right| \psi_i^k \right\rangle : k = 1, 2, \ldots, 2^N \right\}$ is a probability distribution. Finally, the fidelity can be expressed as the expectation value

$$\begin{aligned} F\left( \left| \Psi_c^3 \right\rangle, \rho_{\text{lab}} \right) &= \sum_{k=1}^{4^N} q^k \frac{\rho_{\text{lab}}^k}{\sigma^k} = \sum_{k=1}^{4^N} q^k \frac{\sum_{i=1}^{2^N} \lambda_i^k p_i^k}{\sigma^k} \\ &= \sum_{k=1}^{4^N} \sum_{i=1}^{2^N} q^k p_i^k \frac{\lambda_i^k}{\sigma^k} = \sum_{k=1}^{4^N} \sum_{i=1}^{2^N} u_i^k \frac{\lambda_i^k}{\sigma^k}, \end{aligned} \tag{58}$$

of the random variable $\frac{\lambda_i^k}{\sigma^k}$ over the probability distribution $\left\{ u_i^k \overset{\text{def}}{=} q^k p_i^k : k = 1, 2, \ldots, 4^N, i = 1, 2, \ldots, 2^N \right\}$.

In the experiment, we choose $M$ observables $\{P^1, P^2, \ldots, P^M\}$. Each observable $P^m$ is measured $N_m$ times, with each measurement returning outcome value $\lambda_{i_n}^{k_m}$. The returned measurement outcomes are used to obtain expectation values as

$$\overline{\rho}_{\text{lab}}^{k_m} = \frac{1}{N_m} \sum_{n=1}^{N_m} \lambda_{i_n}^{k_m}. \tag{59}$$

Thus, the fidelity estimate (54) is obtained from measurement outcomes as

$$\overline{F}\left(\left|\Psi_c^3\right\rangle, \rho_{\text{lab}}\right) = \frac{1}{M} \sum_{m=1}^{M} \frac{\frac{1}{N_m} \sum_{n=1}^{N_m} \lambda_{i_n}^{k_m}}{\sigma^{k_m}}. \tag{60}$$

The variance of $\overline{F}\left(\left|\Psi_c^3\right\rangle, \rho_{\text{lab}}\right)$ is obtained using estimator (55), which we express in terms of measurement outcomes by the following simplification. Consider

$$\overline{\text{var}}[F] \stackrel{\text{def}}{=} \frac{1}{M(M-1)} \sum_{m=1}^{M} \left(\frac{\overline{\rho}_{\text{lab}}^{k_m}}{\sigma^{k_m}} - \overline{F}\right)^2, \tag{61}$$

$$= \frac{1}{M(M-1)} \sum_{m=1}^{M} \left(\frac{\left(\overline{\rho}_{\text{lab}}^{k_m}\right)^2}{\left(\sigma^{k_m}\right)^2} - 2\frac{\overline{\rho}_{\text{lab}}^{k_m}}{\sigma^{k_m}}\overline{F} + \left(\overline{F}\right)^2\right), \tag{62}$$

$$= \frac{1}{M(M-1)} \sum_{m=1}^{M} \frac{\left(\overline{\rho}_{\text{lab}}^{k_m}\right)^2}{\left(\sigma^{k_m}\right)^2} - \frac{M}{M(M-1)} \left(\overline{F}\right)^2, \tag{63}$$

where we have used the definition (54) to obtain (63) from (62). Substituting the expression for estimators $\overline{F}$ and $\overline{\rho}_{\text{lab}}^k$, we obtain

$$\overline{\text{var}}[F] = \frac{1}{M(M-1)} \sum_{m=1}^{M} \frac{\left(\frac{1}{N_m} \sum_{n=1}^{N_m} \lambda_{i_n}^{k_m}\right)^2}{\left(\sigma^{k_m}\right)^2}$$
$$- \frac{M}{M(M-1)} \left(\frac{1}{M} \sum_{m=1}^{M} \frac{\frac{1}{N_m} \sum_{n=1}^{N_m} \lambda_{i_n}^{k_m}}{\sigma^{k_m}}\right)^2 \tag{64}$$

$$= \frac{1}{M(M-1)} \sum_{m=1}^{M} \frac{1}{N_m^2} \sum_{n,n'=1}^{N_m} \frac{\lambda_{i_n}^{k_m}}{\sigma^{k_m}} \frac{\lambda_{i_{n'}}^{k_m}}{\sigma^{k_m}}$$
$$- \frac{1}{M^2(M-1)} \sum_{m,m'=1}^{M} \frac{1}{N_m N_{m'}} \sum_{n,n'=1}^{N_m} \frac{\lambda_{i_n}^{k_m}}{\sigma^{k_m}} \frac{\lambda_{i_{n'}}^{k_{m'}}}{\sigma^{k_{m'}}} \tag{65}$$

$$= \frac{1}{M^2} \sum_{m=1}^{M} \frac{1}{N_m^2} \sum_{n,n'=1}^{N_m} \frac{\lambda_{i_n}^{k_m}}{\sigma^{k_m}} \frac{\lambda_{i_{n'}}^{k_m}}{\sigma^{k_m}}$$
$$- \frac{1}{M^2(M-1)} \sum_{\substack{m,m'=1 \\ m \neq m'}}^{M} \frac{1}{N_m N_{m'}} \sum_{n,n'=1}^{N_m} \frac{\lambda_{i_n}^{k_m}}{\sigma^{k_m}} \frac{\lambda_{i_{n'}}^{k_{m'}}}{\sigma^{k_{m'}}}, \tag{66}$$

which is the variance estimator in terms of measurement outcomes

Now we show that the fidelity estimator (60) is an unbiased estimator. Consider the expectation value of the fidelity

$$\mathbb{E}[\overline{F}] = \mathbb{E}\left[\frac{1}{M} \sum_{m=1}^{M} \frac{\frac{1}{N_m} \sum_{n=1}^{N_m} \lambda_{i_n}^{k_m}}{\sigma^{k_m}}\right] \tag{67}$$

$$= \frac{1}{M} \sum_{m=1}^{M} \frac{1}{N_m} \sum_{n=1}^{N_m} \mathbb{E}\left[\frac{\lambda_{i_n}^{k_m}}{\sigma^{k_m}}\right]. \tag{68}$$

We note that the expectation value of $\frac{\lambda_{i_n}^{k_m}}{\sigma^{k_m}}$ is equal to the true fidelity

$$\mathbb{E}\left[\frac{\lambda_{i_n}^{k_m}}{\sigma^{k_m}}\right] = \sum_{k_m=}^{4^N} \sum_{i_n=1}^{2^N} u_{i_n}^{k_m} \frac{\lambda_{i_n}^{k_m}}{\sigma^{k_m}} = F\left(\left|\Psi_c^3\right\rangle, \rho_{\text{lab}}\right) \tag{69}$$

because each of the $\lambda_{i_n}^{k_m}$ values are drawn from the same distribution for each value of $m$ and $n$. Substituting Equation (69) in the fidelity expectation value (68), we obtain

$$\mathbb{E}[\overline{F}] = F\left(\left|\Psi_c^3\right\rangle, \rho\right), \tag{70}$$

which implies that $\overline{F}$ is an unbiased estimator of the fidelity.

Finally, we show $\overline{\mathrm{var}}[F]$ is an unbiased estimator, i.e., that the expectation value of $\overline{\mathrm{var}}[F]$ is the same as the true variance

$$\mathrm{var}[F] \stackrel{\text{def}}{=} \mathbb{E}\left[\overline{F}^2\right] - \left(\mathbb{E}\left[\overline{F}\right]\right)^2 \tag{71}$$

of the estimator $\overline{F}$. From Equation (66), we have the expectation value,

$$\mathbb{E}\left(\overline{\mathrm{var}}[F]\right) = \mathbb{E}\left( \frac{1}{M^2} \sum_{m=1}^{M} \frac{1}{N_m^2} \sum_{n,n'=1}^{N_m} \frac{\lambda_{i_n}^{k_m}}{\sigma^{k_m}} \frac{\lambda_{i_{n'}}^{k_m}}{\sigma^{k_m}} \right.$$
$$\left. - \frac{1}{M^2(M-1)} \sum_{\substack{m,m'=1 \\ m \neq m'}}^{M} \frac{1}{N_m N_{m'}} \sum_{n,n'=1}^{N_m} \frac{\lambda_{i_n}^{k_m}}{\sigma^{k_m}} \frac{\lambda_{i_{n'}}^{k_{m'}}}{\sigma^{k_{m'}}} \right), \tag{72}$$

which we simplify as

$$\mathbb{E}\left(\overline{\mathrm{var}}[F]\right) = \mathbb{E}\left( \frac{1}{M^2} \sum_{m=1}^{M} \frac{1}{N_m^2} \sum_{n,n'=1}^{N_m} \frac{\lambda_{i_n}^{k_m}}{\sigma^{k_m}} \frac{\lambda_{i_{n'}}^{k_m}}{\sigma^{k_m}} \right)$$
$$- \mathbb{E}\left( \frac{1}{M^2(M-1)} \sum_{\substack{m,m'=1 \\ m \neq m'}}^{M} \frac{1}{N_m N_{m'}} \sum_{n,n'=1}^{N_m} \frac{\lambda_{i_n}^{k_m}}{\sigma^{k_m}} \frac{\lambda_{i_{n'}}^{k_{m'}}}{\sigma^{k_{m'}}} \right) \tag{73}$$

$$= \mathbb{E}\left( \frac{1}{M^2} \sum_{m=1}^{M} \frac{1}{N_m^2} \sum_{n,n'=1}^{N_m} \frac{\lambda_{i_n}^{k_m}}{\sigma^{k_m}} \frac{\lambda_{i_{n'}}^{k_m}}{\sigma^{k_m}} \right)$$
$$- \frac{1}{M^2(M-1)} \sum_{\substack{m,m'=1 \\ m \neq m'}}^{M} \mathbb{E}\left( \sum_{n=1}^{N_m} \frac{1}{N_m} \frac{\lambda_{i_n}^{k_m}}{\sigma^{k_m}} \right) \mathbb{E}\left( \sum_{n'=1}^{N_m} \frac{1}{N_{m'}} \frac{\lambda_{i_{n'}}^{k_{m'}}}{\sigma^{k_{m'}}} \right) \tag{74}$$

$$= \mathbb{E}\left( \frac{1}{M^2} \sum_{m=1}^{M} \frac{1}{N_m^2} \sum_{n,n'=1}^{N_m} \frac{\lambda_{i_n}^{k_m}}{\sigma^{k_m}} \frac{\lambda_{i_{n'}}^{k_m}}{\sigma^{k_m}} \right) - \frac{1}{M^2(M-1)} \sum_{\substack{m,m'=1 \\ m \neq m'}}^{M} \left( \mathbb{E}\left[\overline{F}\right] \right)^2, \tag{75}$$

where in the last step we have used Equation (60) and that each $\lambda_{i_n}^{k_m}$ is drawn from the same distribution. We add and subtract $\left(\mathbb{E}\left[\overline{F}\right]\right)^2$ to obtain

$$\mathbb{E}\left(\overline{\mathrm{var}}[F]\right) = \mathbb{E}\left( \frac{1}{M^2} \sum_{m=1}^{M} \frac{1}{N_m^2} \sum_{n,n'=1}^{N_m} \frac{\lambda_{i_n}^{k_m}}{\sigma^{k_m}} \frac{\lambda_{i_{n'}}^{k_m}}{\sigma^{k_m}} \right) -$$
$$\frac{1}{M^2(M-1)} \sum_{\substack{m,m'=1 \\ m \neq m'}}^{M} \left( \mathbb{E}\left[\overline{F}\right] \right)^2 + \left( \mathbb{E}\left[\overline{F}\right] \right)^2 - \left( \mathbb{E}\left[\overline{F}\right] \right)^2 \tag{76}$$

$$= \mathbb{E}\left( \frac{1}{M^2} \sum_{m=1}^{M} \frac{1}{N_m^2} \sum_{n,n'=1}^{N_m} \frac{\lambda_{i_n}^{k_m}}{\sigma^{k_m}} \frac{\lambda_{i_{n'}}^{k_m}}{\sigma^{k_m}} \right) - \frac{1}{M^2(M-1)} \sum_{\substack{m,m'=1 \\ m \neq m'}}^{M} \left( \mathbb{E}\left[\overline{F}\right] \right)^2$$
$$+ \frac{1}{M(M-1)} \sum_{\substack{m,m'=1 \\ m \neq m'}}^{M} \left( \mathbb{E}\left[\overline{F}\right] \right)^2 - \left( \mathbb{E}\left[\overline{F}\right] \right)^2 \tag{77}$$

$$= \mathbb{E}\left( \frac{1}{M^2} \sum_{m=1}^{M} \frac{1}{N_m^2} \sum_{n,n'=1}^{N_m} \frac{\lambda_{i_n}^{k_m}}{\sigma^{k_m}} \frac{\lambda_{i_{n'}}^{k_m}}{\sigma^{k_m}} \right)$$
$$+ \frac{1}{M^2} \sum_{\substack{m,m'=1 \\ m \neq m'}}^{M} \left( \mathbb{E}\left[\overline{F}\right] \right)^2 - \left( \mathbb{E}\left[\overline{F}\right] \right)^2. \tag{78}$$

Performing the simplification of Equation (74)–(75), we combine the summations of

the first two terms as

$$
\mathbb{E}\left(\overline{\mathrm{var}}[F]\right) = \mathbb{E}\left(\frac{1}{M^2}\sum_{m=1}^{M}\frac{1}{N_m^2}\sum_{n,n'=1}^{N_m}\frac{\lambda_{i_n}^{k_m}}{\sigma^{k_m}}\frac{\lambda_{i_{n'}}^{k_m}}{\sigma^{k_m}}\right)
$$
$$
+ \frac{1}{M^2}\sum_{\substack{m,m'=1\\m\neq m'}}^{M}\mathbb{E}\left(\sum_{n=1}^{N_m}\frac{1}{N_m}\frac{\lambda_{i_n}^{k_m}}{\sigma^{k_m}}\right)\mathbb{E}\left(\sum_{n'=1}^{N_m}\frac{1}{N_{m'}}\frac{\lambda_{i_{n'}}^{k_{m'}}}{\sigma^{k_{m'}}}\right) - \left(\mathbb{E}\left[\overline{F}\right]\right)^2 \quad (79)
$$

$$
= \frac{1}{M^2}\sum_{m,m'=1}^{M}\mathbb{E}\left(\sum_{n=1}^{N_m}\frac{1}{N_m}\frac{\lambda_{i_n}^{k_m}}{\sigma^{k_m}}\sum_{n'=1}^{N_{m'}}\frac{1}{N_{m'}}\frac{\lambda_{i_{n'}}^{k_{m'}}}{\sigma^{k_{m'}}}\right) - \left(\mathbb{E}\left[\overline{F}\right]\right)^2 \quad (80)
$$

$$
= \mathbb{E}\left[\overline{F}^2\right] - \left(\mathbb{E}\left[\overline{F}\right]\right)^2, \quad (81)
$$

which is the same as the variance (71) of the fidelity estimate. Thus, we conclude that the variance estimator $\overline{\mathrm{var}}[F]$ is an unbiased estimator.

In summary, we have presented a procedure for estimating error bars on the DFE and have shown that the procedure returns an unbiased estimator of the variance. Using this procedure and the DFE procedure described above we obtain fidelity estimate $0.74 \pm 0.05$.

## VIII. CERTIFIED MPS TOMOGRAPHY IS EFFICIENT FOR 1D LOCAL QUENCH DYNAMICS

### A. Summary of the results

In this section, we show that certified MPS tomography can be used to characterise local quench dynamics with resources that scale efficiently in system size. Specifically, we provide upper bounds to the resources, both experimental and computational, required to characterise a state obtained by evolving a pure product state under a nearest-neighbour Hamiltonian [25]. These required resources grow no faster than polynomially in the size $N$ of the system, inverse polynomially with the tolerated infidelity $\mathcal{I}$ of characterisation, and exponentially in the time $t$ of evolution. That is, at any given time during the quench dynamics $t$, the resources to characterise the state scale efficiently (polynomially) in system size.

To show that certified MPS tomography is efficient for quenched states, a necessary condition is that these states admit an efficient (in $N$) MPS representation. This follows from simple arguments in addition to Corollary 3 of Reference [21]. However, the existence of the MPS is not sufficient to guarantee the existence of a parent Hamiltonian with suitable spectral properties, which is essential in our certification procedure. In this section, we show that such a Hamiltonian for quenched states indeed exists, and this existence enables the use of the certified MPS tomography procedure for these states.

Our argument is structured as follows. First, we show that pure product states have parent Hamiltonians that have a unit gap and that comprise local terms acting on single sites only. Next we generalise the pure product state result to quenched states, i.e., states that start out as pure product states and undergo a time evolution under a nearest-neighbour Hamiltonian. We show that such states too can be well approximated by states which have a gapped parent Hamiltonian. These parent Hamiltonians comprise local terms that act on subsystems whose sizes $2\Omega$ scale linearly in time and logarithmically in $N$ and in $1/\mathcal{I}$.

Physically, the existence of gapped parent Hamiltonians implies that the quenched states can be uniquely identified using only their local reductions because these Hamiltonians are local and have a unique ground state. Furthermore, by showing the existence of $\Omega$-sized gapped parent Hamiltonians, we impose upper bounds on the required resources (number of measurements and computational time) required to characterise the state. Characterising a ground state with $\Omega$-sized parent Hamiltonian requires measuring and classically processing a linear (in $N$) number of $L$-sized local reductions on a 1D chain. This characterisation task requires resources (number of measurements and classical post-processing time) that scale exponentially in $\Omega$ and linearly in $N$ via certified MPS tomography [6, 7]. From this, and the scaling

of $\Omega$ in the parameters $N$, $\mathcal{I}$ and $t$, we obtain the mentioned scaling of the experimental and computational cost in terms of these parameters. In the next subsection, we recall results regarding gapped parent Hamiltonians of pure quantum states.

### B. Background: parent Hamiltonian certificates

Here we recall briefly relevant notation regarding the parent Hamiltonian of a pure quantum state of $N$ qubits on a linear chain as introduced in Section IV A 3. The parent Hamiltonian of a pure state $|\psi\rangle$ is any Hermitian linear operator $H$ such that $|\psi\rangle$ is the nondegenerate ground state of $H$. We set the ground state energy, i.e., the lowest eigenvalue $\langle\psi|H|\psi\rangle$, of $H$ to zero and $E_1 > 0$ be the energy of the first excited state.

Now we consider the energy of any arbitrary, possibly mixed, state $\rho$ with respect to $H$. Then the fidelity $F(|\psi\rangle,\rho) = \langle\psi|\rho|\psi\rangle$ of $\rho$ with respect to $\psi$ is bounded below according to (9) with $E_0 = 0$. That is,

$$F(|\psi\rangle,\rho) \geq 1 - \frac{\mathrm{tr}(\rho H)}{E_1}. \tag{82}$$

Thus, the energy of $\rho$ in terms of $H$ provides a lower bound to the fidelity between $|\psi\rangle$ and $\rho$; we call a lower bound to the fidelity a certificate.

The certification of the lab state using $H$ is efficient, that is, requires number of measurements and computation-time that scale polynomially in the number of qudits. Suppose that $H$ is a sum

$$H = \sum_i^{N-k+1} h_i \tag{83}$$

it the terms which act non-trivially

$$h_i = \mathbb{1}_1 \otimes \mathbb{1}_2 \otimes \cdots \otimes \mathbb{1}_{i-1} \otimes h_{i,i+1,\ldots,i+k-1}^\ell \otimes \mathbb{1}_{i+k} \otimes \cdots \otimes \mathbb{1}_N \tag{84}$$

only on small (i.e., size $k \in \mathcal{O}\log N$) subsets of the complete system. Then, only the matching local reductions of $\rho$ are necessary to obtain the energy $\mathrm{tr}(\rho H)$:

$$\mathrm{tr}(\rho H) = \sum_i \mathrm{tr}(\rho h_i) = \sum_i \mathrm{tr}(\rho_i h_i), \tag{85}$$

where $\rho_i$ are the reduced density matrices that act on subsystem $\{i, i+1, \ldots, i+k-1\}$. In this case the certificate can be obtained from a number of measurements that scales linearly in the number of particles (and polynomially in the subsystem size $k$). This is an exponential improvement over the number of measurements required for estimating fidelity by performing standard quantum state tomography. Furthermore, the summation of Eq. (85) can be performed in linear (in $N$) computational time as compared to the exponentially large computation time required if the output from full tomography is used to obtain fidelity. In summary, determining the fidelity certificate is efficient with respect to measurement and computation time.

### C. Product states have simple parent Hamiltonians

In this section, we show that pure product states admit a parent Hamiltonian that has unit gap and only single-site local terms (Lemma 4). This result is a special case of prior work involving matrix product states [6]. First, we provide two elementary statements used in the proof of this lemma.

**Lemma 2.** *Let $P$ be a positive semidefinite linear operator with $\langle\varphi|P|\varphi\rangle = 0$. Then $P|\varphi\rangle = 0$.*

*Proof.* There is an eigendecomposition $P = \sum_i \lambda_i |u_i\rangle\langle u_i|$ of $P$ (with $\lambda_i \geq 0$) because it is positive semidefinite and thus Hermitian. Then

$$0 = \langle\varphi|P|\varphi\rangle = \sum_i \lambda_i |\langle\psi|u_i\rangle|^2 \tag{86}$$

is a sum of non-negative terms, which means that $\lambda_i \langle u_i | \varphi \rangle = 0$ for all $i$. As a consequence,

$$P|\psi\rangle = \sum_i \lambda_i |u_i\rangle \langle u_i|\psi\rangle = 0, \tag{87}$$

which is the required relation. $\qquad\square$

Now we introduce another lemma that is required in the proof of Lemma 4.

**Lemma 3.** *Let a system be partitioned into subsystems A and B. Let*

$$P = \mathbb{1}_A \otimes (\mathbb{1}_B - |u\rangle_B \langle u|_B) \tag{88}$$

*act identically on subsystem A and map any state to a subspace orthogonal to normalized state $|u\rangle_B$. Then $P|w\rangle_{AB} = 0$ implies that $|w\rangle_{AB}$ is of tensor product form $|w\rangle_{AB} = |v\rangle_A \otimes |u\rangle_B$.*

*Proof.* Let $P|w\rangle_{AB} = 0$. We observe that

$$|w\rangle_{AB} = (\mathbb{1}_A \otimes \mathbb{1}_B)|w\rangle_{AB} \tag{89}$$

$$= P|w\rangle_{AB} + \left[\mathbb{1}_A \otimes |u\rangle_B \langle u|\right]|w\rangle_{AB} \tag{90}$$

$$= [\mathbb{1}_A \otimes |u\rangle_B \langle u|]|w\rangle_{AB}, \tag{91}$$

which is equivalent to

$$[_B\langle u|w\rangle_{AB}] \otimes |u\rangle_B. \tag{92}$$

Thus, $|w\rangle_{AB}$ is of tensor product form $|w\rangle_{AB} = |v\rangle_A \otimes |u\rangle_B$ with $|v\rangle_A = {_B}\langle u|w\rangle_{AB}$. $\qquad\square$

Finally, we prove that pure product states admit a parent Hamiltonian that has unit gap and only single-site local terms.

**Lemma 4.** *Let $|\varphi\rangle = |\varphi_1\rangle \otimes |\varphi_2\rangle \otimes \cdots \otimes |\varphi_N\rangle$ a product state on n qudits. Let $\langle\varphi_i|\varphi_i\rangle = 1$ for all site indices i. Define*

$$H = \sum_{i=1}^{N-k+1} h_i, \quad h_i = \mathbb{1}_{1,\dots,i-1} \otimes P_{\ker(\rho_i)} \otimes \mathbb{1}_{i+1,\dots,N} \tag{93}$$

*where $P_{\ker(\rho_i)}$ is the orthogonal projection onto the null space of the reduced density operator $\rho_i = |\varphi_i\rangle\langle\varphi_i|$ of $|\varphi\rangle$ on site i. Then the eigenvalues of H are given by $\{0, 1, 2, \dots, N\}$, the smallest eigenvalue zero is non-degenerate and $|\varphi\rangle$ is an eigenvector of eigenvalue zero.*

*Proof.* Expand the null space projectors $h_i$ in terms of the single-site pure states $\{|\varphi_i\rangle\}$ as

$$h_i = \mathbb{1}_{1,\dots,i-1} \otimes P_{\ker(\rho_i)} \otimes \mathbb{1}_{i+1,\dots,N}$$
$$= \mathbb{1}_{1,\dots,i-1} \otimes (\mathbb{1}_i - |\varphi_i\rangle\langle\varphi_i|) \otimes \mathbb{1}_{i+1,\dots,N}. \tag{94}$$

For each site $i$, construct an orthonormal basis $|\phi_{i,1}\rangle, \dots, |\phi_{i,\ell}\rangle, \dots, |\phi_{i,d_i}\rangle$ such that $|\phi_{i,1}\rangle = |\varphi_i\rangle$.

First, we show that each of the product basis states is an eigenstates of $H$. Specifically, consider the set

$$\left\{|\Phi_L\rangle \stackrel{\text{def}}{=} |\phi_{1,\ell_1}\rangle \otimes |\phi_{2,\ell_2}\rangle \otimes \cdots \otimes |\phi_{N,\ell_N}\rangle : L \stackrel{\text{def}}{=} \{\ell_1, \ell_2, \dots, \ell_N\} \in \{1, 2, \dots, N\}^{\otimes N}\right\}. \tag{95}$$

Each of the states $|\phi_{i,\ell_i}\rangle$ in the product is an eigenstate of the operators $\mathbb{1}$ and of $\mathbb{1} - |\varphi_i\rangle\langle\varphi_i|$. This implies that their product $|\Phi_L\rangle$ (95) is an eigenstate of the product $h_{i'}$ (94) of the operators for each $i'$. Hence, each $|\Phi_L\rangle$ is an eigenstate of the sum $H$ of $h_i$.

Now we show that $H$ has the eigenvalues $\{0, 1, 2, \dots, N\}$. Notice that $h_i$ have eigenvalues $\{0, 1\}$ and commute pairwise. Thus, the eigenvalues of $H$ are limited to the set of possible summations of $n$ terms each either zero or unity. Thus, the eigenvalues of $H$ take only integral values between 0 and $n$, both included. In particular, we observe

$h_i |\varphi\rangle = 0$ and therefore $H |\varphi\rangle = 0$, i.e., $|\varphi\rangle$ is an eigenvector of $H$ with the eigenvalue zero.

Finally, we show that the zero eigenvalue is non-degenerate. We assume that $|\phi\rangle$ is a state with $H |\phi\rangle = 0$. This implies that

$$0 = \langle\phi| H |\phi\rangle = \sum_{i=1}^{n} \langle\phi| h_i |\phi\rangle. \tag{96}$$

Because the $h_i$ are all positive semidefinite, this is a sum of non-negative terms such that all terms must vanish. From Lemma 2 we obtain that $h_i |\phi\rangle = 0$ for all $i$. Using Lemma 3, $h_1 |\phi\rangle = 0$ implies that $|\phi\rangle = |\varphi_1\rangle \otimes |\phi_2\rangle$ with $|\phi_2\rangle = \langle\varphi_1|\phi\rangle$. Apply Lemma 3 again on $h_2 |\phi\rangle = 0$ to obtain $|\phi\rangle = |\varphi_1\rangle \otimes |\varphi_2\rangle \otimes |\phi_3\rangle$ with $|\phi_3\rangle = \langle\varphi_2|\phi_2\rangle$. Using Lemma 3 repeatedly ($N-1$ times), we obtain $|\phi\rangle = c |\varphi_1\rangle \otimes \cdots \otimes |\varphi_N\rangle$ with $c = (\langle\varphi_1| \otimes \cdots \otimes \langle\varphi_N|) |\phi\rangle$. This show that any ground state $|\phi\rangle$ of $H$ obeys $|\phi\rangle = c |\varphi\rangle$ for some complex number $c$. Hence, the eigenvalue zero of $H$ is non-degenerate. $\qquad\square$

### D. Parent Hamiltonian for a locally time evolved state

This section presents the main results regarding the parent Hamiltonians of quenched states. We show that the experimental and computational cost of characterising quenched states scales no faster than polynomially in $N/\mathcal{I}$ and exponentially in the quench time $t$.

Our proof is in two parts. First, we follow [22] to construct a state which closely approximates our time-evolved state. This approximate state is obtained by starting with a tensor product of pure states on at most $\Omega$ neighbouring sites and acting a single unitary operation that is a tensor product of unitaries on at most $\Omega$ sites. Here, $\Omega$ depends on the time of evolution under the local Hamiltonian.

Next, in Theorem 6 we show that this approximate state is the ground state of a suitable parent Hamiltonian. From Lemma 4, we know that the pure product state has a parent Hamiltonian with terms acting on at most $\Omega$ sites. Specifically, the unitary operation does not increase the range of the terms in the pure-state parent Hamiltonian from $\Omega$ to more than $2\Omega$, which is small, i.e., $O(\log N)$, for suitably low quench time $O(\log N)$. This parent Hamiltonian enables the usual certification procedure, which is described in the main text.

We construct the approximate state using the following theorem from Reference [22].

**Theorem 5** ($\epsilon$-QCA decomposition, [22]). *Let $H$ be a nearest-neighbour Hamiltonian on $N$ qubits in a linear chain, i.e., $H = \sum_{i=1}^{N-1} h_{i,i+1}$. We fix a positive integer $\Omega$ and partition the linear chain into $\mathcal{N} = 2N/\Omega$ contiguous blocks each containing at most $\Omega/2$ qubits (Figure 22). There is an approximation of the time evolution operator $U = e^{-iHt}$ of the form*

$$U' = [U_{12} \otimes U_{34} \otimes \cdots \otimes U_{N-1,\mathcal{N}}][V_1 \otimes V_{23} \otimes V_{34} \otimes \cdots \otimes V_{N-2,N-1} \otimes V_{N-1}] \tag{97}$$

*where $U_{j,j+1}$, $V_{j,j+1}$ and $V_j$ are unitaries acting on the blocks specified by the subscripts. This approximation satisfies $\|U - U'\| \le \epsilon$ under the restriction that*

$$\Omega \ge c_0|t| + c_1 \log(N/\epsilon) \tag{98}$$

*where $\|\cdot\|$ is the operator norm and $c_0$ and $c_1$ are constants.*

*Let $|\psi(0)\rangle = |\psi_1\rangle \otimes \cdots \otimes |\psi_N\rangle$ be an initial product state [26]. The state $|\psi'\rangle = U' |\psi(0)\rangle$ is an approximation of the time-evolved state $|\psi(t)\rangle = U |\psi(0)\rangle$ with $\||\psi'\rangle - |\psi(t)\rangle\| \le \epsilon$. The approximation $|\psi'\rangle$ has a matrix product state representation with bond dimension no more than $2^{\Omega}$.*

Thus, the state $|\psi'\rangle$ closely approximates our time-evolved state. Specifically, is $\Omega$ is required to scale logarithmically with $N$, then the norm-distance between $|\psi\rangle$ and $|\psi'\rangle$ scales as as inverse polynomial in $N$. Now we prove the existence of the parent Hamiltonian of $|\psi'\rangle$ and find an upper bound to the number of sites that the parent Hamiltonian terms act on.

**Theorem 6.** *For $|\psi'\rangle$ as described in Theorem 5, there is a Hermitian linear operator $G = \sum_{j=1}^{N/\Omega+1} g_j$ with non-degenerate smallest eigenvalue zero and eigenvector $|\psi'\rangle$, second smallest eigenvalue one and largest eigenvalue $N/\Omega + 1$. Each local term $g_j$ acts on no more than $2\Omega$ consecutive qubits.*

*Proof.* We define the intermediate product state

$$|\phi\rangle = [V_1 \otimes V_{23} \otimes \cdots \otimes V_N] |\psi(0)\rangle \tag{99}$$

$$=: |\phi_{01}\rangle \otimes |\phi_{23}\rangle \otimes \cdots \otimes |\phi_{N,N+1}\rangle, \tag{100}$$

i.e., $|\phi_{j,j+1}\rangle$ is a state on blocks $j, j+1$. Blocks 0 and $N+1$ are empty and have been introduced for notational convenience. From Lemma 4, we have a parent Hamiltonian

$$F = f_{01} + f_{23} + \cdots + f_{N,N+1} \tag{101}$$

of $|\phi\rangle$ with

$$f_{j,j+1} = \mathbb{1}_{1,\ldots,j-1} \otimes P_{\ker(|\phi_{j,j+1}\rangle\langle\phi_{j,j+1}|)} \otimes \mathbb{1}_{j+2,\ldots,N}. \tag{102}$$

Furthermore, $F$ has a non-degenerate smallest eigenvalue zero with eigenvector $|\phi\rangle$, second smallest eigenvalue one and largest eigenvalue $\mathcal{N}/2 + 1 = N/\Omega + 1$.

We define

$$\tilde{U} = U_{12} \otimes U_{34} \otimes \cdots \otimes U_{N-1,N} \tag{103}$$

and we define $|\psi'\rangle = \tilde{U}|\phi\rangle$. Because $\tilde{U}$ is unitary, the operator $G = \tilde{U}F\tilde{U}^\dagger$ has the same eigenvalue spectrum as $F$. That is, $G$ has a non-degenerate smallest eigenvalue zero with eigenvector $|\psi'\rangle$ and second smallest eigenvalue one. We obtain the following representation of $G$

$$G = g_{12} + g_{1234} + g_{3456} + \cdots + g_{N-1,N}, \tag{104}$$

where the identity operators are omitted and

$$g_{jklm} = [U_{jk} \otimes U_{lm}] f_{kl} [U_{jk} \otimes U_{lm}]^\dagger. \tag{105}$$

The border terms are given by $g_{12} = U_{12}f_{01}U_{12}^\dagger$ and $g_{N-1,N} = U_{N-1,N}f_{N,N+1}U_{N-1,N}^\dagger$. There are $\mathcal{N}/2 + 1$ terms in the sum and each term acts on at most four blocks, i.e., at most $2\Omega$ qubits. $\qquad\square$

This completes our proof regarding the parent Hamiltonian of the approximate time-evolved state. In the following corollary, we use the parent Hamiltonian to obtain a fidelity certificate (Section VIII B) for the lab state $\rho$ with respect to the approximate state.

**Corollary 7.** *Consider $|\psi'\rangle$ as described in Theorem 6 and define $\psi' \stackrel{\text{def}}{=} |\psi'\rangle\langle\psi'|$. Denote by $\|\cdot\|_1$ the trace norm of an operator. Let $\rho$ be an arbitrary quantum state and let $\delta = \|\rho - \psi'\|_1$. Then*

$$\langle\psi'|\rho|\psi'\rangle \geq 1 - \operatorname{tr}(\rho G) \geq 1 - (N/\Omega + 1)\delta. \tag{106}$$

*Proof.* As $G$ has unit gap the fidelity lower bound becomes

$$\langle\psi'|\rho|\psi'\rangle \geq 1 - \operatorname{tr}(\rho G)/1$$
$$= 1 - \operatorname{tr}(\rho G), \tag{107}$$

which is the first inequality of (106). Consider the energy $\operatorname{tr}(\rho G)$ of $\rho$ with respect to $G$. Using $\operatorname{tr}(\psi'G) = 0$, we have

$$\operatorname{tr}(\rho G) = \left|\operatorname{tr}(\rho G) - \operatorname{tr}(\psi'G)\right| \tag{108}$$

Thus, the energy

$$\operatorname{tr}(\rho G) \leq \|\rho - \psi'\|_1 \|G\|_\infty$$
$$= (N/\Omega + 1)\delta \tag{109}$$

is at most $\operatorname{tr}(\rho G) \leq (N/\Omega + 1)\delta$, where $\|\cdot\| = \|\cdot\|_\infty$ denotes the operator norm. Combining Equations (107) and (109), we obtain the required inequalities. $\qquad\square$

The final Theorem requires the following Lemma:

**Lemma 8.** *Let $\|\cdot\|_1$ the trace norm, $\psi = |\psi\rangle\langle\psi|$ and $\psi' = |\psi'\rangle\langle\psi'|$. If $\||\psi\rangle - |\psi'\rangle\| \le \epsilon$, then $\|\psi - \psi'\|_1 \le 2\epsilon$.*

*Proof.* Assume that $\||\psi\rangle - |\psi'\rangle\| \le \epsilon$ holds. This gives us

$$\epsilon^2 \ge 2(1 - \Re(\langle\psi|\psi'\rangle)) \ge 2(1 - \sqrt{F}) \tag{110}$$

where $F = |\langle\psi|\psi'\rangle|^2 = F(|\psi\rangle, |\psi'\rangle)$. This gives $\sqrt{F} \ge 1 - \epsilon^2/2$ and $1 - F = \epsilon^2 - \epsilon^4/4 \le \epsilon^2$, completes the proof of the inequality $\|\psi - \psi'\|_1 \le 2\epsilon$. □

Our final theorem considers a state $\rho$ close to a locally time-evolved state $|\psi(t)\rangle$; as before, $|\psi'\rangle$ is an approximation of $|\psi(t)\rangle$. The theorem states the conditions under which the fidelity $\langle\psi'|\rho|\psi'\rangle$ can be lower bounded by at least $1 - \mathcal{I}$, for some infidelity $\mathcal{I}$:

**Theorem 9.** *Let $H$ be a nearest-neighbour Hamiltonian on $N$ qubits in a linear chain, i.e., $H = \sum_{i=1}^{N-1} h_{i,i+1}$. Let $|\psi(0)\rangle = |\psi_1\rangle \otimes \ldots |\psi_N\rangle$ be an initial product state [27] and let $|\psi(t)\rangle = e^{-iHt}|\psi(0)\rangle$ be the time-evolved state. Define $\psi(t) = |\psi(t)\rangle\langle\psi(t)|$.*

*Let $\gamma = \|\rho - \psi(t)\|_1$ be the trace distance between an unknown state $\rho$ and the time-evolved state. Fix an infidelity $\mathcal{I}$ that satisfies $\mathcal{I} > 2N\gamma$. Choose an integer $\Omega \ge 1$ such that*

$$\Omega \ge c_0|t| + c_1 \log\left(\frac{2N}{\mathcal{I}/2N - \gamma}\right), \tag{111}$$

*with $c_0$, $c_1$ from Theorem 5. $\psi' = |\psi'\rangle\langle\psi'|$ is the approximation of $|\psi(t)\rangle$ from the same theorem. We also use the parent Hamiltonian $G$ from Theorem 6.*

*Then, the fidelity lower bound between $|\psi'\rangle$ and $\rho$ will be at least*

$$\langle\psi'|\rho|\psi'\rangle \ge 1 - \mathrm{tr}(\rho G) \ge 1 - \mathcal{I} \tag{112}$$

*Proof.* Theorem 5 applies for

$$\epsilon = \frac{1}{2}\left(\frac{\mathcal{I}}{2N} - \gamma\right) \tag{113}$$

and it guarantees $\||\psi(t)\rangle - |\psi'\rangle\| \le \epsilon$. As a consequence, we have $\|\psi(t) - \psi'\|_1 \le 2\epsilon$ (Lemma 8) and

$$\|\rho - \psi'\|_1 \le \|\rho - \psi(t)\|_1 + \|\psi(t) - \psi'\|_1 \le \gamma + 2\epsilon. \tag{114}$$

In addition, we observe (using $N \ge 1$ and $\Omega \ge 1$)

$$\mathcal{I} = 2N(2\epsilon + \gamma) \ge (N+1)(2\epsilon + \gamma) \ge \left(\frac{N}{\Omega} + 1\right)(2\epsilon + \gamma). \tag{115}$$

We use Corollary 7 with $\delta = 2\epsilon + \gamma$. It provides

$$\langle\psi'|\rho|\psi'\rangle \ge 1 - \mathrm{tr}(\rho G) \ge 1 - (N/\Omega + 1)(2\epsilon + \gamma) \ge 1 - \mathcal{I}, \tag{116}$$

which is the required result.

□

If the experimental state $\rho$ is the same as the quenched state $|\psi(t)\rangle$, then the requirement (111) for $\Omega$ changes to

$$\Omega \ge c_0|t| + c_2 \log(N) + c_3 \log\left(\frac{1}{\mathcal{I}}\right) + c_4 \tag{117}$$

which enables us to quantify the resources required for certification.

### E. Conclusion

In summary, the time-evolved state $|\psi(t)\rangle$ from Theorem 9 can be certified up to infidelity $\mathcal{I}$ with respect to a state $|\psi'\rangle$, which has a parent Hamiltonian $G$ with unique ground state and unit gap. The local terms in $G$ act on at most $2\Omega$ sites. $\Omega$ can be chosen as the lowest integer that satisfies Equation (117) depending on the evolution time $t$, number of qubits $N$ and infidelity $\mathcal{I}$. Note that $\Omega$ grows no faster than linearly with time and logarithmically with $N/\mathcal{I}$.

The existence of gapped parent Hamiltonians with $2\Omega$-sized terms means that the quenched states can be uniquely identified, in principle, using only $2\Omega$-sized local reductions. Whether such a state can actually be obtained using existing numerical algorithms is discussed in References [6, 7]. Although, no formal proofs for the convergence of these algorithms are available, we observe (main text) that these algorithms perform well in practice. Theorem 9 complements these discussions by ensuring that the fidelity of any reconstructed state with respect to the lab state can always be bounded from below. If this lower bound is smaller than desired, then the numerical algorithms can be run again, perhaps with more measurements to reduce random error or with measurements on larger-sized subsystems to capture all correlations.

Theorem 9 also imposes upper bounds on the required number of measurements and the required computational time for characterising the state. Specifically, the experimental and computational costs for performing certified MPS tomography of quenched states scale no faster than polynomially in $N$, inverse polynomially in $\mathcal{I}$ and exponentially in the quench time $t$. Thus, certified MPS tomography is efficient in the size of the system and in the inverse infidelity tolerance for quenched states.

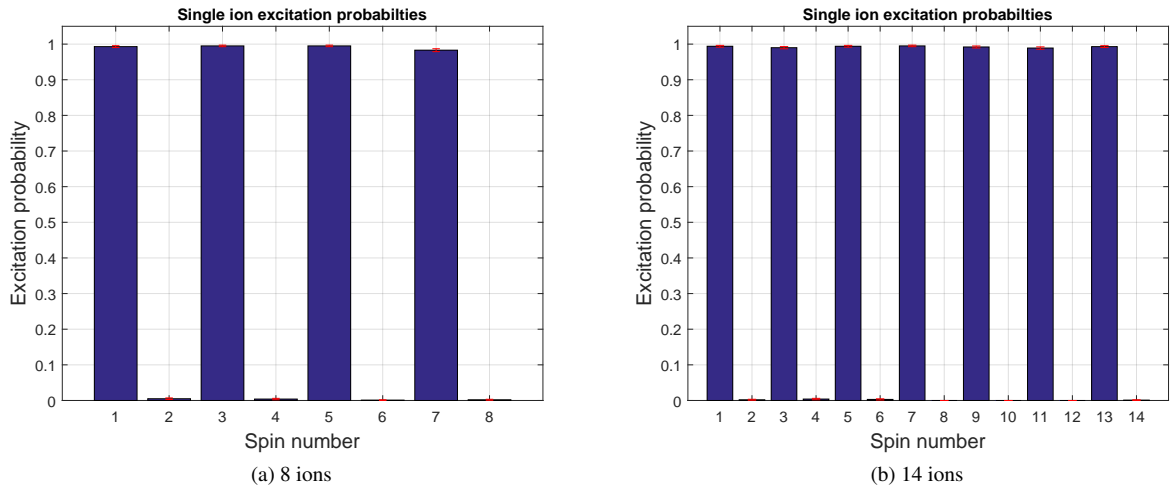

FIG. 5: **Excitation probability of individual ions at the preparation of the Néel-ordered initial state.** Initial state preparation for 8 (a) and 14 (b) ions. The plots show data, averaged over ~1000 measurements with errorbars (red) derived from quantum-projection noise.

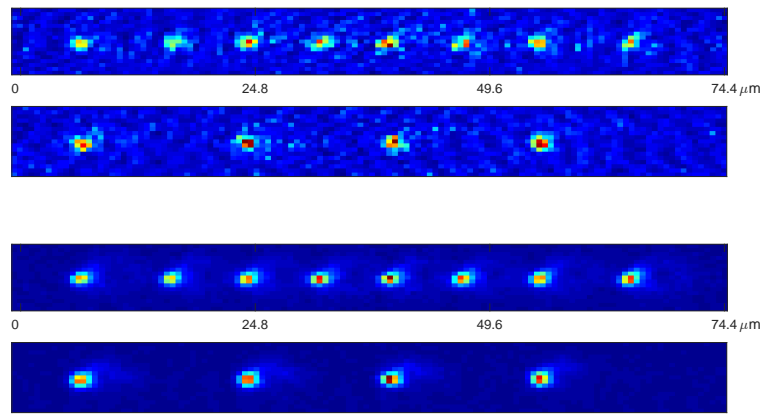

FIG. 6: **CCD camera images of 8-spin states. Line 1:** Single camera shots with 1 ms detection time of an 8-spin chain with all ions in the fluorescing ground state $|\downarrow\rangle$. The spin chain extends over ~ 60 $\mu m$. **Line 2:** Single camera shots with 1 ms detection time of the Néel-ordered state: Spins 1, 3, 5, 7 in the fluorescing state $|\downarrow\rangle$, spins 2, 4, 6, 8 are in the non-fluorescing $|\uparrow\rangle$ state. **Line 3+4:** Camera pictures averaged from 100 single shots with 1 ms detection time each, showing 8 ions in the ground state (line 3) and in the Néel-ordered state (line 4).

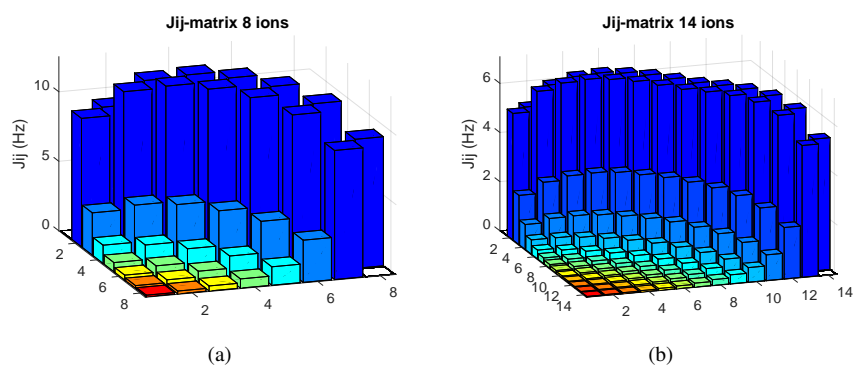

FIG. 7: **Theoretically constructed coupling matrices $J_{ij}$.** The coupling strengths were calculated obeying the experimental parameters for 8 (a) and 14 (b) ions.

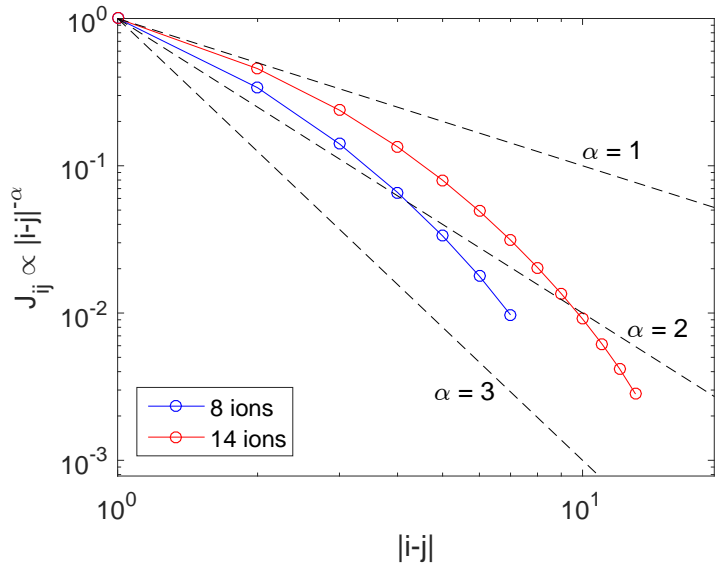

FIG. 8: **Comparing the experimental coupling strengths with ideal power-law dependencies.** Experimental coupling strengths $J_{ij}$ are calculated from a theoretical model obeying the experimental parameters. Coloured lines show $J_{ij}$ for 8 (blue) and 14 (red) ions as a function of the distance $|i - j|$ in a double-logarithmic plot. The black dashed lines show real power-law decays $|i - j|^{-\alpha}$ for $\alpha = 1, 2, 3..$

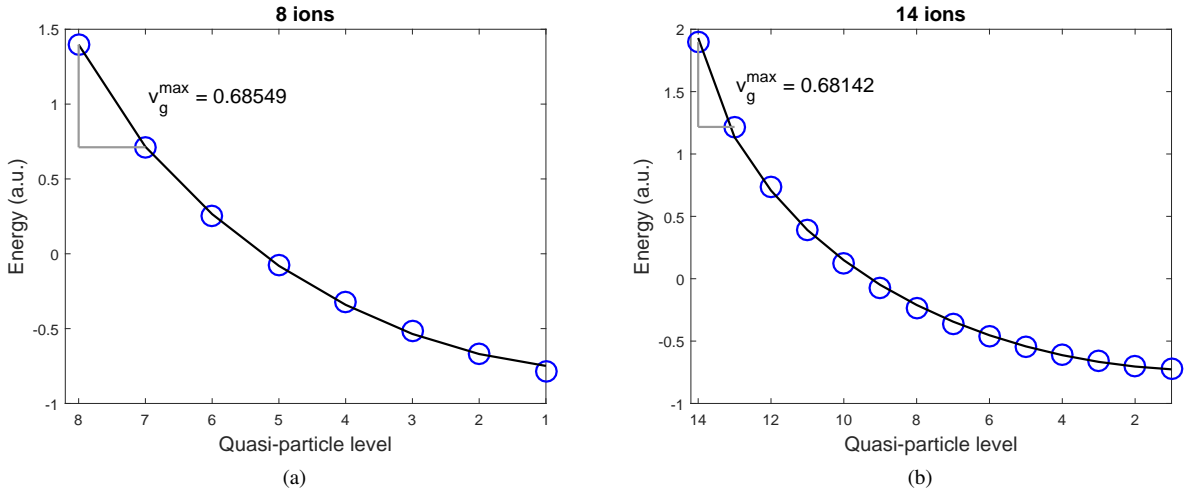

FIG. 9: **Eigenmode spectrum of the implemented Hamiltonian.** The blue circles show the Eigenmode spectrum for 8 (a) and 14 (b) ions, calculated theoretically from experimental parameters. The black lines are the spectra for power-law interactions with best-fit exponents $\alpha = 1.58$ (8 ions) and $\alpha = 1.27$ (14 ions). The maximum group velocity of the energy dispersion is given by $v_g^{max}$.

33

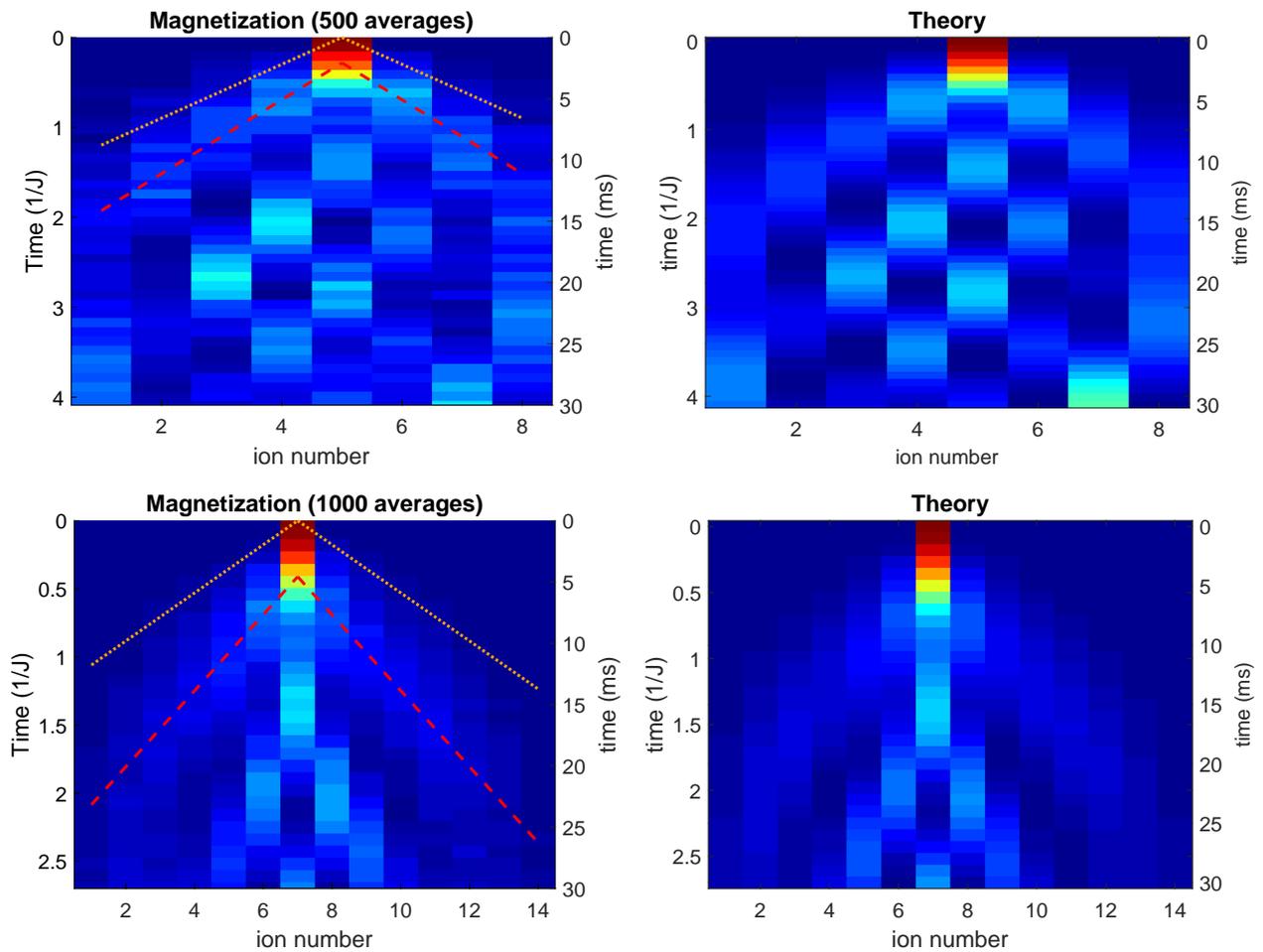

FIG. 10: **Magnetization dynamics of a single excitation under Ising interaction.** Time evolution of a single initial spin excitation for an 8- (first row) and 14- (second row) ion string. The two time axes distinguish between the real time passed in the laboratory (right axis) and time renormalised by the mean nearest-neighbour interaction (equation (2)). Experimental data is shown in the left column, theory in the right column. The colours identify the spin state: Dark blue indicates a $|\downarrow\rangle$ state, red a $|\uparrow\rangle$ state. The spin-excitations, and with it quantum correlations, spread out in light-like cones. Red dashed lines are fits to the observed excitation spread on the experimental data. Orange dotted lines show the maximum expected velocity at which correlations spread out, estimated by renormalising the mean nearest-neighbour interaction strength by the algebraic tail (equation (1)).

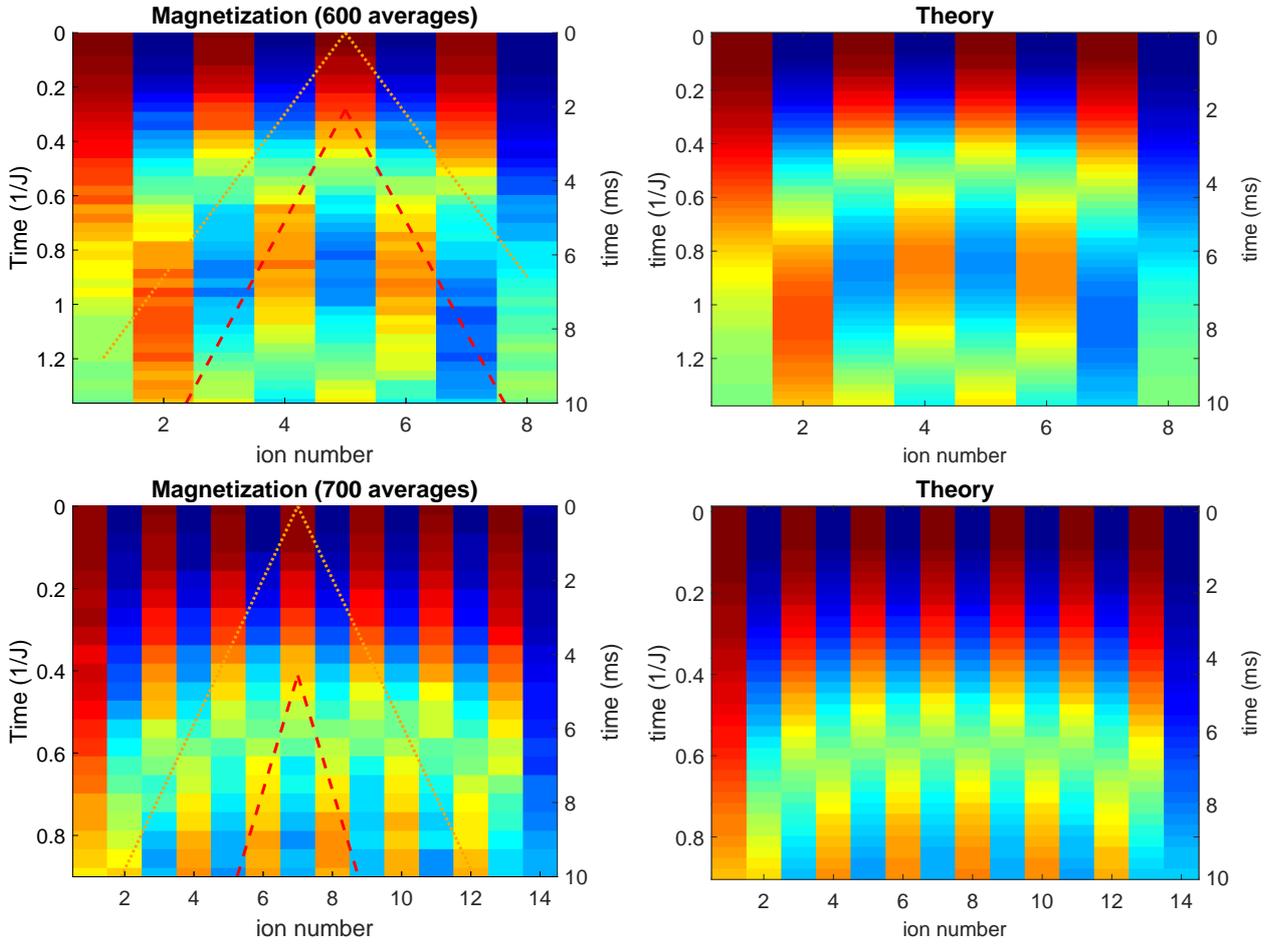

FIG. 11: **Magnetization dynamics of a Néel-ordered initial state under Ising interaction** for an 8- (first row) and 14- (second row) ion string. The two time axes distinguish between the real time passed in the laboratory (right axis) and time renormalized by the mean nearest-neighbour interaction (equation (2)). Experimental data is shown in the left column, theory in the right column. The colors identify the spin state: Dark blue indicates a $|\downarrow\rangle$ state, red a $|\uparrow\rangle$ state. The spin-excitations, and with it quantum correlations, spread out in light-like cones. Orange dotted lines show the maximum expected velocity at which correlations spread out, estimated by renormalizing the mean nearest-neighbour interaction strength by the algebraic tail (equation (1)).

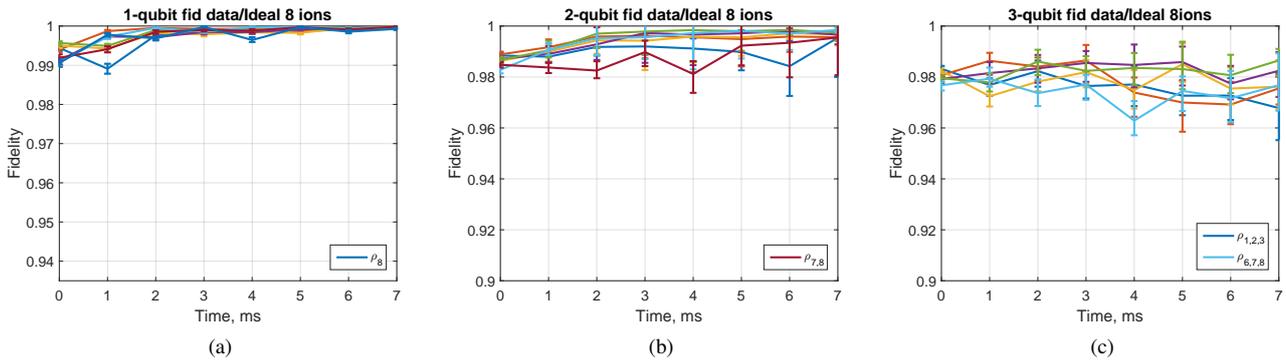

FIG. 12: **Overlap between the 8-spin state in the laboratory and the ideal state.** Time dependent overlap of the measured reduced single-qubit (a), two-qubit (b) and three-qubit (c) density matrices with the theoretical, ideal density matrices. Error bars are derived with the Monte Carlo method using 100 samples.

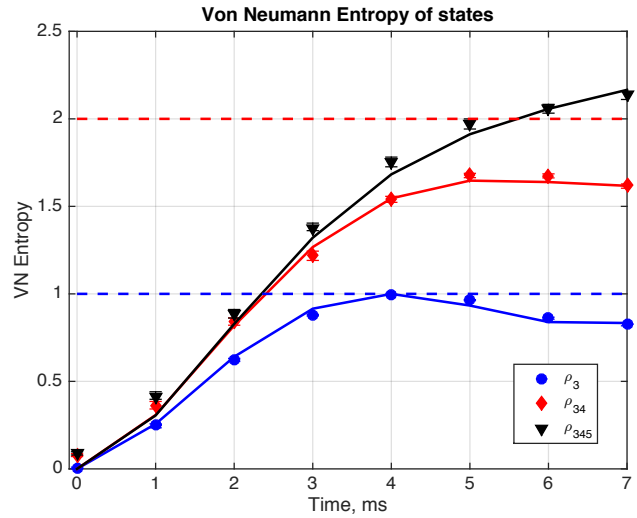

FIG. 13: **Entropy in example local reductions, during 8-spin quench**. Blue: single spin state. Red: two spin state. Black: three spin state (see legend). Shapes: data, entropy of experimentally-reconstructed local reductions via full QST. Solid lines: model based on ideal quantum simulator states. Dashed lines: the maximum entropy for fully mixed state of $N$ spins is $N$ (qubits). Error bars in data are almost smaller than the symbol size and are 1 standard deviation in distributions derived from Monte Carlo simulations of quantum projection noise.

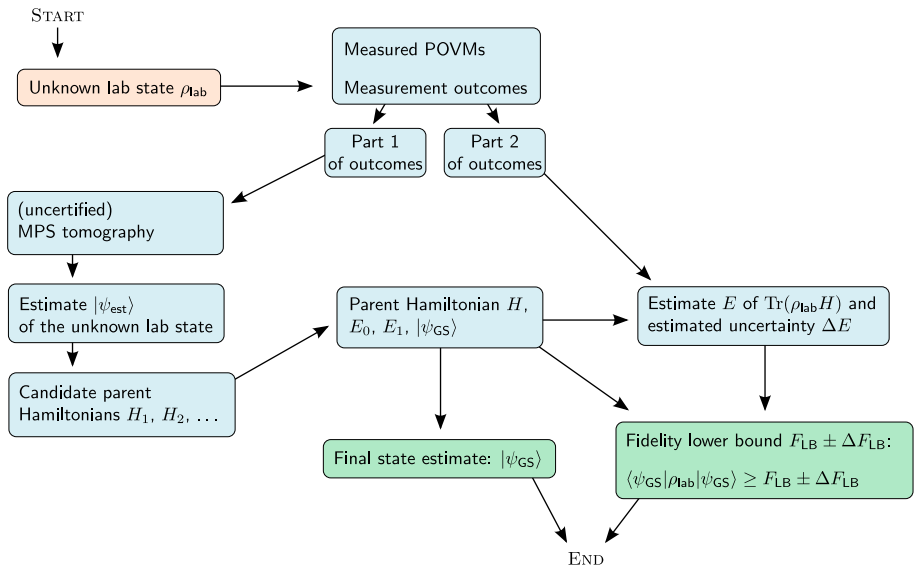

FIG. 14: Schematic of our scheme for certified MPS tomography discussed in Sec. IV A.

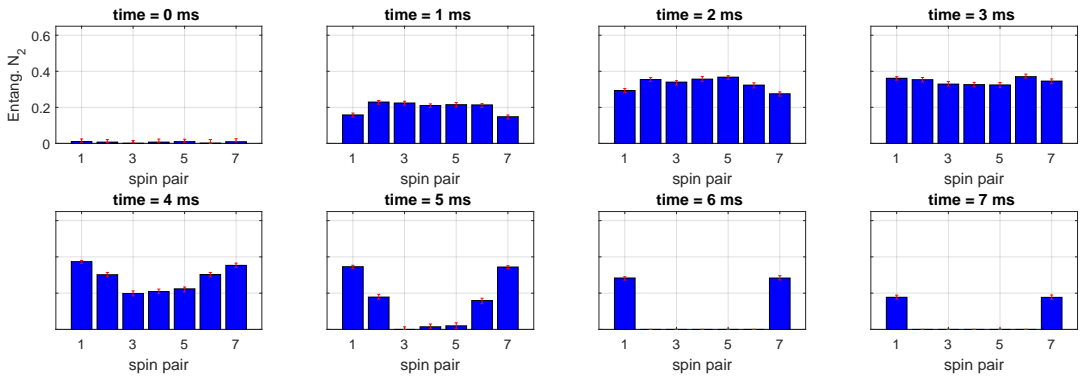

(a) Bipartite negativity $N_2$

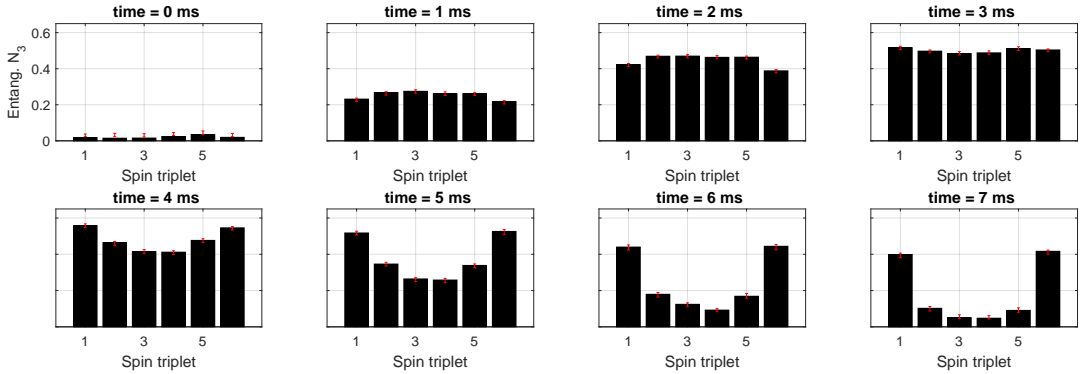

(b) Tripartite negativity $N_3$

FIG. 15: **Time evolution of entanglement for an 8 spin system.** (a) Bipartite negativity $N_2$ for neighbouring spin pairs and (b) tripartite negativity $N_3$ for neighbouring spin triplets. The entanglement measure $N_3$ is defined as the geometric mean of all three bipartite negativity splittings of a triplet. The values are calculated from the measured reduced density matrices. Error bars are derived with the Monte Carlo method using 100 samples. For clarity, values from an ideal simulator model are not shown: the data closely matches the ideal model.

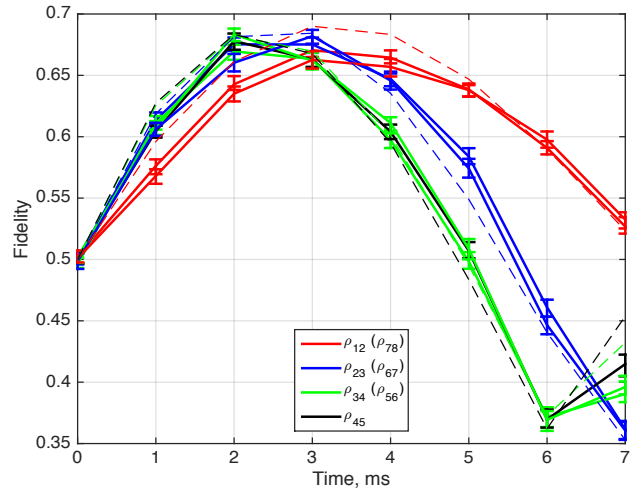

FIG. 16: **Bell state fidelity during 8-spin quench.** Overlap of the $|\Psi^+\rangle = (|01\rangle + |10\rangle)/\sqrt{2}$ Bell state with the absolute value of the experimentally reconstructed neighbouring 2-spin density matrices. Spin pairs symmetrically distributed around the centre of the string are shown in the same color. Solid lines connecting points with error bars: data. Dashed lines: values from ideal model of the simulator. Error bars are derived with the Monte Carlo method using 100 samples and are based on quantum projection noise.

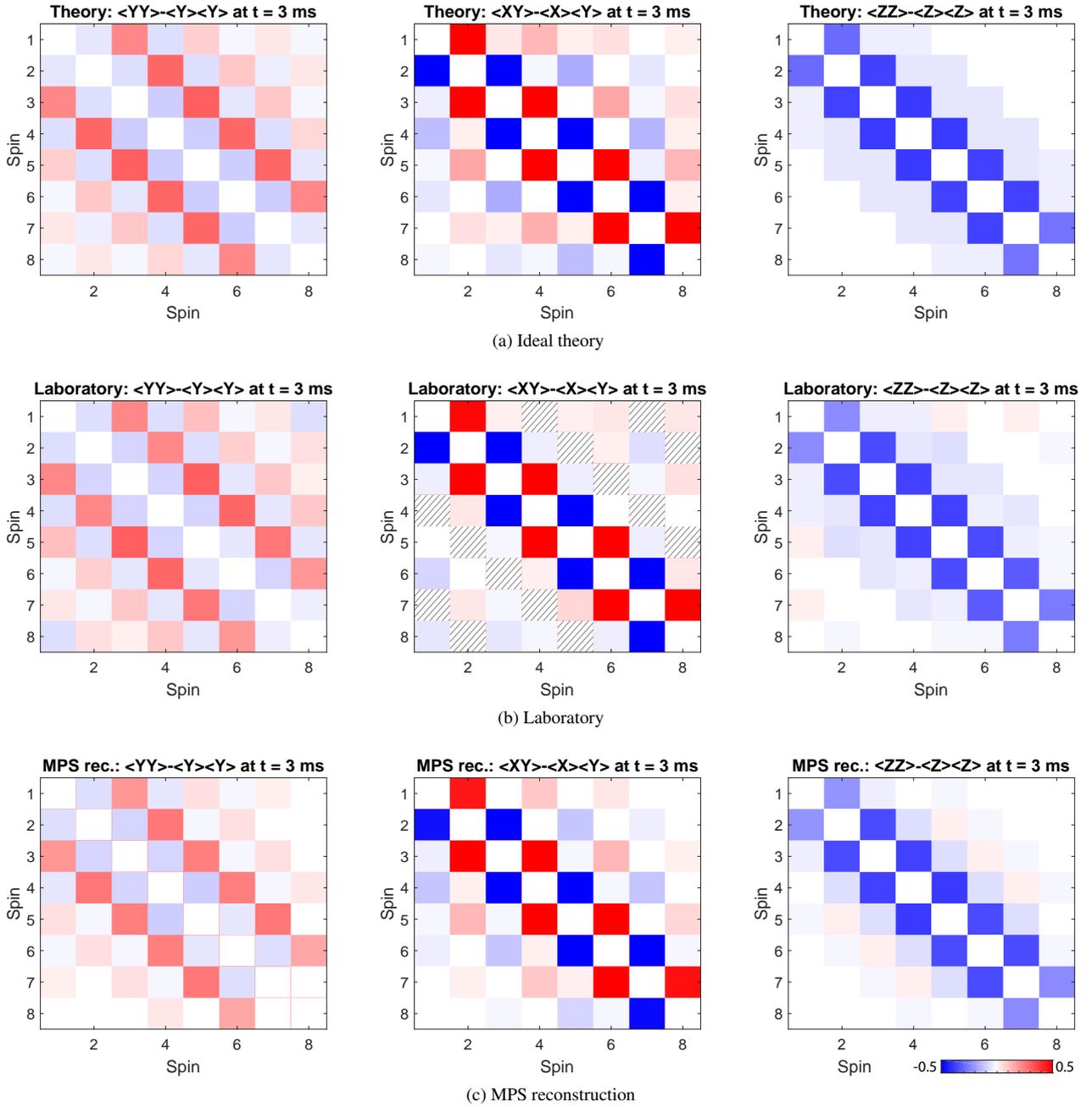

(a) Ideal theory

(b) Laboratory

(c) MPS reconstruction

FIG. 17: **Connected two-point correlation matrices for 8 spins.** $\langle Y_i(t)Y_j(t)\rangle - \langle Y_i(t)\rangle\langle Y_j(t)\rangle$ and $\langle X_i(t)Y_j(t)\rangle - \langle X_i(t)\rangle\langle Y_j(t)\rangle$ and $\langle Z_i(t)Z_j(t)\rangle - \langle Z_i(t)\rangle\langle Z_j(t)\rangle$ after $t = 3\,\mathrm{ms}$ time evolution. Row (a): Connected two-point correlations of the theoretical ideal state, row (b) the measured state in the laboratory, row (c) the reconstructed MPS state for the 8 ion Néel state after 3 ms evolution under Ising interaction. The dashed squares denote correlations which were not measured.

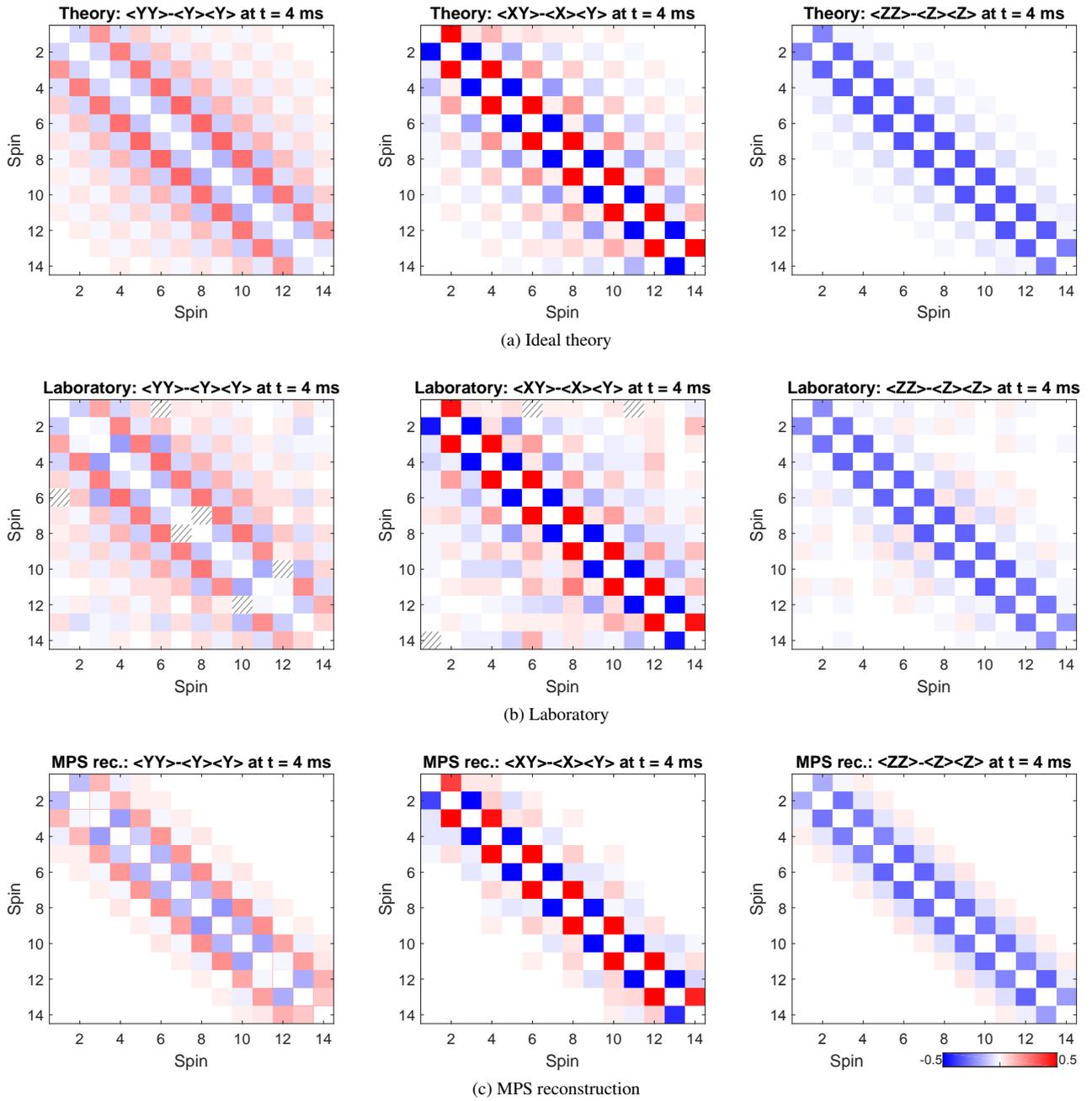

(a) Ideal theory

(b) Laboratory

(c) MPS reconstruction

FIG. 18: **Connected two-point correlation matrices for 14 spins.** $\langle Y_i(t)Y_j(t)\rangle - \langle Y_i(t)\rangle\langle Y_j(t)\rangle$ and $\langle X_i(t)Y_j(t)\rangle - \langle X_i(t)\rangle\langle Y_j(t)\rangle$ and $\langle Z_i(t)Z_j(t)\rangle - \langle Z_i(t)\rangle\langle Z_j(t)\rangle$ after $t = 4$ ms time evolution. Row (a): Connected two-point correlations of the theoretical ideal state, row (b) the measured state in the laboratory, row (c) the reconstructed MPS state for the 14 ion Néel state after 4 ms evolution under Ising interaction. The dashed squares denote correlations which were not measured.

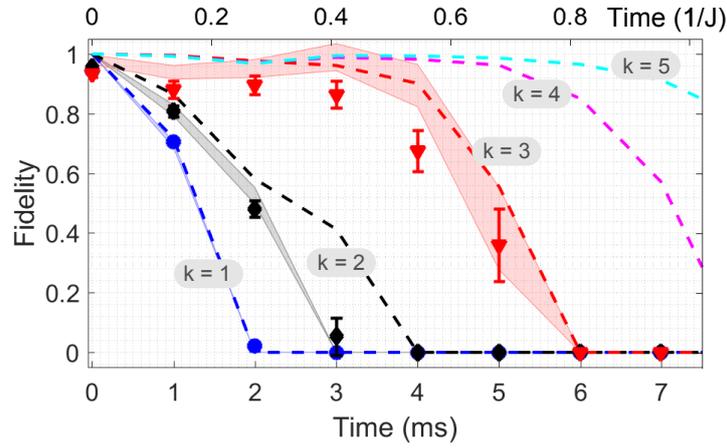

FIG. 19: **Fidelity lower bounds for the 8-spin quench: comparison with theory.** Certified lower bounds on the fidelity between MPS $|\Psi\rangle_c^k$, reconstructed from measurements over $k$ sites, and the quantum simulator state $\rho_{lab}$. Shapes: data points with errors (uncertainty due to finite measurement numbers). Data is compared to two theoretical models (dashed lines and shaded areas). Both models consider ideal simulator states $|\Psi(t)\rangle$. Dashed lines: exact knowledge of local reductions (e.g. infinite measurements per local observable). Shaded areas: outcome allowing for 1000 measurements per local observable, as per the experiments. Color: Blue, black, red, magenta and cyan represent local reductions of length k=1,2,3,4,5 sites, respectively.

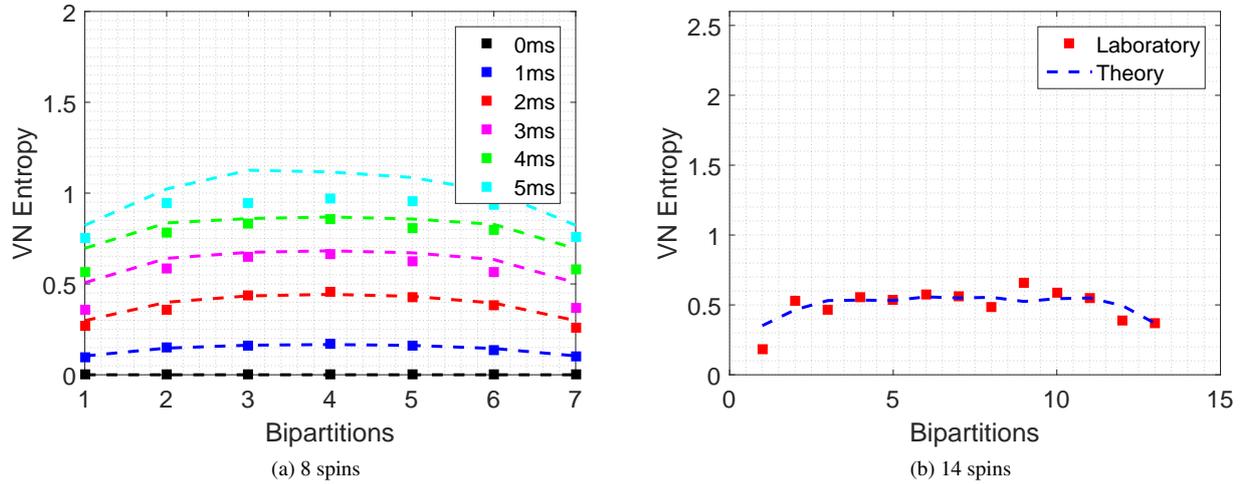

(a) 8 spins

(b) 14 spins

FIG. 20: **Von Neumann Entropy over all bipartitions:** (a) Time evolution of the VN Entropy for a 8-spin system (time encoded in the colors). Squares: Reconstructed MPS from experimental 3-qubit tomography data. Dashed lines: Entropies of the theoretical state evolved under ideal conditions and reconstructed via MPS tomography from exact local reductions. (b) VN Entropy for a 14-spin system after 4 ms in time dynamics. Red squares: Reconstructed MPS from experimental 3-qubit tomography data. Blue dashed line: Entropies of the theoretical state evolved under ideal conditions and reconstructed via MPS tomography from exact local reductions.

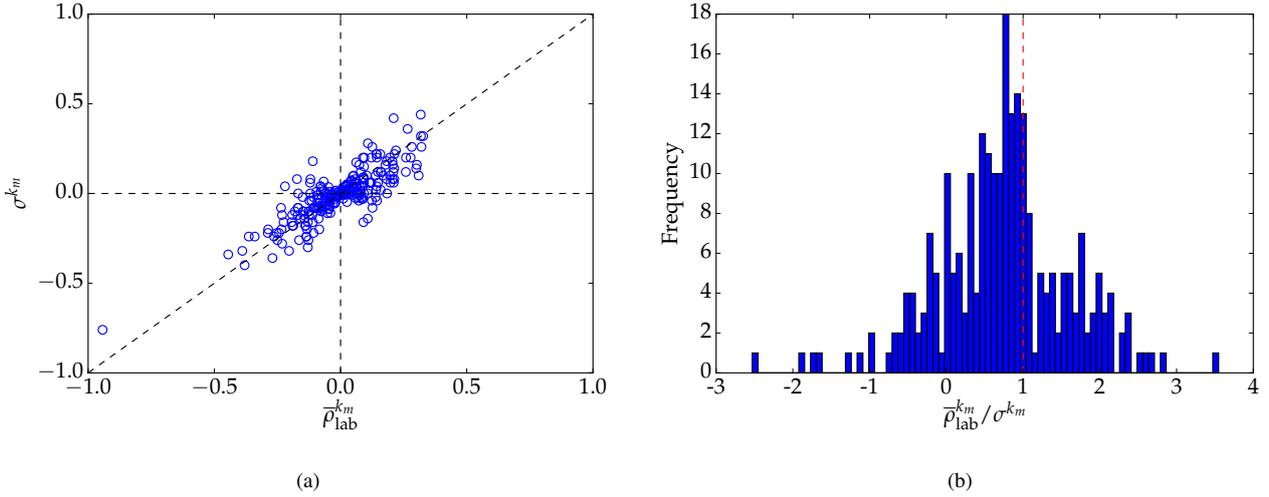

(a)                                                                (b)

FIG. 21: **Expectation values used in direct fidelity estimation** (a) A scatter plot of the expected (for MPS state) and observed (for lab state) expectation values, $\sigma^{k_m}$ and $\overline{\rho}^{k_m}_{\text{lab}}$ respectively, for the chosen $M = 250$ observables. (b) The distribution of the random variable $\overline{\rho}^{k_m}_{\text{lab}}/\sigma^{k_m}$ for the different observables. The mean and standard deviation of this distribution are the respective estimators of the fidelity estimate and its error. For our experiment, the obtained fidelity estimate is $0.74 \pm 0.05$.

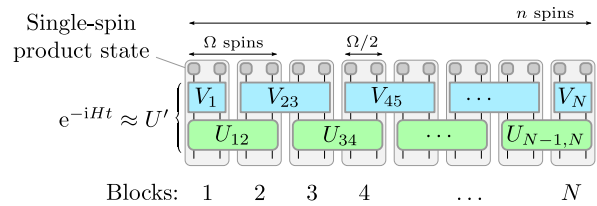

FIG. 22: The $\epsilon$-QCA decomposition from Theorem 5. A linear chain of $n$ spins is divided into $N = 2n/\Omega$ blocks, such that each blocks contains at most $\Omega/2$ spins. The local terms of the parent Hamiltonian $G$ from Theorem 6 act on four blocks, i.e., on $2\Omega$ spins.

---



%
%
%\onecolumngrid
%\appendix
%
%%\newgeometry{top=52pt,left=130pt,right=130pt,bottom=70pt}
%
%\section{appendix}
%boo

%%%%%%%%%%
\end{document}